\documentclass[pra,10pt,aps,twocolumn,preprintnumbers,amsmath,amssymb,longbibliography]{revtex4-1}
\usepackage{graphicx}
\usepackage{dcolumn}
\usepackage{bm}
\usepackage{url}
\allowdisplaybreaks

\begin{document}

\title{Quantum Density Mechanics:\\Accurate, purely density-based \textit{ab initio} implementation \\
of many-electron quantum mechanics}

\author{James C. Ellenbogen}
\email{ellenbgn@mitre.org}

\affiliation{%
Emerging Technologies Division, The MITRE Corporation, McLean, VA, USA 22102
}

\date{\today}

\begin{abstract}
This paper derives and demonstrates a new, purely density-based \textit{ab initio} approach for the calculation of the energies and properties of many-electron systems.  It is based upon the discovery of previously unappreciated relationships that govern the ``mechanics" of the electron density---i.e., relations that connect its behaviors at different points in space. Unlike wave mechanics or prior electron-density-based implementations, such as density functional theories, this density-mechanical implementation of quantum mechanics involves no many-electron or one-electron wave functions.  For this reason, for example, there is no need to calculate exchange energies, because there are no one-electron wavefunctions (i.e., orbitals) to permute or ``exchange" within two-electron integrals used to calculate electron-electron repulsion energies.  For practical purposes, exchange does not exist within quantum density mechanics.  In fact, no two-electron integrals need be calculated at all, beyond a single coulomb repulsion integral associated with the two-electron system.  Instead, a ``radius expansion method" is introduced that permits accurate determination of a parameter embodying the two-electron interaction for an $N$-electron system from that of an $(N\!-\!1)$-electron system.  Another departure is that the method does not rely upon  a Schr{\"o}dinger-like equation or the variational principle for the determination or refinement of accurate energies and densities.  Rather, the above-described methods and results follow from the derivation and solution of a ``governing equation" for each number of electrons to obtain a screening relation that connects the behavior at the ``tail" of a one-electron density, where $r\!\rightarrow\!\infty$, to its behavior at the Bohr radius, where $r\!=\!(\langle 1/r \rangle)^{-1}$. It is shown, as well, that the solution of these equations produces simple analytical expressions that deliver a total energy for a neutral two-electron atom that is nearly identical to the experimental value, plus similarly accurate energies for neutral 3, 4, and 5-electron atoms, along with formulas for accurate one-electron densities of all these atoms.  Further, these methods scale in complexity only as $N$, not as a power of $N$, as do most other accurate many-electron methods. 
\end{abstract}

\pacs{TBD}

\maketitle

\section{\label{sec:Intro} Introduction}

This paper derives and demonstrates a new, very simple, purely density-based \textit{ab initio} approach for the direct determination of the energy and the one-electron density of a many-electron system.  This new approach, which focuses upon and derives equations that govern the mechanics of how the density must vary throughout space, does not make any reference to the one-electron wavefunctions (i.e., orbitals) or many-electron wavefunctions that are intrinsic to and add much complexity to prior quantum mechanical approaches.  Only the one-electron density or representatives of its behavior are involved in the equations that describe the structure and energy of the system.  Another departure is that, for the most part, this new, density-mechanical approach or implementation for many-electron quantum mechanics does not make use of an energy functional, nor require its variational optimization, to determine an accurate energy or electron density.  Further, this new approach does not require the calculation of any two-electron integrals or functionals to represent the interelectron interaction, beyond a single coulomb repulsion integral associated with the two-electron system.  These characteristics are very unlike those of both wave-mechanical and density-functional-theory (DFT) approaches for describing atoms, molecules, and solids.

In developing this new, much simpler approach, and demonstrating it here for atoms, several other fundamental problems of atomic physics are solved, as well.  Especially, it is shown how to accurately determine the magnitude of the two-electron interaction in an $N$-electron system from that in an $(N\!-\!1)$-electron system, thereby eliminating the need for calculating any two-electron integrals, except for a single one in the case of the two-electron system. This is accomplished via a ``radius expansion method" that follows from the derivation of relations that show the precise manner in which the behavior in the ``middle" of the electron density, at finite positive radius $r\!=\!(\langle 1/r \rangle)^{-1}$ from the nucleus, connects to its well-understood bounding behavior as it approaches infinite radius from the nucleus (i.e., at its ``tail").  It is knowledge of this relationship between the middle of the density and its tail that also enables the accurate determination of the kinetic energy for a many-electron atom without any reference to a wavefunction or its second derivatives, which is otherwise demanded by the postulates of wave mechanics. 

It is further shown how these insights permit one to readily determine, with \textit{ab initio} accuracy, the total energies and one-electron densities for the four atoms considered here, He, Li, Be, and B.  This is done with very simple formulas that can be evaluated on a hand calculator or in a spreadsheet, in calculations that scale in complexity only with $N$.  This is unlike conventional, computationally intensive, \textit{ab initio}, wavefunction-based calculations that previously have been been required, which scale in complexity as a power of $N$ and which generate long, complex wavefunction and energy expansions to achieve similar accuracy.

To provide perspective on these developments, we observe that almost from the dawn of quantum mechanics, a century ago, shortly after de Broglie's assertion of wave-particle duality for electrons~\cite{debroglie1923} and Schr\"odinger's derivation of the associated wave equation~\cite{Schrodinger1926a}, it was recognized that the complicated many-electron wavefunction that results for an $N$-electron system might not be required to determine all the system's measurable properties.  That wavefunction varies in a large, abstract 3$N$-dimensional space and contains more information than is necessary.  Instead, it was appreciated that the much simpler function $\rho(\vec{r})$, describing the mean density of those electrons as it varies throughout just familiar 3-space, might be sufficient~\cite{thomas1927,fermi1928,dirac1930,Wilson1962}.  This intuition was validated and formalized three decades later via the well-known theorems of Hohenberg and Kohn~\cite{Hohenberg_and_Kohn1964}.  These showed, first, the existence of a universal energy functional $\mathcal{E}^{HK}\{\rho(\vec{r})\}$ for a many-particle quantum system.  Second, they showed that minimizing this functional would deliver the correct density function $\rho(\vec{r})$ and total energy $E_N$ for the system.  However, the theorems did not describe the form of the functional. Thus, its form remained a bit of a mystery~\cite{Epstein_and_Rosenthal1976}, especially in the case of a rapidly varying electron density, as is found in an atom or molecule.

Instead of taking on directly the difficult problem of deriving $\mathcal{E}^{HK}\{\rho(\vec{r})\}$ for such a system of interacting electrons, in the year following the publication of the Hohenberg-Kohn theorems, Kohn and Sham~\cite{Kohn_and_Sham1965} derived a clever shortcut.  They showed that, in principle, one only need find the density for a fictitious \textit{non}-interacting system of electrons, constructed or engineered so that it would have the same density as the interacting one.  Also, to implement this approach they introduced into what might have been a purely density-based theory, a set of one-electron wave equations, known as the Kohn-Sham orbital equations.  These are used to determine an intermediate set of orbitals, from which the electron density can be constructed.  The Kohn-Sham orbital equations, after several decades of work engineering energy functionals of the density that the Kohn-Sham orbitals produce, became the foundation for DFT, the widespread use of which has led to many valuable contributions~\cite{Parr_and_Yang1989,Kohn_etal1996,becke2014}.

One drawback of this approach, though, is that the choice of one of the many engineered DFT functionals is somewhat arbitrary. Different approximations necessarily are incorporated into each one and unforeseen inaccuracies can result~\cite{perdew2009}.  Thus, unlike \textit{ab initio} Hartree-Fock self-consistent field methods~\cite{roothaan1951,froese-fischer1977}, for example, there is not a single canonical set of DFT results for any system.  DFT is not an \textit{ab initio} method.

The Kohn-Sham orbitals themselves also present a fundamental problem.  Orbitals were introduced, in large part, because of the great challenge~\cite{ligneres_and_carter2005} of accurately determining the kinetic energy of interacting electrons in the absence of a wavefunction or orbital, from which one can evaluate its second derivative.  In order to ensure that the Pauli Principle is obeyed, though, one must implicitly incorporate the orbitals in a Slater determinant~\cite{slater1929} (i.e., a single-determinant, antisymmetric many-electron wavefunction).  Apart from the incongruity and inelegance of this within a supposedly density-based method, this also introduces the challenge of calculating the two-electron exchange energies that arise from the antisymmetrization, a problem so omnipresent and difficult that work addressing it~\cite{becke1988density,Perdew_et_al1992,Perdew_et_al1996} is among the most widely referenced in all modern science.

Here, however, \textit{ab initio} quantum density mechanical methods solve the first of these major challenges by permitting the accurate calculation of the kinetic energy of interacting electrons solely from their density.  Then, since it is not necessary to introduce orbitals, one sidesteps completely the second of these challenges.  There can be no exchange energy terms contributing to a total energy if there are no orbitals to permute or ``exchange" within two-electron integrals.  For practical purposes, exchange does not exist within quantum density mechanics (QDM).

Consequently, what results is an accurate, purely density-based quantum mechanical formalism, such as was implicitly promised by the Hohenberg-Kohn theorems, but not previously delivered.  The development of this formalism proceeds in two major stages.  First, in Section~\ref{sec:Derivations}, below, screening relations are derived and applied to produce simple accurate formulas that yield the ionization potentials and the dimensions for several atoms, He, Li, Be, and B.  These atoms are treated one at a time, but it is also shown how their structures are related and how key parameters for a neutral atom with $N$ electrons can be very simply calculated from those for the neutral atom with only $N\!-\!1$ electrons.  Then, in Section~\ref{sec:Densities}, those results and well-known constraints upon the form of the electron density are applied to easily construct accurate density functions for all of these atoms.  Before we can explain in detail how all this is accomplished, though, we must define some terms and present other essential foundational ideas and concepts.

\section{\label{sec:Foundations} Foundations}

\subsection{\label{sec:Ansatz} Density $Ansatz$}

For an atom having $N$ electrons and nuclear charge $Z$, the radial electron density can be represented via an $ansatz$
\begin{equation}
	\rho(r,Z) = \frac{1}{N}\sum_{i=1}^N \rho_i(r,Z)                                                        \label{eq:DensityAnsatz}
\end{equation}
that consists of a sum of radial density components $\rho_i$, one for each electron, each of which is itself is expanded in a set $\{\rho_{i,\nu}(r,Z)\}$ of $M_i$ normalized, one-electron exponential basis functions
\begin{subequations}
\label{eq:EBC_sum}
\begin{eqnarray}
\lefteqn{\rho_i(r,Z)}  \nonumber \\
& = & \sum_{\nu=1}^{M_i} c_{i,\nu} \rho_{i,\nu}(r,Z)  \label{eq:EBC_sum_a}\\
& = & \sum_{\nu=1}^{M_i} c_{i,\nu} \mathcal{N}_{i,\nu} r^{n_{i,\nu}} \exp(-2\xi_{i,\nu}r)\label{eq:EBC_sum_b} \, .
\end{eqnarray}
\end{subequations}
Note here, however, that we are expanding the $density$ in basis functions $\rho_{i,\nu}(r,Z)$, $not$ a wavefunction.  This is unlike the usual practice in standard quantum mechanical treatments~\cite{roothaan1951,Pauling_and_Wilson1935,Karplus_and_Porter1970,Pilar1990} of atoms and molecules, including DFT~\cite{Parr_and_Yang1989}.
Note also that we have incorporated the nuclear charge $Z$, in addition to radius $r$, as an argument of the density $\rho$ and its components $\rho_i$.  This is to account explicitly for the fact that these functions will be different in atoms with different nuclear charges (e.g., the function $\rho_2$ describing the density of the paired electrons in the $1s$ shell of an atom will be different in, say, an He atom with $Z\!=\!2$ than it is in an Li atom with $Z\!=\!3$). 

The quantity
\begin{equation}
\label{eq:NormConst}
\mathcal{N}_{i,\nu} = \frac{1}{4\pi}\frac{(2\xi_{i,\nu})^{n_{i,\nu} + 1}}{n_{i,\nu}!} 
\end{equation}
is a normalization constant, evaluated to ensure that the norm of each exponential density basis function is:
\begin{equation}
	\int_{0}^{\infty}\rho_{i,\nu}(r,Z) 4 \pi r^2dr = 1 \, .  \label{eq:Normalization}
\end{equation}
Expansion coefficients $c_{i,\nu}$ in Eq.~(\ref{eq:EBC_sum}) are to be determined in a manner that ensures each density component $\rho_i$ likewise has a unit norm, so that a unit norm will apply, as well, to the sum of components given in Eq.~(\ref{eq:DensityAnsatz}).

In practice, for each electron it will usually require more than one basis function, such as appear in Eq.~(\ref{eq:EBC_sum}), to accurately expand a density component $\rho_i(r,Z)$.  For simplicity, however, in our initial development we take there to be just one such basis function $\rho_{i,1}(r,Z)$ per electron (i.e, $M_i=1$ for all $i=1$ to $N$):
\begin{equation}
\label{eq:OneTermEBC}
\rho_i(r,Z) = \rho_{i,1}(r,Z) \,.
\end{equation}
Correspondingly, to start with, we will drop the second subscript $\nu$ from the exponents etc.\!\! that describe the electron density: e.g, $\xi_{i,1} \equiv \xi_i$ and $n_{i,1} \equiv n_i$.

In this simple case where there are just $N$ exponential basis functions, one per electron, that implies each electron is assumed to be under the influence of a Coulombic nuclear, ``external" potential component:
\begin{equation}
v_{i}(r) = -\xi_{i}/r.	 \label{eq:ScreenedPotential}
\end{equation}
These would be consistent with the exact quantum mechanical solution of Schr\"{o}dinger~\cite{Schrodinger1926a,Pauling_and_Wilson1935} for a single electron in an atom experiencing an effective or ``screened" nuclear charge $\xi_{i}$ and having principal quantum number $n_{i}$.  Thus, the density expansion of Eqs.~(\ref{eq:DensityAnsatz}) and (\ref{eq:EBC_sum}) is consistent with the wave-mechanical solution for an atom having a number of such effectively noninteracting electrons.

\subsection{\label{sec:Screening} Two Types of Screening Parameters}

In this simplified treatment, as well, the screened nuclear charge for the $i$th component or $i$th electron, as experienced by that electron at radius $r_i=(\langle 1/r \rangle_i)^{-1}$, will be given as
\begin{equation}
\xi_i = Z - s_i.	  \label{eq:ExponentFormula}
\end{equation}
Then, the expectation value of $1/r$ for the density function representing this electron in the atom is calculated:
\begin{eqnarray}
\label{eq:MeanRecipRadius}
\langle 1/r \rangle_i & = & \int_{0}^{\infty}\frac{1}{r}\rho_i(r,Z) 4 \pi r^2dr \nonumber \\
                      & = & \xi_i/{n_i^2}         \\
                      & = & \frac{(Z - s_i)}{n_i^2}     \nonumber                                                       
\end{eqnarray}
For later descriptive convenience, we shall term this quantity the mean reciprocal radius or reciprocal radius.  If $i\!\!=\!\!N$, corresponding to the highest energy, ``outermost", $N$th electron, the reciprocal radius is the reciprocal of the atom's most probable or Bohr radius, $r_B\!\!=\!\!(\langle 1/r \rangle_N)^{-1}$.  Thus, from Eq.~(\ref{eq:MeanRecipRadius}) it is seen that determining $s_N$ also determines an atom's dimensions.


\begin{table*}[ht]  

\caption{\label{tab:ParameterTable} Experimental first ionization potentials~\cite{NIST_AtomicIPs} $I_N$, \textit{ab initio} Hartree-Fock reciprocal radii~\cite{Clementi_and_Roetti1974,bunge1993} $\langle 1/r \rangle_N$, plus screening parameters $s_N$ and $\sigma_N$ calculated from Eqs.~(\ref{eq:s_formula}) and (\ref{eq:sigma_formula}), respectively, are tabulated for neutral atoms with electron numbers $N\!=\!Z\!=$\!1 through 10. $Z$ is the nuclear charge and $n$ is the principal quantum number of the highest energy, ``outermost" electron on an atom. Significantly, $s_N$ and $\sigma_N$ are evaluated using only \textit{ab initio} and experimental values, as noted in the text.}
\begin{ruledtabular}
\begin{tabular}{ccccccccc}
Atom & $N=Z$ & $n$ & $I_N$  & $\langle 1/r \rangle_N$ & $s_N$ &  $\sigma_N$ & $\Delta s_N$  &  $\Delta \sigma_N$\\
\hline
  &    &    &  (Hartree)  &  (Bohr$^{-1}$)  & \\
\hline
\noalign{\vskip 0.1cm} 
H  &  1  &  1  &  0.4997  &  1  &  0.0000  &  0.0005  &     &    \\
He  &  2  &  1  &  0.9036  &  1.6873  &  0.3127  &  0.9290  &  0.3127  &  0.9284 \\
Li  &  3  &  2  &  0.1981  &  0.3454  &  1.6184  &  1.8527  &  1.3057  &  0.9237 \\
Be  &  4  &  2  &  0.3426  &  0.5225  &  1.9099  &  2.6887  &  0.2915  &  0.8360 \\
B  &  5  &  2  &  0.3049  &  0.6050  &  2.5800  &  3.9919  &  0.6701  &  1.3033 \\
C  &  6  &  2  &  0.4138  &  0.7835  &  2.8660  &  4.9437  &  0.2860  &  0.9518 \\
N  &  7  &  2  &  0.5341  &  0.9577  &  3.1692  &  5.8846  &  0.3033  &  0.9409 \\
O  &  8  &  2  &  0.5005  &  1.1111  &  3.5556  &  7.0992  &  0.3863  &  1.2146 \\
F  &  9  &  2  &  0.6403  &  1.2717  &  3.9133  &  7.9930  &  0.3577  &  0.8938 \\
Ne  &  10  &  2  &  0.7925  &  1.4354  &  4.2586  &  8.8958  &  0.3453  &  0.9027 \\

\end{tabular}
\end{ruledtabular}
\end{table*}


In the expressions above, the screening parameter $s_i$ approximates and encapsulates the effect upon the $i$th electron of its Coulomb repulsion interactions with the other $N\!-\!1$ elections.  The use of this screening parameter permits the density of the $i$th electron and quantities arising from it to be approximated using equations of the same form as would apply for a single-electron system or for that electron in a system of $N$ independent electrons, each of which only interacts with the screened nuclear charge, but not explicitly with each other.

Screening parameters, such as the one we use above, arise from electrostatic considerations and have been employed in atomic and molecular physics by many, many investigators, dating back to their introduction in the 1920s and 1930s by Pauling~\cite{pauling1927}, Slater~\cite{slater1930}, Eckart~\cite{eckart1930}, and Zener~\cite{zener1930}. More recent notable contributions, influential upon the thinking of the present author, were made by Clementi and Raimondi~\cite{Clementi_and_Raimondi1963}, who calculated atomic screening parameters from \textit{ab initio} Hartree-Fock results, as well as by Bessis and Bessis~\cite{bessis1981}, who invented a technique to approximate the parameters from first principles, based upon some simple assumptions about the two-electron interaction.  Almost always, though, screening parameters have been employed as a patch, kluge, or shortcut to enable investigators to develop intuitions or to perform approximate quantum calculations, especially in an earlier era when computers were unavailable or less powerful than they are today. 

Here, though, we take screening parameters to have formal significance and build our \textit{ab initio}, density-based quantum theory around them.  Screening parameters are the representatives of the one-particle potential, as is suggested by Eqs.~(\ref{eq:ScreenedPotential}) and (\ref{eq:ExponentFormula}), for example.  Thus, founding a density-based theory around them is at least consistent with, if not mandated by the Hohenberg-Kohn theorem~\cite{Hohenberg_and_Kohn1964}, since it is based upon the proof of a one-to-one correspondence between one-particle densities and one-particle potentials.

Consequently, here we construct a kind of rigorous calculus for screening parameters in the potential, which also govern or represent the behavior of the density as it experiences different values of the potential at different distances from the nucleus of an atom.  To that end, first we shall $define$ the reciprocal radius screening parameter $s_i$, used above, via the formula
\begin{equation}
	s_i \equiv Z -n_i^2 \langle 1/r \rangle_i  \label{eq:s_formula},
\end{equation}
which follows from Eq.~(\ref{eq:MeanRecipRadius}).  From Eq.~(\ref{eq:MeanRecipRadius}), as well, it is clear that the definition in Eq.~(\ref{eq:s_formula}) establishes each screening parameter $s_i$ to be functional of the density for the $i$-th electron:
\begin{equation}
	s_i \equiv Z -n_i^2 \int_{0}^{\infty}\frac{1}{r}\rho_i(r,Z) 4 \pi r^2dr  \label{eq:s_functional}.
\end{equation}

In addition, for the highest energy $N$th electron, we define the ionization screening parameter 
\begin{equation}
	\sigma_N = Z -2I_N (\langle 1/r \rangle_N)^{-1}  \label{eq:sigma_formula}.
\end{equation}
We assume that $\sigma_N$ always delivers an exact or very near exact experimental first ionization potential~\cite{NIST_AtomicIPs}, $I_N$, when used in the equation
\begin{equation}
\label{eq:IP_formula}
\begin{aligned}
	I_N & = \frac{1}{2}(Z - \sigma_N)\langle 1/r \rangle_N   \\
	    & = \frac{1}{2n_i^2}(Z - \sigma_N)(Z - s_N). 
\end{aligned}
\end{equation}

Experimental values of $I_N$ are given in Table~\ref{tab:ParameterTable} for the first 10 atoms in the periodic table.  Just for reference, also given in the table are values of the screening parameters $s_N$ and $\sigma_N$, as determined  by means of Eqs.~(\ref{eq:s_functional}) and (\ref{eq:sigma_formula}) from accurate \textit{ab initio} Hartree-Fock calculations~\cite{Clementi_and_Roetti1974,bunge1993} of $\langle 1/r \rangle_N$ and the experimental ionization potentials.

Other quantities that figure prominently in the development below are the $changes$ in the two screening parameters for the highest energy $N$th electron, as $N$ and $Z$ each increase by 1:
\begin{subequations}
\label{eq:delta_ScreeningParameters}
\begin{eqnarray}
	\Delta s_N = s_N - s_{N-1} 	\label{eq:delta_s} \\
	\Delta \sigma_N = \sigma_N - \sigma_{N-1}.  	\label{eq:delta_sigma}
\end{eqnarray}
\end{subequations}
Reference values of these changes in the screening parameters also are given in Table~\ref{tab:ParameterTable} for the first 10 atoms in the periodic table.  

\begin{figure}[t]   
\includegraphics[height=3.0in]{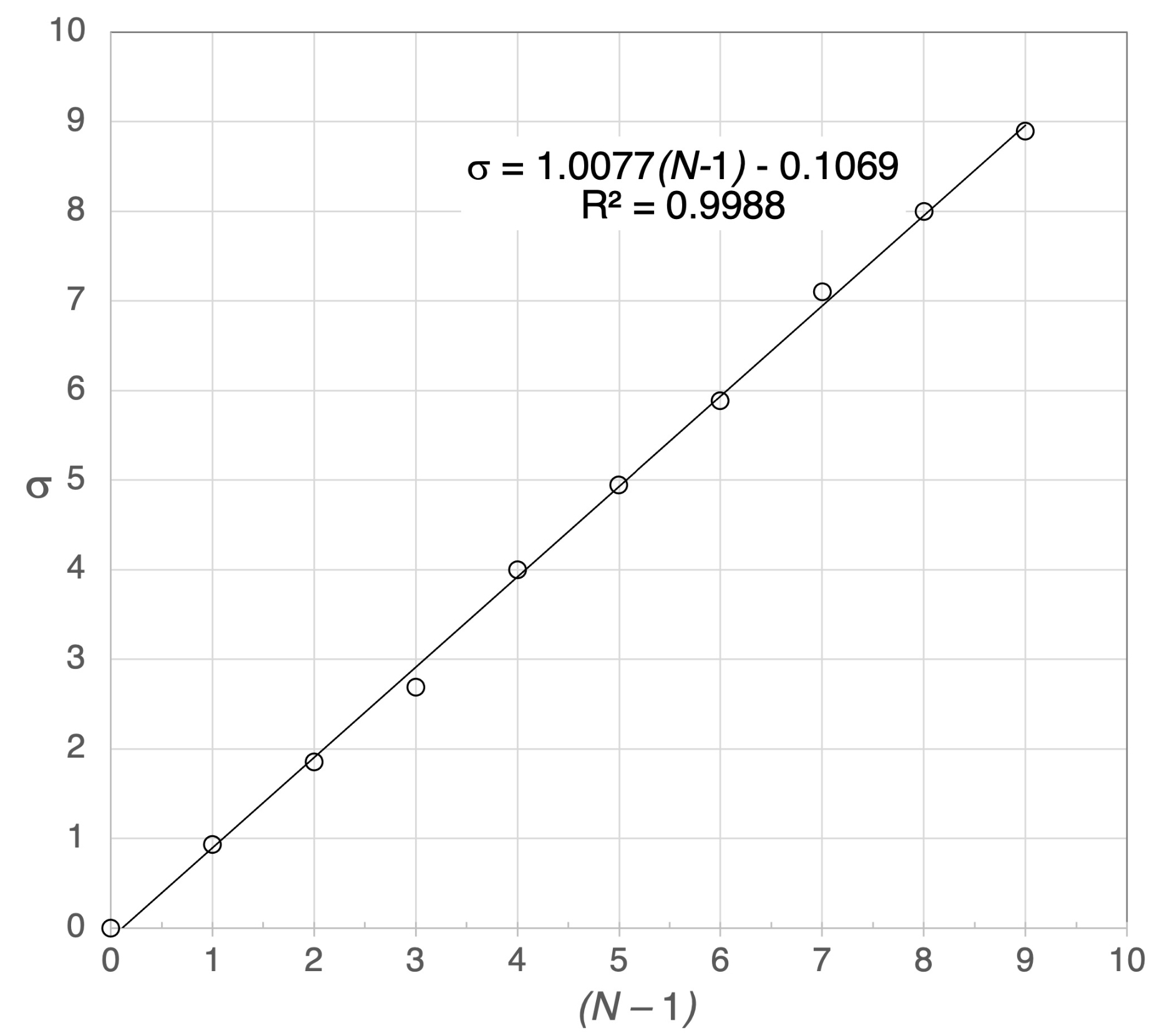}
\caption{\small Values of screening parameters $\sigma_N$, as given in Table~\ref{tab:ParameterTable}, are plotted versus $N\!-\!1$ for neutral atoms having $N$=1 through 10 electrons. The regression line with a slope so nearly equal to 1 and its small intercept, plus its large $R^2$ value, verify the validity of approximations $\sigma_N\! \approx\! N\!-\!1$ and $\Delta \sigma_N \approx 1$, which are suggested in the text.}
\label{fig:sigma_vs_N-1}
\end{figure}

In considering the behavior of $\sigma_N$ and $\Delta \sigma_N$, it proves to be important here to be aware of the dominant role of simple principles of electrostatics.  Especially, since $\sigma_N$ represents screening experienced by an electron at very large $r$, where the other $N\!\!-\!\!1$ electrons lie mostly between it and the nucleus, Gauss's theorem~\cite{Halliday_and_Resnick2001} of electrostatics suggests the common approximation:
\begin{equation}
\label{eq:SigmaApproxn}
    \sigma_N \approx N-1.	
\end{equation}
Then, Eq.~(\ref{eq:delta_sigma}) dictates the further approximation:
\begin{equation}
\label{eq:DeltaSigmaApproxn}
	\Delta \sigma_N \approx 1.
\end{equation}
The regression line in the plot of $\sigma_N$ versus $N\!-\!1$ and its large $R^2$ value, displayed within Fig.~\ref{fig:sigma_vs_N-1}, using \textit{ab initio} values of $\sigma_N$ from Table~\ref{tab:ParameterTable}, shows that these are strong approximations.  They prove useful in the analytical development described in subsequent sections, even though Table~\ref{tab:ParameterTable} also shows that in certain cases there are significant departures from these simple approximations.


\subsection{\label{sec:Constraints} Boundary Conditions and Constraints
                                    \\Upon the Density \textit{Ansatz}}

As might be inferred from Eq.~(\ref{eq:s_formula}), $s_N$ represents the screening experienced by the $N$th electron in the vicinity of the atom's Bohr radius
\begin{equation}
\label{eq:BohrRadius}
r_{B,N} \equiv (\langle 1/r \rangle_N)^{-1} \, , 
\end{equation}
in the ``middle" of its density distribution.  Also, from Eqs.~(\ref{eq:EBC_sum}) or (\ref{eq:OneTermEBC}) and (\ref{eq:ExponentFormula}), it is seen that $s_N$ will govern the behavior of the radial density in that vicinity, as well.

On the other hand, as might be inferred from Eq.~(\ref{eq:sigma_formula}), the other screening parameter $\sigma_N$ helps represent the screening experienced by that same electron far from the nucleus, at the ``tail" of the radial density where the electron is nearly ionized or removed from the atom.  Thus, $\sigma_N$ governs the behavior of the radial density in that limit.  This can be shown more clearly, starting from the well-known limiting form of the density as as $r \rightarrow \infty$
\begin{equation}
	\rho(r,Z) \sim r \exp[-2(2I_N)^{1/2}r] \, ,  \label{eq:LongRangeFunctionOfI}
\end{equation}
which first was proven by Morell, Parr, and Levy~\cite{Morrell_etal1975}, then also in further work by Katriel and Davidson~\cite{Katriel_and_Davidson1980}.
Applying Eq.~(\ref{eq:IP_formula}) to this limiting function, one obtains a limiting form that is explicitly dependent upon $\sigma_N$:
\begin{equation}
\rho(r,Z) \sim \exp\{-(2/n_i)[(Z - \sigma_N)(Z - s_N)]^{1/2}\}r \, .  \label{eq:LongRangeFunctionOfsigma}
\end{equation}

In addition to being constrained to take on this limiting form at its tail, far from the nucleus, an atomic density function like that given in Eq.~(\ref{eq:DensityAnsatz}) also must be constructed to be sure it simultaneously fulfills a constraint at its ``cusp" near the nucleus, as $r\rightarrow 0$.  This cusp constraint, due to Kato~\cite{kato1957,march1986}, is written: 
\begin{equation}
-\frac{1}{2}\biggl[ \frac{1}{\rho(r,Z)} \frac{d\rho(r,Z)}{dr} \biggr]_{r \rightarrow 0} = Z.           
                                                        \label{eq:CuspCond}
\end{equation}
These two boundary conditions or constraints, in Eqs.~(\ref{eq:LongRangeFunctionOfsigma}) and (\ref{eq:CuspCond}), are readily fulfilled by the density \textit{ansatz} of Eq.~(\ref{eq:DensityAnsatz}) and are seen in the development below to  assist in the construction of atomic densities.

 The variation in the behavior of the density in different regions of space that is mentioned immediately above can be accounted for using the density $ansatz$ of Eq.~(\ref{eq:DensityAnsatz}).  Thus, implicitly, from Eq.~(\ref{eq:EBC_sum}), this corresponds to the external potential function for the atom being represented by a sum of several different terms like the one in Eq.~(\ref{eq:ScreenedPotential}).  That fulfills the requirement that the density be $V$-representable~\cite{Parr_and_Yang1989}, explicitly realizing the requisite 1:1 matching between the density and the external potential shown by Hohenberg's and Kohn's proof~\cite{Hohenberg_and_Kohn1964}. 

We observe, too, that atomic densities constructed as a sum of exponential components, as in Eq.~(\ref{eq:DensityAnsatz}), also are automatically $N$-representable~\cite{Coleman1963,Davidson1976}. That is, we are assured that $\rho(r,Z)$ could be derived from an antisymmetric $N$-electron wavefunction, in principle, even though we do not specify that wavefunction.  Thus, $\rho(r,Z)$ represents electrons distributed through space in a way that must be in conformity with the Pauli Principle.  This is because densities from Eq.~(\ref{eq:DensityAnsatz}) are everywhere positive, differentiable, and also normalized via Eq.~(\ref{eq:Normalization}), thereby fulfilling the one-electron-density $N$-representability conditions given in a theorem due to Gilbert~\cite{gilbert1975}.

\subsection{\label{sec:One-electronEnergies} Approximate One-electron Energies\\
                                       and Total Energies}

In Section~\ref{sec:Derivations}, the screening parameters defined above are applied to calculate very accurate energies for atoms.  Taking a key foundational step in that direction, it is shown here that screening parameters $s_i$ readily permit the expression of approximate one-electron and total energies, as well as approximate one-electron kinetic energies and total kinetic energies, solely as functionals of the density.  These energies are approximate for the reason that they account for the interaction with other electrons only approximately, via a screened nuclear potential given by Eqs.~(\ref{eq:ScreenedPotential}) and (\ref{eq:ExponentFormula}).

As a very important aspect of this development, we observe (as did Thomas and Fermi~\cite{thomas1927,fermi1928,Parr_and_Yang1989}) that even though the kinetic energy of a particle in quantum mechanics is most often calculated from the second derivative, or curvature, of its wavefunction, that curvature arises from and is tightly linked to the level of spatial constraint imposed upon the particle's motion by the external potential in which it moves. Thus, the kinetic energy of a non-interacting particle can be calculated exactly from that external dimensional constraint.  For example, in the simplest of exactly soluble quantum systems, the particle in a box~\cite{Pauling_and_Wilson1935,Karplus_and_Porter1970} of length $L$, the kinetic energy goes as $1/L^2$. In atomic units, for a particle of unit mass, it is exactly $T\!\!=\!\!(n^2 \pi^2)/(2L^2)$, where $n$ is a quantum number.

Analogously, in the case of a 1-electron atom, its single electron moving in a Coulomb potential or a screened Coulomb potential, like that in Eq.~(\ref{eq:ScreenedPotential}), has a kinetic energy that may be written exactly in terms of the reciprocal of the square of the Bohr radius for the $i$th electron:
\begin{equation}
\label{eq:BasicKEeqn}
\begin{aligned}
	\langle t \rangle_i 
	    & = \frac{n_i^2}{2r_{B,i}^{2}} = \frac{n_i^2}{2}(\langle 1/r \rangle_i)^2.     \\
	    & = \frac{n_i^2}{2} \bigg[\int_{0}^{\infty}\frac{1}{r}\rho_i(r,Z) 4 \pi r^2dr \bigg]^2, 
\end{aligned}
\end{equation}
where $n_i$ is a principal quantum number.  Note that the second line of this equation expresses the kinetic energy of a noninteracting particle as a functional of its density.

The expression for the kinetic energy in Eq.~(\ref{eq:BasicKEeqn}) may be rationalized or derived directly from the basic precepts of wave mechanics, as set forth by de Broglie.  From his 1923 work~\cite{debroglie1923}, the momentum of a particle is given by $p_i\!=\!h/\lambda_i$, where $h$ is Planck's constant and $\lambda_i$ is the wavelength of the wave associated with the particle's motion.  Also, he asserted that, in an atom, a stable ``orbit" for a single electron at a Bohr radius $r_{B,i}$ must be associated with a standing wave such that the orbit's circumference consists of an integral number $n_i$ of these wavelengths, or $n_i \lambda_i = 2\pi r_{B,i}$, so that:
\begin{equation}
	p_i = \frac{h}{\lambda_i} = \frac{n_ih}{2 \pi r_{B,i}}.
\end{equation}
Then, calculating the kinetic energy for the electron of mass $m_e$ as $t_i = p_i^2/2 m_e$, and using atomic units where $m_e = h/2\pi = 1$, one arrives at the first line of Eq.~(\ref{eq:BasicKEeqn}).  Its second line, which expresses the 1-electron kinetic energy explicitly in terms of the density, then follows from Eq.~(\ref{eq:MeanRecipRadius}).

A calculation of the kinetic energy for that electron, such as by the means above, also can be used to evaluate a system's total one-electron energy $\varepsilon_i$ via the virial theorem.  This states that~\cite{Johnson_and_Pedersen1974,Pilar1990}:
\begin{equation}
    \langle v \rangle_i = (2/m)\langle t \rangle_i \, , \label{eq:VirialEq}
\end{equation}
if the force between any two particles of the system (e.g, the nucleus and an electron) results from a potential energy function of the form
\begin{equation}
\label{eq:GeneralPotl}
v(x) = kx^m,	
\end{equation}
where $x$ is a general interparticle distance, while $k$ and $m$ are constants.  In the case of the screened coulombic potential of Eq.~(\ref{eq:ScreenedPotential}), $x\!=\!r$ and $m\!=\!-1$, so that
\begin{equation}
    \langle v \rangle_i \!=\!-2 \langle t \rangle_i   	\label{eq:CoulombVfromT}
\end{equation}
and the total one-electron energy is:
\begin{equation}
	\varepsilon_i =  \langle t \rangle_i +  \langle v \rangle_i = -\langle t \rangle_i.
	                              	        \label{eq:EfromT}
\end{equation} 

Applying this equation, in combination with Eqs.~(\ref{eq:BasicKEeqn}) and (\ref{eq:MeanRecipRadius}), in the specific instance of an independent $1s$ electron, with principal quantum number $n_i\!\!=\!\!1$, we may write
\begin{equation}
\label{eq:1sKE}
\varepsilon_i = -\langle t \rangle_i = -\frac{1}{2}(Z - s_i)^2 \, .	
\end{equation}
For an independent $2s$ electron, with principal quantum number $n_i\!=\!2$, we may write
\begin{equation}
\label{eq:2sKE}
\varepsilon_i = -\langle t \rangle_i = -\frac{1}{8}(Z - s_i)^2 \, .		
\end{equation}
More generally, we observe that a set of $\varepsilon_i$ for $N$ effectively noninteracting electrons may be used to calculate an $approximate$ total energy for an atom
\begin{equation}
\label{eq:ApproxTotalEnergy}
\begin{aligned}
\widetilde{E}_N(Z) = & \sum_{i=1}^N \varepsilon_i \\
                   = & -\sum_{i=1}^N  \frac{n_i^2}{2}(\langle 1/r \rangle_i)^2 \, .
\end{aligned}
\end{equation}

Note that the development above---esp., Eqs.~(\ref{eq:CoulombVfromT}) through (\ref{eq:ApproxTotalEnergy})---assumes that electrons are in $s$-shells, with zero angular momentum.  It must be modified somewhat for $p$-type electrons, having nonzero angular momentum.  In particular, a $2p$ valence electron, such as is in the 5-electron boron atom, will have a one-electron energy with a much smaller coefficient than would be predicted by the equations above: 
\begin{equation}
\label{eq:2pKE}
\varepsilon_5 = -\frac{2}{3}(\langle 1/r \rangle_5)^2 = -\frac{1}{24}(Z - s_5)^2 \, .	
\end{equation}
The algebraic explanation of this, while straightforward, is a bit lengthy and, thus, is reserved for Appendix~\ref{sec:AppxB}.

The foundational precepts set forth above, throughout Section~\ref{sec:Foundations}, now will be applied to derive and demonstrate for two-electron, three-electron, four-electron, and five-electron atoms the accurate, purely density-based \textit{ab initio} quantum mechanical method promised in the Introduction.


\begin{table*}[ht]  

\caption{\label{tab:ScreeningVsZTable} Values from \textit{ab initio} calculations that demonstrate  (a) the invariance with $Z$ of the 2-electron screening parameter $s_2$ for atoms with $N\!=\!2$, and also (b) the near constancy of the ratio $\sigma_2/s_2\approx 3$ for all those species, both of which are commented upon in the text.  These values are extracted from the Hartree-Fock SCF calculations of Clementi and Roetti~\cite{Clementi_and_Roetti1974} upon 2-electron atoms, including the 2-electron anion with $Z\!\!=\!\!1$, the neutral 2-electron atom with $Z\!\!=\!\!2$, and 2-electron atomic cations with $Z\!>\!2$. Ionization potentials $I_2(Z)$ are expressed in Hartree units and $\langle 1/r \rangle_2$ in Bohr$^{-1}$.}
\begin{ruledtabular}
\begin{tabular}{crcccccccccc}
Footnotes & \\
to Sources  &  $Z=$  &  1  &  2  &  3  &  4  &  5  &  6  &  7  &  8  &  9  &  10 \\
\hline
\noalign{\vskip 0.1cm} 
$a$  &  $s_2$  &  0.3125  &  0.3127  &  0.3124  &  0.3126  &  0.3125  &  0.3125  &  0.3124  &  0.3124  &  0.3125  &  0.3125 \\
$b$  &  $\sigma_2$  &  0.9194  &  0.9290  &  0.9314  &  0.9325  &  0.9331  &  0.9331  &  0.9325  &  0.9319  &  0.9297  &  0.9273 \\
  &  (1/3)$\sigma_2$  &  0.3065  &  0.3097  &  0.3105  &  0.3108  &  0.3110  &  0.3110  &  0.3108  &  0.3106  &  0.3099  &  0.3091 \\
  &  $\sigma_2/s_2$  &  2.9420  &  2.9706  &  2.9814  &  2.9836  &  2.9856  &  2.9862  &  2.9846  &  2.9836  &  2.9750  &  2.9668 \\
$c$  &  $\langle 1/r \rangle_2$   &  0.6875  &  1.6873  &  2.6876  &  3.6874  &  4.6875  &  5.6875  &  6.6876  &  7.6876  &  8.6875  &  9.6875 \\
 $d$  &  $I_2(Z)^{exp'tal}$  &  0.0277  &  0.9036  &  2.7797  &  5.6556  &  9.5319  &  14.4089  &  20.2883  &  27.1684  &  35.0556  &  43.9459 \\
$e$  &  $I_2(Z)^{theory}$   &  0.0215  &  0.8958  &  2.7719  &  5.6461  &  9.5213  &  14.3966  &  20.2722  &  27.1487  &  35.0217  &  43.8956  \\
\end{tabular}

\footnotetext[1] {Calculated via Eq.~(\ref{eq:s_formula}), using values given in the table, with $n_i$=1.}
\footnotetext[2] {Calculated via Eq.~(\ref{eq:sigma_formula}), using the experimental ionization potentials and other values given in the table, with $n_i$=1.}
\footnotetext[3] {$\langle 1/r \rangle_2$ values calculated by the author from the Hartree-Fock results of Clementi and Roetti~\cite{Clementi_and_Roetti1974}.}
\footnotetext[4] {Experimental values from reference~\cite{CRC_Handbook2004}.}
\footnotetext[5]{Calculated via Eq.~(\ref{eq:ExactI2Expansion}), a formula derived in this work, using values of $s_2$ given in the table.}

\end{ruledtabular}
\end{table*}

\section{\label{sec:Derivations} Accurate\\ Energies and Dimensions\\
From Screening Relations}
\vspace{-0.25 cm}

This section introduces, derives, and applies screening relations for each of the neutral atoms He, Li, Be, and B, which are treated one at time, in Sections~\ref{sec:N=2} through \ref{sec:N=5}, respectively, with a summary of all the results presented in Section~\ref{sec:AtomResultsSummary}.  Each of the screening relations connects the behavior of an atom's density at long range to that in the middle, at the most probable radius, where $r\!=\!(\langle 1/r \rangle)^{-1}$.  It is shown that screening relations permit the derivation of simple, accurate formulas that yield the energies and dimensions for each of the atoms, while dispensing almost entirely with many of the ingredients thought to be essential to many-electron quantum calculations upon atoms: e.g., the Schr\"odinger equation, many-electron wavefunctions, orbitals, and two-electron integrals.  Instead, only screening parameters and electron densities are involved.

Although the atoms with $N\!\!=\!\!2$ to 5 electrons each are treated individually, the analysis for each atom with more than two electrons builds upon the treatment of the atom with one less electron and shows the relationship between them.  This is permitted, in large part, by the introduction of the novel and very important radius expansion method within the treatment of the Li atom in Section~\ref{sec:N=3}.  There, it enables calculation of values for the screening parameters and interelectron interaction energy of Li from those of He.  The method is then used for Be and B, as well.  This sequence of interlinked analyses for the four atoms begins with that for smallest of them, He, which also establishes the steps for treating them all, as are summarized in Section~\ref{sec:LessonsFm2e}.

\subsection{\label{sec:N=2} Helium ($N\!=\!2$)}

\subsubsection{\label{2e-Observation} Empirical Observation of a 2-electron Screening Relation}

We start by considering the He atom, primarily because it is the simplest neutral, many-electron atom.  However, it also is an appropriate place to start, because a consideration of the screening parameters for this atom was the genesis of the work described throughout this paper.  Following publication of a 2006 article~\cite{Ellenbogen_PRA2006} describing empirically-derived scaling relations for the quantum capacitances of atoms, the author undertook an effort to derive these relations from first principles, a problem that proved more difficult than anticipated.  As part of that effort, though, he built an extensive table containing atomic screening parameters  $s_N$ and $\sigma_N$, as well as other atomic properties, for many atoms, of which Table~\ref{tab:ParameterTable} is just a small part.

In that table, it was noticed that $\sigma_2$ appeared to be almost exactly three times $s_2$, as is also seen here in Table~\ref{tab:ParameterTable}.  Further calculations, based upon Hartree-Fock data for atomic cations~\cite{Clementi_and_Roetti1974} and their experimental ionization potentials~\cite{CRC_Handbook2004}, revealed that this relation held true, as well, for a number of 2-electron atomic cations with $Z>2$ and the 2-electron anion with $Z\!=\!1$.  (See Table~\ref{tab:ScreeningVsZTable}.) Thus, on the face of it, the relation
\begin{equation}
\sigma_2 \approx 3 s_2	\label{eq:BasicScreeningRatio}
\end{equation}
appeared to be fundamental, not merely incidental.  Also, it was almost immediately recognized that, if it were grounded in basic principles, the screening relation in Eq.~(\ref{eq:BasicScreeningRatio}) would have further important consequences.

\subsubsection{\label{2e-Proof} Proof of the 2-electron Screening Relation \\ 
                                $\sigma_2 \approx 3 s_2$}

However, could this ratio between the screening parameters be proven on fundamental grounds? After some struggle, it was.  All variants of the proof depend upon our expressing the ionization potential $I_2$ for the He atom in two different ways.

Preliminarily, we remind ourselves of the well-known~\cite{Bohr1913a,Schrodinger1926a,Pauling_and_Wilson1935,Karplus_and_Porter1970,Pilar1990} exact quantum energy formula for a 1-electron atom or ion with nuclear charge $Z$,
\begin{equation}
    E_1= -I_1 = -\frac{1}{2}Z^2 ,	   \label{eq:E_1}
\end{equation}
then use Eqs.~(\ref{eq:MeanRecipRadius}) or (\ref{eq:1sKE}), along with Eq.~(\ref{eq:ApproxTotalEnergy}), to approximate the total energy of the two-electron atom:
\begin{equation}
     E_2 \approx \widetilde{E_2} = -(\langle 1/r \rangle_2)^2 = -(Z - s_2)^2.	\label{eq:AproxE_2}
\end{equation}
From these expressions, an approximation $\widetilde{I}_2$ to the ionization potential of the 2-electron atom is derived, written in terms of $s_2$ as follows:
\begin{equation}
\label{eq:Approx2electronIP}
\begin{aligned}
	 I_2 \approx \widetilde{I_2} & = E_1 - \widetilde{E_2}     \\
	                             & =  - \frac{1}{2}Z^2 + (Z - s_2)^2.      
\end{aligned}
\end{equation}
A second, different formula for $I_2$, written in terms of $\sigma_2$, follows from Eq.~(\ref{eq:IP_formula}) and is assumed to deliver the exact experimental ionization potential
\begin{equation}
	 I_2 = (1/2)(Z - \sigma_2)(Z - s_2).   \label{eq:Exact2electronIP}
\end{equation}

A key step now is to eliminate $I_2$ between the approximate formula in Eq.~(\ref{eq:Approx2electronIP}) and the exact one in Eq.~(\ref{eq:Exact2electronIP}).  That is, we take $I_2 \approx \widetilde{I}_2$, in order to derive what the author terms a ``governing relation" for the 2-electron atom:
\begin{equation}
	\frac{1}{2}(Z - \sigma_2)(Z - s_2) \approx (Z - s_2)^2 - \frac{1}{2}Z^2.
	                                         \label{eq:2electronGovEqn}
\end{equation}
This relation might be treated immediately as though it were an approximate quadratic ``governing equation" in $(Z - s_2)$ for the two-electron system
\begin{equation}
	0 = (Z - s_2)^2 - \frac{1}{2}(Z - \sigma_2)(Z - s_2) - \frac{1}{2}Z^2,   \label{eq:Quadratic2electronGovEqn}
\end{equation}
and solved accordingly.

However, to obtain an initial, approximate result, in keeping with the approximate nature of the relation in Eq.~(\ref{eq:2electronGovEqn}), we reduce its order by multiplying on its right and left sides by $2/(Z - s_2)$ to obtain:
\begin{equation}
	(Z - \sigma_2) \approx 2(Z - s_2) - \frac{Z^2}{(Z - s_2)} \,.
	                       \label{eq:FirstOrder2electronGovEqn}
\end{equation}
Now, the denominator in the second term on the right is expanded via a binomial series, but only through first order, again in keeping with the approximate nature of the relation.  This yields:
\begin{equation}
	(Z - \sigma_2) \approx 2(Z - s_2) - (Z + s_2) \,.  \label{eq:2eFirstOrderExpansion}
\end{equation}
After cancellation of the terms in $Z$, collecting terms in $s_2$, and slight rearrangement, the expansion above readily reduces to and proves the empirically observed relation of Eq.~(\ref{eq:BasicScreeningRatio}).

Observe that in the proof above we have proceeded carefully, recognizing that the left and right sides of the reduced-order relation in Eq.~(\ref{eq:FirstOrder2electronGovEqn}) are $not$ exactly equal, but only approximately so.  That is, one might say they are just ``equal through first order".  Thus, we only expand the second term on the right through first order.  Recognition of the limits of accuracy of the governing relations or approximate governing equations we derive above and also further along in this work is key to their application.

Notable, as well, in the preceding derivation, is that it makes no reference to any wavefunctions.  It only employs quantities such as $\langle 1/r \rangle_2$ and $s_2$, which are both defined in terms of an electron density, as shown in Section~\ref{sec:Foundations} and Appendix~\ref{sec:AppxA}.

Using the density-based first approximation for the value of $s_2$ derived in  Appendix~\ref{sec:AppxA},
\begin{equation}
\label{eq:s2value}
	s_2 = 5/16 = 0.3125, 
\end{equation}
and taking $Z=2$, from Eqs.~(\ref{eq:AproxE_2}) and (\ref{eq:Approx2electronIP}) we can calculate initial approximations to the total energy and ionization potential of the He atom.  These are:
\begin{equation}
\label{eq:InitialE2Approxn}
	\widetilde{E}_2 = -(Z - 0.3125)^2 = -2.8477 \textrm{\,Hartree}.
\end{equation}	                                                                
and
\begin{equation}
\label{eq:InitialI2Approxn}
	\widetilde{I}_2 = -(Z - 0.3125)^2 - \frac{1}{2}Z^2 = 0.8477 \textrm{\,\,Hartree}.  
\end{equation}
However, these values are far from even the Hartree-Fock values in the literature~\cite{Clementi_and_Roetti1974,bunge1993}, $E_2^{H-F}\!\!=\!\!-2.8617$ Hartree and $I_2^{H-F}\!\!=\!\!0.8617$.   They are even further from the experimental value for the first ionization potential $I_2^{exp'tal}\!\!=\!\!0.903570$ Hartree and for the experimental total energy $E_2^{exp'tal}\!\!=\!\!-2.903386$ Hartree of the helium atom.  The latter is derived from the negative of the sum of the atom's experimentally determined~\cite{NIST_AtomicIPs} first ionization potential and second ionization potential, $I_1^{exp'tal}\!\!=\!\!1.99816$ Hartree.

\subsubsection{\label{sec:BetterE2} Simple, More Accurate, \\
                                    Density-Mechanical Formulas\\ 
                                    for $I_2$ and $E_2$}

By taking advantage of the density-mechanical screening relation proved above and stated in Eq.~(\ref{eq:BasicScreeningRatio}), however, it is easy to develop much more accurate, but still very simple formulas for $I_2$ and $E_2$.  We use the screening relation for the 2-electron atom, Eq.~(\ref{eq:BasicScreeningRatio}), to expand Eq.~(\ref{eq:IP_formula}), the expression for the exact ionization potential:
\begin{eqnarray}
\label{eq:ExactI2Expansion}
    I_2 & = & \frac{1}{2}(Z - \sigma_2)(Z - s_2)          \nonumber  \\
        & \approx & \frac{1}{2}(Z - 3s_2)(Z - s_2)        \nonumber  \\
        & \approx & \frac{1}{2}[Z^2 - 4Zs_2 + 3s_2^2]     \nonumber \\
        & \approx & \big[ (Z - s_2)^2 - \frac{1}{2}Z^2 \big] + \frac{1}{2}s_2^2 
\end{eqnarray}        
Substitution into this last expression of the value $s_2 = 0.3125$, from above, yields the ionization potential 
\begin{equation}
	I_2 \approx 0.8965 \textrm{\,\,Hartree}.   \label{eq:BetterI2value}
\end{equation}
This better approximate value, from the very simple formula in Eq.~(\ref{eq:ExactI2Expansion}), makes up 87$\%$ of the error in the initial approximation to $I_2$, within Eq.~(\ref{eq:InitialI2Approxn}).  Also, the value in Eq.~(\ref{eq:BetterI2value}) accounts for 83$\%$ of the correlation energy missing from $I_2^{H-F}$.


\begin{table*}[th]  

\caption{\label{tab:FormulaTable} Summary of formulas for evaluating from first principles the screening parameters $\sigma_N$ and $s_N$, for $N\!=\!2$ through 5.  These formulas, with the equation numbers shown, are derived throughout Section~\ref{sec:Derivations}.  Formulas to the left, for $\sigma_N$ are derived in Sections~\ref{2e-Proof}, \ref{sec:3eScreeningRelation}, \ref{sec:4eScreeningRelation}, and \ref{sec:5eScreeningRelation} by solution of the governing relations for $N\!=\!2$,3,4,5.  Formulas to the right for $s_N$,  with $N\!=\!3,4,5$, are determined via application of the radius expansion method in Sections~\ref{sec:s3-radius-expansion}, \ref{sec:s4-radius-expansion}, and \ref{sec:s5-radius-expansion}.  Formulas for the values of $s_2$ result, though, from iterative solution of the 2-electron governing relations within Section~\ref{sec:EvenBetterE2}.}
\begin{ruledtabular}
\begin{tabular}{c|cc|l|c|l|c}
      &     &         & \multicolumn{1}{c|}{Screening Relations: Formulas} & Eq.\;\; & \multicolumn{1}{c|}{Formulas} & Eq.\;\;\\
$N\!\!=\!\!Z$ & ~$n$ & $\ell$~~  & \multicolumn{1}{c|}{ for calculating $\sigma_N$ from $s_{N}$ and $s_{N-1}$} & No.\;\; & \multicolumn{1}{c|}{for the \textit{ab initio} evaluation of $s_N$}& No.\;\;\\
\hline
\noalign{\vskip 0.1cm} 
          &     &     &            &                &                            & \\
2         &  ~1  &  0~  &  $\sigma_2 \approx 3s_2$  &  \ref{eq:BasicScreeningRatio}\;\;  &   $s_2^{(1)} = 5/16 = 0.3125$  & \ref{eq:Initial_s2}, \ref{eq:s2valueAppx}\;\; \\
          &     &     &            &                & $s_2^{(2)} = s_2^{(1)} - (1/24)(s_2^{(1)})^2/Z  = 0.31047$ \:\:                      &  \ref{eq:Better_s2_c}\;\;\\
                    &     &     &            &      & $s_2^{(3)} = s_2^{(2)} - (1/36)(s_2^{(1)})^3/Z^2  = 0.31025$ \:\:                         & \ref{eq:EvenBetter_s2_c}\;\;\\
           &     &     &            &                &                            & \\
  \hline
          &     &     &            &                &                            & \\
          &     &     & $\sigma_3 \approx s_3 +  4s_2^2/(Z - s_2)$ &  \ref{eq:3eScreeningRelation}\;\;              &    $s_3 = Z - (Z - \sigma_{2})^2/[(Z - \sigma_{2}) + 1]$  & \ref{eq:s3ExpansionFormula}\;\;          \\
3         &   ~2  &  0~   &   Resulting first-order approximations are:\;\;  &      &     &          \\
    &    &   &  $\sigma_3 \approx s_3$ & \ref{eq:Linear_sigma3_Approxn}\;\;  &   &  \\
           &     &      &  $\Delta \sigma_3 \approx \Delta s_3 -2s_2$  &  \ref{eq:Delta_s3_thru1stOrder}\;\;             &           &  \\
          &     &     &            &                &                            & \\
\hline
          &     &     &            &                &                            & \\
4         &  ~2  &  0~&  $\Delta \sigma_4 \approx 3\Delta s_4$  &  \ref{eq:4eScreeningRelation}  &   $s_4 = Z - (Z - s_{3})^2/[(Z - s_{3}) + (1/3)]$  &  \ref{eq:s4ExpansionFormula} \\
          &     &     &   $\sigma_4 = s_4 + 2 \Delta s_4 + 4 s_2^2/(Z - s_4)	$  &  \ref{eq:2ndOrder4eScreeningRelation}              &                            & \\
          &     &     &            &                &                            & \\
\hline
          &     &     &            &                &                            & \\
5         &  ~2  &  1~&  $\sigma_5 \approx (2/3)Z + (1/3)s_5 - 4s_2^2/(Z - s_5)$  &  \ref{eq:5eScreeningRelation} &  $s_5 = Z - (3/2)(Z - \sigma_{4})^2/[(Z - \sigma_{4}) + 1]$ & \ref{eq:s5ExpansionFormula} \\
          &     &     &            &                &                            & \\
\end{tabular}

\end{ruledtabular}
\end{table*}



\begin{table*}[th]  

\caption{\label{tab:TotalEnergyFormulas} Summary of total energy formulas for atoms with $N\!=\!1$ to 5 electrons.  In each case where an ionization potential $I_N$ is included in an expression, it is taken to be calculated via the formula $I_N\!=\!(1/2n_i^2)(Z - \sigma_N)(Z - s_N)$.  Values of screening parameters $s_i$ and $\sigma_N$ are taken to be those provided in the Table~\ref{tab:ScreeningParameter_Results}, as calculated via the formulas in Table~\ref{tab:FormulaTable}.}
\begin{ruledtabular}
\begin{tabular}{c|ll|l|l|l|c}
      &     &         & \multicolumn{1}{c|}{Formulas for calculating} & Eq.\;\; & \multicolumn{1}{c|}{Formulas for calculating} & Eq.\;\; \\
$N\!\!=\!\!Z$ & ~$n$ & $\ell$~~  & \multicolumn{1}{c|}{approximate total energies $\widetilde{E}_N$}& No.\;\; & \multicolumn{1}{c|}{corrected, accurate total energies $E_{N}$}& No.\;\;\\
\hline
\noalign{\vskip 0.1cm} 
1         &  ~1. &. 0~  & \quad - &.   &    $E_1$ - (1/2)$Z^2$   & \ref{eq:E_1}\;\;  \\
2         &  ~1  &  0~  &  $\widetilde{E}_2 = -(Z - s_2)^2$  &  \ref{eq:AproxE_2}  &   $E_2 = \widetilde{E}_2 - (1/2)s_2^2$  & \ref{eq:BetterE2Approxn}\;\; \\
3         &  ~2  &  0~  &  $\widetilde{E}_3 = -(Z - s_2)^2 - (1/8)(Z - s_3)^2$\;\;  &  \ref{eq:3eApproxEnergy}  &   $E_3 = \widetilde{E}_3$  & \ref{eq:3eExactEnergy}\;\; \\                       
4         &  ~2  &  0~  &  $\widetilde{E}_4 = -(Z - s_2)^2 - (1/4)(Z - s_4)^2$  &  \ref{eq:4eApproxEnergy}  &   $E_4 = \widetilde{E}_4 + (1/2)s_2^2 - (1/8)(\Delta s_4)^2$  & \ref{eq:4eTotalEnergy}\;\; \\
5         &  ~2  &  1~  &  $\widetilde{E}_5 = \widetilde{E}_4 - (1/24)(Z - s_5)^2$  & \ref{eq:5eTotalEnergy}  &   $E_5 = \widetilde{E}_5$  & \ref{eq:5eTotalEnergy} \; \\
\end{tabular}

\end{ruledtabular}
\end{table*}



\begin{table}[ht]  

\caption{\label{tab:IPFormulas} Ionization potential formulas for atoms with $N\!=\!1$ to 5 electrons, as derived here in Sections~\ref{sec:N=2} through \ref{sec:N=5}.}
\begin{ruledtabular}
\begin{tabular}{c|ll|l|l|}
      &     &         & \multicolumn{1}{c|}{Formulas for calculating} & Eq.\;\; \\
$N\!\!=\!\!Z$ & ~$n$ & $\ell$~~  & \multicolumn{1}{c|}{approximate ionization potentials $I_N$}& No.\;\; \\
\hline
\noalign{\vskip 0.1cm} 
1         &. ~1  &  0~  & $I_1 - (1/2)Z^2$ &  \ref{eq:E_1}\footnotemark[1] \\  
2         &  ~1  &  0~  &  $I_2 = \big[(Z - s_2)^2 - (1/2)Z^2\big] + (1/2)s_2^2$ \,  &  \ref{eq:ExactI2Expansion} \\
3         &  ~2  &  0~  &  $I_3 = (1/8)(Z - s_3)^2 - (1/2)s_2^2$   &  \ref{eq:3eExactIPApproxn}   \\                       
4         &  ~2  &  0~  &  $I_4 = (1/4)(Z - s_4)^2 - (1/8)(Z - s_3)^2$ & \ref{eq:4eExactIPExpansion_Two} \\
          &     &     &   \qquad \, \, $- (1/2)s_2^2 + (1/8)(\Delta s_4)^2$  & \\
5         &  ~2  &  1~  &  $I_5 = (1/24)(Z - s_5)^2 + (1/2)s_2^2$  & \ref{eq:5eExactIPApproxn}  \\
\end{tabular}
\end{ruledtabular}
\footnotetext[1] {Well-known formula first derived by Bohr and by Schr\"odinger~\cite{Bohr1913a,Schrodinger1926a,Pauling_and_Wilson1935,Karplus_and_Porter1970,Pilar1990}.}
\end{table}


In addition, we may now use the expansion of Eq.~(\ref{eq:ExactI2Expansion}), along with the well-known formula of Eq.~(\ref{eq:E_1}), in a sum of the He ionization potentials.  After a small amount of algebraic manipulation, the result is a better expression for the total energy of the He atom, one that has an amazingly simple form:
\begin{equation}
\label{eq:BetterE2Approxn} 
\begin{aligned}
    E_2 = -\langle T \rangle_2 & = -(I_1 + I_2)                            \\    
        & = -\frac{1}{2}Z^2 - \frac{1}{2}[Z^2 - 4Zs_2 + 3s_2^2]      \\
        & = -(Z - s_2)^2 - \frac{1}{2}s_2^2 \, .
\end{aligned}
\end{equation}        
Also, as would be expected from the above-stated accuracy of Eqs.~(\ref{eq:ExactI2Expansion}) and (\ref{eq:BetterI2value}), this very simple formula is fairly accurate, too, delivering the value $E_2 \approx -2.896484$ Hartree, versus the experimental value of $E_2^{exp'tal} = -2.903386$ Hartree, an error of only $-0.006902$ Hartree.

Of importance in Eq.~(\ref{eq:BetterE2Approxn}), as well, the total kinetic energy $\langle T \rangle_2$ of the fully interacting two-electron atom, including correlation, has been calculated with much accuracy.  Further, this has been accomplished without any reference to one-electron wavefunctions, many-electron wavefunctions, or their second derivatives.

In that connection, it is to be emphasized that although the accurate total two-electron energy for He, including electron correlation, is expressed in terms of a screening parameter in Eq.~(\ref{eq:BetterE2Approxn}), implicitly it is written as a functional of the one-electron density $\rho(r,Z)$ for the two-electron system:
\begin{equation}
\label{eq:E2_Density_Functional}
\begin{aligned}
E_2 & = -\bigg(\int_{0}^{\infty}\frac{1}{r}\rho(r,Z) 4 \pi r^2dr\bigg)^2  \\
	&    \quad \quad -\frac{1}{2} \bigg(Z - \int_{0}^{\infty}\frac{1}{r}\rho(r,Z) 4 \pi r^2dr\bigg)^2.
\end{aligned}
\end{equation}
This transformation of Eq.~(\ref{eq:BetterE2Approxn}) follows directly from Eq.~(\ref{eq:s_functional}), recognizing that for the ground state of the He atom $\rho(r,Z)=\rho_2(r,Z)$.

The screening relations, energy formulas, and ionization potential formulas derived above for the 2-electron atom also are presented in Tables~\ref{tab:FormulaTable}, \ref{tab:TotalEnergyFormulas}, and \ref{tab:IPFormulas}.  Analogous formulas derived below for atoms with $N\!=\!3$, 4, and 5 electrons are summarized in those tables, as well.

In both Eqs.~(\ref{eq:ExactI2Expansion}) and (\ref{eq:BetterE2Approxn}), the term $(1/2)s_2^2$ accounts for the error in the initial approximations to the ionization potential $\widetilde{I}_2$ and total energy $\widetilde{E}_2$, making up the difference between those estimators and the more accurate formulas for the energies.  That is, the term $(1/2)s_2^2$ accounts for the correlation energy (and a little more).  Further, as is seen below, this term or expression recurs to serve a similar role within accurate energy formulas for atoms with greater numbers of electrons.  For this reason, as well as for the calculation of the energies in the 2-electron case, it is important to obtain an improved estimate for the parameter $s_2$.


\begin{table*}[t]  

\caption{\label{tab:E2Results} Total energies $E_2$ calculated for the He atom in this work compared with those from experiment, from canonical Hartree-Fock calculations, and from correlated wavefunction-based calculations by prior investigators. In the tabulated results it can be seen that the best of the simple density-based formulas derived in this work delivers a value for the He atom total energies and correlation energies $E_2$(Correlation) that are near exact and more accurate than those from much more complex wavefunction-based methods. For each value of $E_2$, the correlation energy is calculated $E_2$(Correlation) $= E_2 - E_2^{H-F}$, with the maximum or ``total" correlation energy defined as $(E_2^{exp'tal} - E_2^{H-F})$, as is customary. Energies are given in Hartrees.}
\begin{ruledtabular}
\begin{tabular}{rcccl}
  & He Total   &   Difference  &  Pct. of Total\\
  &    Energy  &  from  &  Correlation\\
\multicolumn{1}{c}{Nature of Calculation} &  $E_2$  &  Experiment  &  Energy & Sources \\
\hline
\noalign{\vskip 0.1cm}
\multicolumn{5}{c}{\underline{From This Work, with $E_2=-(Z - s_2)^2 - (1/2)s_2^2$}} \\
\noalign{\vskip 0.1cm}
Using $s_2 = s_2^{(1)} = 0.3125$  &  -2.896484  &  0.006902  &  83.45 & Eqs.~(\ref{eq:s2value}),  (\ref{eq:BetterE2Approxn}), $\&$ Appendix~\ref{sec:AppxA} \\
Using $s_2 = s_2^{(2)} = 0.31047$  &  -2.902707  &   0.000679  &  98.37 & Eqs.~(\ref{eq:Better_s2}) $\&$  (\ref{eq:BetterE2Approxn}) \\
Using $s_2 = s_2^{(3)} = 0.31025$  &  -2.903383  &  0.000003  &  99.99 & Eqs.~(\ref{eq:EvenBetter_s2}) $\&$  (\ref{eq:BetterE2Approxn}) \\
\noalign{\vskip 0.1cm}
\multicolumn{5}{c}{\underline{From Experiment\footnotemark[1]}} \\
\noalign{\vskip 0.1cm}
Experimental Total Energy, $E_2^{exp'tal}$  &  -2.903386  &     &  100 & Ref.~\cite{NIST_AtomicIPs}\\
\noalign{\vskip 0.1cm} 
\multicolumn{5}{c}{\underline{From Hartree-Fock Calculations}} \\
\noalign{\vskip 0.1cm}
Hartree-Fock SCF, $E_2^{H-F}$  &  -2.861679  &  0.041706  &  0 & Refs.~\cite{Clementi_and_Roetti1974} $\&$~\cite{bunge1993} \\
\noalign{\vskip 0.1cm}
\multicolumn{5}{c}{\underline{From Correlated Wavefunction Calculations by Prior Investigators}} \\
\noalign{\vskip 0.2cm}
5-term CI Wavefunction of Taylor $\&$ Parr (1952)  &  -2.8974 \,\,\,\,  &  ~0.005956  &  86.05 & Ref.~\cite{Taylor_and_Parr1952}\\
35-term CI Wavefunction of Weiss (1961)  &  -2.90320 \, &  0.00019  &  99.55 & Ref.~\cite{Weiss1961} \\
1078-term expansion of Pekeris (1959)\footnotemark[2]  &  -2.903724  &  -0.000338  &  100.81 & Ref.~\cite{pekeris1959} \\
22000-term expansion of Aznabev \textit{et al.} (2018)\footnotemark[2]  &  -2.903724  &  -0.000338  &  100.81 & Ref.~\cite{aznabaev_etal2018} \\
\end{tabular}
\footnotetext[1]{The experimental total energy $E_2^{exp'tal}$ is calculated from the negative sum of the experimental first ionization potential of He, $E_2^{exp'tal}=0.903570$ Hartree with the experimental ionization potential of the He$^+$ cation, $E_1^{exp'tal}=1.999816$ Hartree, as they are listed online in the NIST Atomic Spectra Database~\cite{NIST_AtomicIPs}.}
\footnotetext[2]{Note that the accurate non-relativistic values of $E_2$ calculated by Pekeris (1959) and by Aznabev~\textit{et al.} (2018) are actually lower and \textit{more negative} than $E_2^{exp'tal}$.}
\end{ruledtabular}
\end{table*}

\subsubsection{\label{sec:EvenBetterE2} Iteration Method for Improvement in Calculations \\
                                        of $s_2$, $I_2$, and $E_2$} 

In a standard wave-mechanical treatment, this would usually involve a further variational optimization of an energy function or functional.  That is not necessary here.  From the results above, we already have the elements of the solution that such a procedure would deliver.

Here, to improve the approximation for $s_2$ given in Eq.~(\ref{eq:s2value}), we use the improved formula for $I_2$ given in the last line of Eq.~(\ref{eq:ExactI2Expansion}) and set it equal to the formula for $I_2$ in Eq.~(\ref{eq:Exact2electronIP}). Rearranging that result slightly yields the improved quadratic governing equation:
\begin{equation}
	0 \approx (Z - s_2)^2 - \frac{1}{2}(Z - \sigma_2)(Z - s_2) - \frac{1}{2}(Z^2 - s_2^2)\,.
											\label{eq:ImprovedQuad2eGovEqn}
\end{equation}
Now, instead of linearizing this more nearly exact equation and reducing its order in $(Z - s_2)$, as was done above to the more approximate Eq.~(\ref{eq:Quadratic2electronGovEqn}) in proving the result in Eq.~(\ref{eq:BasicScreeningRatio}), we will solve it as a quadratic equation.  First, though, we transform it by using the screening relation of Eq.~(\ref{eq:BasicScreeningRatio}) to eliminate the term in $s_2^2$, in favor of one in $\sigma_2^2$:
\begin{equation}
	0 \approx (Z - s_2)^2 - \frac{1}{2}(Z - \sigma_2)(Z - s_2) - \frac{1}{2}(Z^2 - \frac{1}{9}\sigma_2^2).
											\label{eq:TransformedQuad2eGovEqn}
\end{equation}
Then, the (+)-solution of this quadratic equation expresses the unknown value of the variable $(Z - s_2)$ in terms of 
quantities in $\sigma_2$:
\begin{eqnarray}
(Z - s_2) & \approx & \frac{1}{2}\bigg\{\frac{1}{2}(Z - \sigma_2)  \nonumber \\
	   	  &         & + \bigg[\frac{1}{4}(Z - \sigma_2)^2 + 2(Z^2 
	   	               - \frac{1}{9}\sigma_2^2)\bigg]^{1/2}\bigg\}, \;\;\;\;
	                  \label{eq:2eQuadSolution}
\end{eqnarray}

By means of a first-order binomial expansion, the square root expression in this solution is evaluated:
\begin{eqnarray}
\lefteqn{\bigg[\frac{1}{4}(Z - \sigma_2)^2 + 2(Z^2 - \frac{1}{9}\sigma_2^2)\bigg]^{1/2}} \nonumber \\
& \approx & \frac{3}{2}Z\bigg[1 - \frac{1}{9}\bigg(\frac{\sigma_2}{Z}\bigg) + \frac{1}{162}\bigg(\frac{\sigma_2}{Z}\bigg)^2\bigg]. \label{eq:SqRootExpansion}
\end{eqnarray}
Then, substitution of this expansion back in Eq.~(\ref{eq:2eQuadSolution}) and collecting terms yields the second-order screening relation:
\begin{subequations}
\label{eq:Better_s2}
\begin{eqnarray}
s_2 & \approx & \frac{1}{3}\sigma_2 - \frac{1}{216}\bigg(\frac{\sigma_2^2}{Z}\bigg)  \label{eq:Better_s2_a}  \\
    & \approx & 0.31047 \, .                    \label{eq:Better_s2_b}
\end{eqnarray}
\end{subequations}
The improved value in the second line of the Eqs.~(\ref{eq:Better_s2}), immediately above, derives from substituting into the right side of the first line the first approximation $\sigma_2 \approx 3s_2 = 0.9375$ from Eqs.~(\ref{eq:BasicScreeningRatio}) and (\ref{eq:s2value}).

This better value for $s_2$ in Eq.~(\ref{eq:Better_s2_b}) is very close to the value $s_2 = 0.3104$ obtained from a few-term configuration interaction (CI) wavefunction by Taylor and Parr~\cite{Taylor_and_Parr1952}. Further, when this improved value for $s_2$ is used in the very simple Eq.~(\ref{eq:BetterE2Approxn}) for the total 2-electron energy, one obtains:
\begin{equation}
	E_2 = -2.9027 \textrm{\, Hartree}  \label{eq:NearExactE2}
\end{equation}
This result compares favorably to the value $E_2\!=\!-2.9032$ Hartree obtained by Weiss~\cite{Weiss1961}, who employed a 35-term CI wavefunction in a very much more complex calculation than that conducted above, and producing a $much$ more complex energy expression than Eq.~(\ref{eq:BetterE2Approxn}).

Still further improvement in the total energy may be obtained if the result of Eq.~(\ref{eq:Better_s2}) for $s_2$ is used to produce the third-order expansion
\begin{equation}
(1/2)s_2^2 \approx \frac{1}{9}\sigma_2^2 - \frac{1}{324}\bigg(\frac{\sigma_2^3}{Z}\bigg),
\end{equation}
where a small term in $\sigma_2^4$ has been neglected.  If this expansion is substituted in the quadratic governing equation~(\ref{eq:ImprovedQuad2eGovEqn}), the square root expression in its solution has the  first-order binomial expansion
\begin{eqnarray}
\label{eq:ImprovedSqRootExpansion}
\lefteqn{\bigg[\frac{1}{4}(Z - \sigma_2)^2 + 2(Z^2 - \frac{1}{9}\sigma_2^2) + \frac{1}{324}\bigg(\frac{\sigma_2^3}{Z}\bigg)\bigg]^{1/2}} \nonumber \\
& \approx & \frac{3}{2}Z\bigg[1 - \frac{1}{9}\bigg(\frac{\sigma_2}{Z}\bigg) + \frac{1}{162}\bigg(\frac{\sigma_2}{Z}\bigg)^2 + \frac{1}{729}\bigg(\frac{\sigma_2}{Z}\bigg)^3\bigg] \;\;\;\;\; 
\end{eqnarray}
and the resulting solution of the quadratic governing equation is the further refined result 
\begin{subequations}
\label{eq:EvenBetter_s2}
\begin{eqnarray}
s_2 & \approx & \frac{1}{3}\sigma_2 - \frac{1}{216}\bigg(\frac{\sigma_2^2}{Z}\bigg)
                 - \frac{1}{972}\bigg(\frac{\sigma_2^3}{Z^2}\bigg)  \label{eq:EvenBetter_s2_a}  \\
    & \approx & 0.31025 .                   \label{eq:EvenBetter_s2_b}
\end{eqnarray}
\end{subequations}
In evaluating Eq.~(\ref{eq:EvenBetter_s2_a}), we have once again taken $\sigma_2 \approx 3s_2 = 0.9375$, as was done in Eq.~(\ref{eq:Better_s2_a}).

Now, using the value immediately above in Eq.~(\ref{eq:BetterE2Approxn}), one obtains
\begin{equation}
	E_2 = -2.903383 \textrm{\, Hartree}  \label{eq:EvenCloserToExactI2},
\end{equation}
a result that differs by only $3 \times 10^{-6}$ Hartree from the experimental~\cite{NIST_AtomicIPs} value, $I_2^{exp'tal}\!=\!-2.903386$ Hartree.  This simply obtained, density-based result now accounts for 99.5$\%$ of the correlation energy. In Table~\ref{tab:E2Results}, results from all the He total energy calculations in this section are summarized and compared with those by others and with the experimental value.

Another way to think about the solution process demonstrated above is as a bootstrap or iteration method that starts with an initial estimate for the screening parameter $s_2$ from the variational procedure in Appendix~\ref{sec:AppxA} and Eq.~(\ref{eq:s2value}):
\begin{equation}
s_2^{(1)}=0.3125.	\label{eq:Initial_s2}
\end{equation}
Then, using $\sigma_2 \approx 3s_2^{(1)}$, as in Eq.~(\ref{eq:BasicScreeningRatio}), the two iterative solutions of the 2-electron quadratic governing equation described above are seen to produce two successively better estimates
\begin{equation}
\label{eq:Better_s2_c}
	s_2^{(2)} = s_2^{(1)} - \frac{1}{24}\frac{(s_2^{(1)})^2}{Z} = 0.31047
\end{equation}
and
\begin{equation}
 \label{eq:EvenBetter_s2_c}
 s_2^{(3)} = s_2^{(2)} - \frac{1}{36}\frac{(s_3^{(1)})^3}{Z^2} = 0.31025 \, ,
\end{equation}
but without further use of a variational procedure.
These yield successive estimates, via Eq.~(\ref{eq:BetterE2Approxn}), for the total energy $E_2^{(1)}\!=\!-2.896484$, $E_2^{(2)}\!=\!-2.902707$, and $E_2^{(3)}\!=\!-2.903383$ that are closer and closer to the experimental He total energy $E_2^{exp'tal}\!=\!-2.903386$.  See also Table~\ref{tab:E2Results} in this connection.

Correspondingly, as the successively smaller values of $s_2^{(i)}$ yield lower and more accurate energy estimates, they also produce progressively larger values of $\langle 1/r \rangle_2$, in accordance with Eq.~(\ref{eq:MeanRecipRadius}), and thus smaller and smaller values of its reciprocal, the atomic radius.  This shrinkage of the radius is precisely what one would expect as the energy of the atom gets lower because more and more electron correlation is incorporated and permits an increasingly compact electronic structure.

Further, from the results above and in Table~\ref{tab:E2Results}, one may see that, by using the refined value of $s_2$ given in Eq.~(\ref{eq:EvenBetter_s2_c}) within Eqs.~(\ref{eq:ExactI2Expansion}) and (\ref{eq:BetterE2Approxn}), these give near-exact analytic formulas for the ionization potential and total energy of the He atom.  They also give near-exact analytic formulas for the atom's total kinetic energy $\langle T_2 \rangle\!=\!-E_2$, via the virial theorem.

In Section~\ref{sec:Densities} below, the results from this QDM treatment of the energetics and dimensions of the He atom will be applied, as well, to produce an accurate analytic expression for the atom's density.  First, however, we summarize the key steps in the above derivations for the 2-electron case, then apply them to treat atoms with more electrons.

\subsubsection{\label{sec:LessonsFm2e} Summary of the Method: \\
Observations and Lessons from the $N\!\!=\!\!2$ Case}

In concluding the QDM treatment of the energetics and dimensions for the 2-electron atom, it is important to observe and emphasize that, just as had been stated in the Introduction, the kinetic energy of this many-electron system has been calculated accurately based only on functions of the density, without any use of wavefunctions or their second derivatives.  From the virial theorem and Eq.~(\ref{eq:BetterE2Approxn}), the total kinetic energy of the system is given:
\begin{equation}
\label{eq:Total2eKE}	
\begin{aligned}
	\langle T \rangle_2 & = \!-E_2 \,=\, 2\langle t \rangle_2 + \frac{1}{2}s_2^2  \\
	                    & = (\langle 1/r \rangle_2)^2 + \frac{1}{2}s_2^2.  \\
	                    & = (Z - s_2)^2 + \frac{1}{2}s_2^2. 	
\end{aligned}
\end{equation}
Note that, just as is $\langle 1/r \rangle_2$, the term in $s_2^2$ is a functional of the density, which can be seen from Eqs.~(\ref{eq:MeanRecipRadius}) and (\ref{eq:s_formula}).
Thus, the density-based expression for the kinetic energy of two effectively $non$-interacting electrons in the atom, given by the first term on the right side in each the expressions immediately above, has been corrected by the density-based interaction or correlation term $(1/2)s_2^2$ to yield an accurate, completely density-based expression for the total kinetic energy of the fully interacting system.

Also, we observe and learn that this has been enabled by three key steps.  The first step is the calculation of an approximate first ionization potential for an N-electron system via an expression
\begin{equation}
\label{eq:ApproxNelectronIP}
	\widetilde{I}_N = E_{N-1}(Z) - \widetilde{E}_N(Z)
\end{equation}
that incorporates an approximate kinetic-energy-based formula, like that in Eq.~(\ref{eq:AproxE_2}), for the total energy $\widetilde{E}_N$ of an $N$-electron system with nuclear charge $Z$, but subtracts from it a near-exact expression $E_{N-1}$ for the $(N\!-\!1)$-electron systems having the same nuclear charge $Z$, as is done above in Eq.~(\ref{eq:Approx2electronIP}).

The second step is to equate this result, but only $approximately$, to an exact or near-exact expression $I_N$, having the form of Eq.~(\ref{eq:IP_formula}), for that same ionization potential.  This yields a governing relation---i.e., an $approximate$ governing equation,
\begin{equation}
\label{eq:GenGovEqn}
0 \approx \widetilde{I}_N(s_N) - I_N(\sigma_N) \, ,	
\end{equation}
as is done in Eq.~(\ref{eq:2electronGovEqn}) or (\ref{eq:Quadratic2electronGovEqn}), for example.  

Then, the final step is to solve the governing equation for a screening relation. That is, one solves for an (approximate) equation that expresses $\sigma_N$ in terms of $s_N$, which implicitly relates the behavior of the potential and density far from the nucleus to the behavior at the $N$th electron's most probable or Bohr radius from the nucleus, like Eq.~(\ref{eq:BasicScreeningRatio}), for example.  The screening relation may be substituted back into one of the original energy expressions to make it more accurate.  In this process, however, it is important to ensure that expansions and applications of the approximate governing equation do not treat it as an exact relation or equation, but only demand from it first-order or second-order accuracy, as is done in Eqs.~(\ref{eq:2eFirstOrderExpansion}) and (\ref{eq:SqRootExpansion}), because that is all it yields.

Specifically, we use just first-order expansions when calculating initial screening relations, but then employ these in second-order expansions when calculating energies.  (This is somewhat akin to the way in which a first-order wavefunction expansion in Rayleigh-Schro{\"o}dinger perturbation theory is sufficient to determine an energy through second or third order in a perturbation parameter.)  A single, direct higher-order expansion of a screening parameter usually will diverge.  Via successive applications of first-order expansions for a screening parameter, though, it may be possible to accurately evaluate that parameter through higher order in another parameter (e.g., as seen in the iteration scheme of Section~\ref{sec:EvenBetterE2}).

However, an exception to the use of just first-order expressions within screening relations is that for atoms with $N\!\!>\!\!2$ that have a single unpaired electron in the valence shell (i.e., doublet states), we shall retain terms in $s_2^2$.  That is because such terms provide for and represent the coupling of the valence electron to the closed shell to which it is attached, as well as to the core of the atom.

 Other terms second-order or higher-order in screening parameters or their differences, which usually represent intra-shell effects, are to be neglected, though, as they are in the $N\!=\!2$ case.  These most often are small and pairs of them often have opposite signs, as well, so their net effect would usually be minor, even if it were within the accuracy of the governing relations to include them.

By applying these lessons and steps, it is also shown below that simple, accurate density-mechanical expressions can be derived for atoms with $N\!>\!2$, with the same success as has been achieved above for the $N\!=\!2$ electron system.  (See Tables~\ref{tab:ScreeningParameter_Results} and \ref{tab:IP_Results}.)


\begin{table*}[ht]  

\caption{\label{tab:ScreeningParameter_Results} Screening parameters calculated in this work are compared with those based upon the detailed \textit{ab initio} Hartree-Fock calculations of Clementi and Roetti~\cite{Clementi_and_Roetti1974} and of Bunge~\cite{bunge1993}, in combination with experimental ionization potentials~\cite{NIST_AtomicIPs}.  Values of $s_N$ in column 5 are all calculated using the radius expansion formulas for Li, Be, and B---i.e., Eqs.~(\ref{eq:s3ExpansionFormula}), (\ref{eq:s4ExpansionFormula}), and (\ref{eq:s5ExpansionFormula})---in succession, after starting with the value for $s_2$ derived in Appendix~\ref{sec:AppxA}.  Note that those results and the corresponding results $\sigma_N$ from this work are very close in their values to those from Hartree-Fock and experiment.  However in contrast to the Hartree-Fock values, which required lengthy calculations involving the evaluation of many one-electron and two-electron integrals, the screening parameters from this work are determined via the very simple formulas just listed and others identified in the table by their numbers, as well as shown in Table~\ref{tab:FormulaTable}. Further, by means of Eq.~(\ref{eq:IP_formula}), the parameters from this work result in very accurate ionization potentials, as shown in Table~\ref{tab:IP_Results}.}
\begin{ruledtabular}
\begin{tabular}{cc|cc|ccccc|c}
& &  \multicolumn{2}{c|}{From Hartree-Fock \,\,} &   &   &   &   &      \\
&  &  \multicolumn{2}{c|}{$\&$ Experiment \,\,}   &  \multicolumn{5}{c|}{From Theory in This Work}    & \\
\hline
\noalign{\vskip 0.1cm}
&  &   &   &   &  Pct. &   &   &   Pct. \, \, &  Employed \, \\
& Neutral \,&   &   &   &  Difference &  & &  Difference \, \, &  Equation \, \\
$N$ & Atom \,&  $s_N^{H-F}$ &  $\sigma_N^{H-F}$ &  $s_N$ &  from $s_N^{H-F}$ &  $\Delta$s &  $\sigma_N$ &   from $\sigma_N^{H-F}$ \, &  Numbers \, \\
\noalign{\vskip 0.1cm} 
\hline
\noalign{\vskip 0.1cm} 
1 & H &  - &  - &  - &  - &  - &  - &  -  \\
2 & He &  0.3127 &  0.9290 &  0.3125 &  -0.1\% &  0.3125 &  0.9375 &  0.9\% & \ref{eq:s2value}, \ref{eq:s2valueAppx}, \ref{eq:BasicScreeningRatio} \,   \\
3 & Li &  1.6184 &  1.8527 &  1.6110 &  -0.5\% &  1.2985 &  1.8922 &  2.1\% & \ref{eq:s3ExpansionFormula}, \ref{eq:3eScreeningRelation}  \,  \\
4 & Be &  1.9099 &  2.6887 &  1.9035 &  -0.3\% &  0.2925 &  2.6748 &  -0.5\% & \ref{eq:s4ExpansionFormula}, \ref{eq:2ndOrder4eScreeningRelation} \, \\
5 & B &  2.5800 &  3.9919 &  2.5612 &  -0.7\% &  0.6577 &  4.0269 &  0.9\% & \ref{eq:s5ExpansionFormula}, \ref{eq:Alt5eScreeningRelation}  \, \\ 

\end{tabular}
\end{ruledtabular}
\end{table*}



\begin{table*}[th]  

\caption{\label{tab:IP_Results} First ionization potentials $I_N$ calculated using the density-mechanical \textit{ab initio} methods described in this work are compared for neutral atoms with $N\!=\!2$ to 5 with those, $I_N^{exp't}$, from experiment~\cite{NIST_AtomicIPs} and those, $I_N^{SCF}$, from $\Delta$SCF calculations.  The $\Delta$SCF values have been determined by the author from the total Hartree-Fock energies of Clementi and Roetti~\cite{Clementi_and_Roetti1974}. The quantum density-mechanical (QDM) values $I_N$ have been determined by substituting in Eq.~(\ref{eq:IP_formula}) the values of the screening parameters $s_N$ and $\sigma_N$ displayed in Table~\ref{tab:ScreeningParameter_Results}, which have calculated from simple formulas derived in this work.  Of particular note is how close the very easily calculated density-mechanical values $I_N$ are to the experimental values $I_N^{exp'tal}$.}
\begin{ruledtabular}
\begin{tabular}{cc|c|ccc|cc}
&  &   &  From &  &   &   & \\ 
&  &   & QDM Theory  &   & $I_N$ Pct.  & From  & $I_N^{SCF}$ Pct.\\ 
& Neutral \, &   & in This Work,  & Difference \, & Difference  & $\Delta$SCF  & Difference\\ 
$N$ &Atom & $I_N^{exp'tal}$ \, \, \, & $I_N$  & from $I_N^{exp'tal}$ \, &  from $I_N^{exp'tal}$  & $I_N^{SCF}$   &  from $I_N^{exp'tal}$\\ 
\noalign{\vskip 0.1cm}
\hline
&  & (Hartree)\, \, \, & (Hartree) \, &   & (Hartree) & (Hartree)  \\
\hline
\noalign{\vskip 0.1cm} 
1 & H  & 0.4997 \, \,  & 0.5  & 0.0003  & 0.1\% & -  & -   \\ 
2 & He  & 0.9036 \, \,  & \,\,0.9034\footnotemark[1]  & -0.0002  & -0.02\% & 0.8617  & -4.6\%   \\ 
3 & Li  & 0.1981 \, \,  & 0.1923  & -0.0058  & -2.9\% & 0.1963  & -0.9\%  \\ 
4 & Be  & 0.3426 \, \,  & 0.3473  & 0.0047  & 1.4\% & 0.2956    & -13.7\%  \\ 
5 & B  & 0.3049 \, \,   & 0.2967  & -0.0083  & -2.7\% & 0.2915  & -4.4\%   \\ 
\end{tabular}
\end{ruledtabular}
\footnotetext[1] {The He ionization potential is calculated via a refined version of the method used 
to determine the other values of $I_N$ in the table from QDM Theory.  See Section~\ref{sec:BetterE2} 
and Table~\ref{tab:E2Results}.}
\end{table*}



\subsection{\label{sec:N=3} Lithium ($N\!=\!3$)}

The density-mechanical analysis of the 2-electron atom above hinged on deriving a screening relation, Eq.~(\ref{eq:BasicScreeningRatio}), that connects $\sigma_2$ to $s_2$.  That is, it connects the behavior of the density at long range to its behavior in the middle of its domain, at the most probable radius.

Similarly, in the case of the neutral 3-electron Li atom, with $N\!=\!Z\!=\!3$, we shall start by deriving a screening relation that connects $\sigma_3$ to $s_3$: 
\begin{equation}
\label{eq:3eScreeningRelation}
\sigma_3 \approx s_3 + \frac{4s_2^2}{(Z - s_3)} \,.	
\end{equation}

\subsubsection{\label{sec:3eScreeningRelation} Derivation of the 3-electron screening relation}

The derivation of this 3-electron screening relation proceeds according to the steps outlined in the lessons learned from the two-electron case, as recounted in Section~\ref{sec:LessonsFm2e}.  First, taking the valence quantum number to be $n_3\!=\!2$, we write an \textit{approximate} expression for the 3-electron total energy
\begin{equation}
\label{eq:3eApproxEnergy}
\begin{aligned}
\widetilde{E}_3 & = - n_2^2(\langle 1/r \rangle_2)^2 -\frac{1}{2} n_3^2(\langle 1/r \rangle_3)_2 \\
                & = - (Z - s_2)^2 -\frac{1}{8}(Z - s_3)^2 \, .
\end{aligned}
\end{equation}
We subtract this approximate 3-electron energy expression from the accurate 2-electron energy energy expression derived above and stated in Eq.~(\ref{eq:BetterE2Approxn}), in order to establish an \textit{approximate} formula for the valence ionization potential for the 3-electron atom: 
\begin{subequations}
\label{eq:3eApproxIP}
\begin{eqnarray}
\widetilde{I}_3 & = & E_2 -\widetilde{E}_3                               \\
                & = & -\big[(Z - s_2)^2 + \frac{1}{2}s_2^2 \big]    \nonumber \\
                &   & \qquad + \big[\frac{1}{8}(Z - s_3)^2 + (Z - s_2)^2 \big]     \\
                & = & \frac{1}{8}(Z - s_3)^2 - \frac{1}{2}s_2^2 \, ,
\end{eqnarray}
\end{subequations}
To write a more accurate formula for that ionization potential we use Eq.~(\ref{eq:IP_formula}) in the 3-electron case:
\begin{equation}
I_3  =  \frac{1}{8}(Z - \sigma_3)(Z - s_3) \label{eq:3eExactIPApproxn_a} .	
\end{equation}

Then, we equate \textit{approximately} the two preceding results.  That is, we take 
$I_3 \approx \widetilde{I}_3$ to obtain the quadratic governing relation for the 3-electron atom:
\begin{equation}
\label{eq:3eGoverningRelation}
\frac{1}{8}(Z - \sigma_3)(Z - s_3) \approx \frac{1}{8}(Z - s_3)^2 - \frac{1}{2}s_2^2	
\end{equation}

Finally, proceeding much as was done in the 2-electron case, we linearize the relation above by multiplying it on the left and right sides by by $8/(Z - s_3)$, to yield the linearized 3-electron governing equation or relation:
\begin{equation}
\label{eq:3eLinearGoverningRelation}
(Z - \sigma_3) \approx (Z - s_3) - \frac{4s_2^2}{(Z - s_3)}
\end{equation}
 This expression is readily simplified to the 3-electron screening relation of Eq.~(\ref{eq:3eScreeningRelation}) by cancelling $Z$ on the left and right, then changing signs throughout. 
 
\subsubsection{\label{sec:3eScreeningRelationApproxns} Useful approximations derived from the 3-electron screening relation}

Of course, from Eqs.~(\ref{eq:3eScreeningRelation}) and (\ref{eq:3eLinearGoverningRelation}), a purely first-order solution to the linearized 3-electron governing relation is simply:
\begin{equation}
\label{eq:Linear_sigma3_Approxn}
\sigma_3 \approx s_3.	
\end{equation}
This constitutes a useful approximation in some circumstances, because the term in $s_2^2$ within Eqs.~(\ref{eq:3eScreeningRelation}) and (\ref{eq:3eLinearGoverningRelation}) is small relative to the 3-electron screening parameters---e.g., using the Hartree-Fock values of $s_2$ and $s_3$ from Table~\ref{tab:ParameterTable}, with $Z\!=\!3$, the value of the term is about 0.28, as compared with 1.62 and 1.85 for $s_3$ and $\sigma_3$, respectively.  (Of course, from Eq.~(\ref{eq:3eScreeningRelation}), the relation in Eq.~(\ref{eq:Linear_sigma3_Approxn}) becomes even more accurate as $Z$ gets larger.)

Using both the 3-electron screening relation of Eq.~(\ref{eq:3eScreeningRelation}) and the 2-electron screening relation of Eq.~(\ref{eq:BasicScreeningRatio}), one also may write, respectively:
\begin{equation}
\label{eq:3eScreeningRelation_Reversed}
	s_3 \approx \sigma_3 - \frac{4s_2^2}{(Z - s_3)}
\end{equation}
and
\begin{equation}
	s_2 \approx \sigma_2 - 2s_2.
\end{equation}  
Subtracting the second equation immediately above from the first and applying the definitions of Eqs.~(\ref{eq:delta_ScreeningParameters}), we find an important relationship between the \textit{changes} in the screening parameters as one moves from the 2-electron to the 3-electron case:
\begin{equation}
\label{eq:Delta_s3_Eqn}
	\Delta s_3 \approx \Delta \sigma_3 + \Big[ 2s_2 - \frac{4s_2^2}{(Z - s_3)}\Big]\, .
\end{equation}
A useful first-order approximation to this last relation is:
\begin{equation}
\label{eq:Delta_s3_thru1stOrder}
	\Delta s_3 \approx \Delta \sigma_3 + 2s_2.
\end{equation}

The accuracy of Eq.~(\ref{eq:Delta_s3_Eqn}) is readily verified using the Hartree-Fock values for the screening parameters and their differences given in Table~\ref{tab:ParameterTable}.  The first-order approximation, Eq.~(\ref{eq:Delta_s3_thru1stOrder}), is somewhat less accurate, though, as might be expected.

\subsubsection{\label{sec:3-electron-energies} Expansion of the ionization potential $I_3$ and total energy $E_3$ for the 3-electron atom}
        
Substitution of the 3-electron screening relation of Eq.~(\ref{eq:3eScreeningRelation}) into the accurate ionization potential equation~(\ref{eq:3eExactIPApproxn_a}) for the 3-electron case leads directly to an accurate valence ionization potential expansion expressed solely in terms of screening parameters $s_3$ and $s_2$:
\begin{equation}
\label{eq:3eExactIPApproxn}
I_3 \approx 	\frac{1}{8}(Z - s_3)^2 - \frac{1}{2}s_2^2 \,. 
\end{equation}

Taking $Z\!=\!3$, the formula of Eq.~(\ref{eq:3eExactIPApproxn}) may be tested, using the Hartree-Fock values of the screening parameters given in Table~\ref{tab:ParameterTable}, to yield the value $I_3=0.1899$ Hartree for the ionization potential of the Li atom.  This value compares favorably with the experimental~\cite{NIST_AtomicIPs} value $0.1981$ Hartree, having only -4.1\% error.

In addition, if we sum this accurate expression for the first ionization potential of the 3-electron atom with those for for the 1-electron, and 2-electron atoms, from Eqs.~(\ref{eq:E_1}) and (\ref{eq:ExactI2Expansion}), respectively, we obtain:
\begin{equation}
\label{eq:3eExactEnergy}
\begin{aligned}
E_3 & = -(I_1 + I_2 + I_3)                                            \\
    & = -\frac{1}{2}Z^2 -\big[ (Z - s_2)^2 - \frac{1}{2}Z^2 \big]     \\
    & \qquad \qquad -\big[ \frac{1}{8}(Z - \sigma_3)^2 - \frac{1}{2}s_2^2 \big]  \\
    & = -(Z - s_2)^2 - \frac{1}{8}(Z - s_3)^2 \, .
\end{aligned}
\end{equation}
Observe in this formula that the correlation terms in $s^2$ that arise from the 2-electron and 3-electron contributions are the same, but of opposite sign.  Thus, the fact that they \textit{exactly cancel}, leaving no net correlation contribution to the total energy of the 3-electron atom, is an important insight arising from this density-mechanical approach. The resulting absence of such a term arising from the two-electron interaction may be seen as a rationale for why the valence electron of Li behaves as though it is nearly independent of the 2-electron core. (We say ``nearly" here because the \textit{value} of the screening parameter $s_3$ that governs behavior of the valence electron is dependent upon that of screening parameter $s_2$ that governs the 2-electron core of the atom, as is shown below.)

Further, because of the above-demonstrated cancellation, $\widetilde{E}_3\!=\!E_3$, as is readily seen by comparing the result of Eq.~(\ref{eq:3eExactEnergy}) with the expression in Eq.~(\ref{eq:3eApproxEnergy}).  Based upon that comparison, as well, it is clear from the expansion of $\widetilde{E}_3$ in terms of the reciprocal radii expectation values, each of which is a functional of the density in accordance with Eq.~(\ref{eq:MeanRecipRadius}), that the accurate total 3-electron energy expressed in terms of the screening parameters in Eq.~(\ref{eq:3eExactEnergy}) is also implicitly a functional of the density components $\rho_2$ and $\rho_3$:
\begin{equation}
\label{eq:E3_Density_Functional}
\begin{aligned}
E_3 & = -\bigg(\int_{0}^{\infty}\frac{1}{r}\rho_2(r,Z) 4 \pi r^2dr\bigg)^2  \\
	&    \quad \quad -2 \bigg(\int_{0}^{\infty}\frac{1}{r}\rho_3(r,Z) 4 \pi r^2dr\bigg)^2.
\end{aligned}
\end{equation}

\subsubsection{\label{sec:s3-radius-expansion} Radius expansion method \\ for calculating the value of $s_3$}

Above, in Section~\ref{sec:3eScreeningRelationApproxns}, for example, in order to evaluate the energies and structure for a 3-electron atom we have made liberal use of a value for $s_3$ determined via detailed wavefunction-based Hartree-Fock calculations~\cite{Clementi_and_Roetti1974,bunge1993}.  These and other wavefunction-based calculations upon many-electron systems require the difficult, time-consuming evaluation of a large number of two-electron integrals~\cite{Pauling_and_Wilson1935,Karplus_and_Porter1970,Johnson_and_Pedersen1974,Pilar1990}.  The need to evaluate the two-electron interaction in detail, over a large orbital basis set, is a primary reason why many-electron calculations upon atoms and molecules have been so challenging and why the time to perform them scales exponentially with $N$~\cite{Kohn_etal1996,Strout_and_Scuseria1995}.

In this subsection, however, we derive and present a solution to this problem, which constitutes one of the most important results of this paper.  Specifically, we show that, beyond the 2-electron case, one need not calculate quantum mechanical two-electron integrals or energy functionals in order to determine the energy and structure of a many-electron atom.  Rather, once having derived the density-mechanical screening relations that connect $s_N$ with $\sigma_N$, and thereby $\Delta s_N$ with $\Delta \sigma_N$, one may use a very simple new method, which we term here the ``radius expansion method", to accurately calculate the $value$ of $s_3$ from the value of $s_2$ (and, more generally, the value of $s_N$ from the value of $s_{N-1}$, as is shown for $N\!>\!3$ further below).

We start by noting that, for the 3-electron atom, the mean radius or Bohr radius $r_3$ of the atom is proportional to the reciprocal of $\langle 1/r \rangle_3$.  Thus, from Eq.~(\ref{eq:MeanRecipRadius}), ignoring the factor due to the quantum numbers, we may write
\begin{equation}
\label{eq:r3_proportinality_relation}
r_3 \propto \frac{1}{Z - s_3},	
\end{equation}
where $Z=3$.  Then, using Eqs.~(\ref{eq:delta_s}) and (\ref{eq:Delta_s3_thru1stOrder}), we may expand the radius by expanding $s_3$ in terms of $s_2$, $\Delta s_3$ and $\Delta \sigma_3$:
\begin{equation}
\label{eq:s_3Expansion}	
\begin{aligned}
	s_3 & = s_2 + \Delta s_3 \\
	    & \approx 3s_2 + \Delta \sigma_3 \\
	    & \approx \sigma_2 + \Delta \sigma_3, 
\end{aligned}
\end{equation}
where the last line immediately above follows from the 2-electron screening relation of Eq.~(\ref{eq:BasicScreeningRatio}).  Note that this result is also consistent with the first-order approximation in Eq.~(\ref{eq:Linear_sigma3_Approxn}).

Now, the result in Eq.~(\ref{eq:s_3Expansion}) is substituted in the right side of Eq.~(\ref{eq:r3_proportinality_relation}), with the additional use of a first-order binomial expansion of the resulting fraction:
\begin{equation}
\label{eq:3-electron radius-expansion_1}	
\begin{aligned}
\frac{1}{Z - s_3} & \approx \frac{1}{(Z - \sigma_2) - \Delta \sigma_3} \\
                  & \approx \frac{1}{(Z - \sigma_2)} \bigg[1 + \frac{\Delta \sigma_3}{Z - \sigma_2} \bigg]
\end{aligned}
\end{equation}
While interesting, the last relation would, at first, appear to express one unknown, $s_3$, in terms of another, $\Delta \sigma_3$.  That would be true, except Eq.~(\ref{eq:DeltaSigmaApproxn}) offers a well-established and fairly accurate approximation $\Delta \sigma_3 \approx 1$.  Applying this approximation in the result of Eq.~(\ref{eq:3-electron radius-expansion_1}) yields the expression:
\begin{equation}
\label{eq:3-electron radius-expansion_2}	
\begin{aligned}
\frac{1}{Z - s_3} & \approx \frac{1}{(Z - \sigma_2)} \bigg[1 + \frac{1}{Z - \sigma_2} \bigg] \\	
                  & \approx \frac{(Z - \sigma_2) + 1}{(Z - \sigma_2)^2}, 
\end{aligned}
\end{equation}
where all the quantities on the right side now are known.

Finally, taking the reciprocal of the right and left sides of this last relation, after slight rearrangement, yields the remarkable formula:
\begin{equation}
\label{eq:s3ExpansionFormula}
\begin{aligned}
s_3 & \approx	Z - \frac{(Z - \sigma_2)^2}{(Z - \sigma_2) + 1} \\
   & \approx	Z - \frac{(Z - 3s_2)^2}{(Z - 3s_2) + 1}.
\end{aligned}
\end{equation}
Using in Eq.~(\ref{eq:s3ExpansionFormula}) the parameter values $Z\!\!=\!\!3$ and $s_2\!=\!0.3125$, as was established in the treatment of the 2-electron atom above, produces the value $s_3\!=\!1.6110$. This value is nearly the same as the value $s_3\!=\!1.6184$ that is obtained from a detailed Hartree-Fock calculation and that is displayed in Table~\ref{tab:ParameterTable}.  It differs by only $0.4\%$.

In addition, when substituted in Eq.~(\ref{eq:3eExactIPApproxn}), the value $s_3\!=\!1.6110$ produces an ionization potential $I_3=0.1923$ Hartree that is closer to the experimental value than the one that results from use of the Hartree-Fock value for $s_3$. This energy differs from the experimental value $I_3^{exp'tal}=0.1981$ Hartree by only -2.9$\%$.

Below, it will be shown that the radius expansion method operates with similar accuracy to evaluate atomic screening parameters $s_N$ for $N\!>\!3$.
Further notable here is that Eq.~(\ref{eq:s3ExpansionFormula}), the expansion formula for $s_3$, shows the dependency of this parameter that governs the structure of the 2$s$ valence shell upon the parameter $s_2$ that governs the structure of the 1$s$ core shell.  This provides much physical insight, in that it reveals specifically and clearly the way in which the valence shell is coupled to the core shell of the Li atom.  In Section~\ref{sec:Densities}, the results above from the QDM treatment of the energetics and dimensions of the Li atom will be embodied within an accurate analytic expression for the atom's density.

\subsection{\label{sec:N=4} Beryllium ($N\!=\!4$)}

The density mechanical analysis of the Be atom utilizes much the same techniques demonstrated above in the cases of the 2-electron and 3-electron atoms.  It is both guided by and affirms basic physical intuitions about a neutral 4-electron atom in its ground singlet state.

In that state, the $2s$ valence shell of the Be atom consists of two electrons paired in the same way as the two electrons in the valence $1s$ shell of the He atom, treated above in Section~\ref{sec:N=2}.  Thus, one might expect the structure of the valence density and the corresponding screening relation for Be to be similar or analogous to the one in Eq.~(\ref{eq:BasicScreeningRatio}) for He.

In fact, this analogy is in evidence in the first-order screening relation for the pair of $2s$ valence electrons in Be, which is:
\begin{equation}
\label{eq:4eScreeningRelation}
	\Delta \sigma_4 \approx 3 \Delta s_4.	
\end{equation}
The quantities $\Delta \sigma_4$ and $\Delta s_4$ express the changes in the 4-electron screening parameters from those for the 3-electron atom, in accordance with Eqs.~(\ref{eq:delta_ScreeningParameters}). The accuracy of this relation can be readily verified using, for example, the values of the Hartree-Fock screening parameters and their differences, given in Table~\ref{tab:ParameterTable}. With those substitutions, the left side has the value $\Delta \sigma_4=0.8360$, while the product of the factors on the right side of Eq.~(\ref{eq:4eScreeningRelation}) is $0.8745$, yielding a difference of just $4.6\%$.

The analogy to Eq.~(\ref{eq:BasicScreeningRatio}) can be made even stronger if we recognize that the 2-electron screening parameters, $s_2$ and $\sigma_2$, actually represent $changes$ in the screening, too.  That is, since there is no interelectron interaction or screening in the 1-electron system, $s_1\!\!=\!\!\sigma_1\!\!=\!\!0$. It follows from Eqs.~(\ref{eq:delta_s}) and (\ref{eq:delta_sigma}) that $s_2\!=\!\Delta s_2$ and $\sigma_2\!=\!\Delta \sigma_2$.  Therefore, Eq.~(\ref{eq:BasicScreeningRatio}) could be rewritten to take the \textit{very same form} as Eq.~(\ref{eq:4eScreeningRelation}):
\begin{equation}
\label{eq:2eScreeningRelationRestated}
	\Delta \sigma_2 \approx 3 \Delta s_2.	
\end{equation}

Using the screening relation of Eq.~(\ref{eq:4eScreeningRelation}), it is proven below in Section~\ref{sec:4eIPformula} that an accurate formula for the first ionization potential of Be is:
\begin{eqnarray}
\label{eq:4eExactIPExpansion_Two}
I_4 & = & \widetilde{I}_4 - \frac{1}{2}s_2^2 + \frac{1}{8}(\Delta s_4)^2  \\
    & = & [\frac{1}{4}(Z - s_4)^2 - \frac{1}{8}(Z - s_3)^2] - \frac{1}{2}s_2^2 + \frac{1}{8}(\Delta s_4)^2 \, , \nonumber
\end{eqnarray}
This formula is exact through second order in the screening parameters and their differences.

As a preliminary test of Eq.~(\ref{eq:4eExactIPExpansion_Two}), we substitute into it Hartree-Fock parameter values from Table~\ref{tab:ParameterTable} for $s_2$, $s_3$ and $\Delta s_4$, also taking $Z\!=\!4$.  This yields the value $I_4=0.3449$ Hartree.  That value compares very favorably with the experimental value $I_4^{exp'tal}=0.3426$ Hartree, having only a +0.7\% error.

This result is considerably better than the less accurate Hartree-Fock $\Delta$SCF result for this ionization potential, which is $0.2956$ Hartree and has a $-13.7\%$ error.  The numerical result from the work here slightly overcorrects, but does capture essentially all of the correlation energy.  See also Table~\ref{tab:IP_Results}.

The ionization potential formula for the 4-electron atom in Eq.~(\ref{eq:4eExactIPExpansion_Two}) can be applied, as well, to develop further the analogy between the density-mechanical equations for the 4-electron system and those for the 2-electron system.  By defining a 3-electron effective nuclear charge
\begin{equation}
Z^{eff}_3 \equiv (Z - s_3),	
\end{equation}
the expansion of the 4-electron ionization potential in Eq.~(\ref{eq:4eExactIPExpansion_Two}) can
be restated as
\begin{equation}
I_4 \approx \frac{1}{4}[(Z^{eff}_3 - \Delta s_4)^2 - \frac{1}{2}(Z^{eff}_3)^2] - \frac{1}{2}s_2^2 \, , 	
\end{equation}
where the small term in $(\Delta s_4)^2$ has been neglected.  This last expression is seen to be analogous in form to that for the 2-electron ionization potential in the last line of Eq.~(\ref{eq:ExactI2Expansion}), except for the sign of the term in $s_2^2$.
It is satisfying to see that the simple relations of Eqs.~(\ref{eq:4eScreeningRelation}) and (\ref{eq:4eExactIPExpansion_Two}) that govern paired electrons in the 2$s$ shell are so similar to the ones the derived above in Section~\ref{sec:N=2} to govern paired electrons in the $1s$ shell.

Moreover, in viewing the formula for $I_4$, it is of much interest to see the way in which the term $(1/2)s_2^2$, which represents correlation effects in the $1s$ shell, penetrates through into the $2s$ shell.  As is shown below in Section~\ref{sec:N=5}, this term and its associated core correlation effects also penetrate through into the $2p$ shell, for the boron atom, with $N\!=\!5$.  First, however, let us derive the key relations stated above in Eqs.~(\ref{eq:4eScreeningRelation}), as well as in Eqs.~(\ref{eq:4eExactIPExpansion_Two}) for this $N\!=\!4$ case.

\subsubsection{\label{sec:4eScreeningRelation} Derivation of the 4-electron screening relations}

As in the 2-electron and 3-electron cases, the 4-electron screening relation in Eq.~(\ref{eq:4eScreeningRelation}) is derived by solving the linearized or first-order governing equation for that number of electrons.  Similar to the procedure followed in the cases with fewer electrons, the first-order 4-electron governing equation is formulated by developing expansions of the quantities $\widetilde{I}_4$ and $I_4$, then asserting them to be approximately equal.

From Eqs.~(\ref{eq:1sKE}) through (\ref{eq:ApproxTotalEnergy}), we have: 
\begin{equation}
\label{eq:4eApproxEnergy}
\begin{aligned}
	\widetilde{E}_4 = & (\langle 1/r \rangle_2)^2 + (\langle 1/r \rangle_4)^2 \nonumber \\
	                = & (Z - s_2)^2 + (1/4)(Z - s_4)^2.
\end{aligned}
\end{equation}
This equation, along with Eq.~(\ref{eq:3eExactEnergy}) permit the calculation of an approximate ionization potential:
\begin{equation}
\label{eq:4eApproxIP}
\begin{aligned}
	\widetilde{I}_4 = & E_3 - \widetilde{E}_4,  \nonumber \\
	                = & (1/4)(Z - s_4)^2 - (1/8)(Z - s_3)^2
\end{aligned}
\end{equation}
while, from Eq.~(\ref{eq:IP_formula}), a more exact first ionization potential formula is
\begin{equation}
\label{eq:4eExactIP}
I_4 = (1/8)(Z - \sigma_4)(Z - s_4).	
\end{equation}

Using the relations
\begin{subequations}
\label{eq:ScreeningParameters_4Expansion}
\begin{eqnarray}
	s_4 & = & s_3 + \Delta s_{4} 	\label{eq:s_4Expansion} \\
	\sigma_4 & = & \sigma_3 + \Delta \sigma_{4}   	\label{eq:sigma_4Expansion}
\end{eqnarray}
\end{subequations}
each of the two preceding equations may be expanded just through first order in the screening parameters and their differences to yield, respectively
\begin{equation}
\label{eq:4eApproxIP_Expanded}
\widetilde{I}_4 \approx \frac{1}{8}\big[(Z - s_3)^2 - 4(Z - s_3)\Delta s_4\big]	
\end{equation}
and
\begin{equation}
\label{eq:4eExactIPExpanded}
\begin{aligned}
I_4 = & \frac{1}{8}\big[(Z - \sigma_3) - \Delta \sigma_4 \big]\big[(Z - s_3) - \Delta s_4 \big]  \\
    \approx & \frac{1}{8}\big[(Z - s_3)^2 - (Z - s_3)(\Delta \sigma_4 + \Delta s_4)\big]. \end{aligned}
\end{equation}
In the second line of Eq.~(\ref{eq:4eExactIPExpanded}) the first-order 3-electron screening relation of Eq.~(\ref{eq:Linear_sigma3_Approxn}) has been used to eliminate $\sigma_3$ in favor of $s_3$, while terms second-order and higher in the screening parameters and their differences have been neglected.

Now, using the preceding two equations to set $\widetilde{I}_4=I_4$, one may write the first-order governing equation:
\begin{eqnarray}
\label{eq:4eLinearGoverningRelation}
\lefteqn{\big[(Z - s_3)^2 - 4(Z - s_3)\Delta s_4\big]} \\ 
             & = & \big[(Z - s_3)^2 - (Z - s_3)(\Delta \sigma_4 + \Delta s_4)\big] \nonumber
\end{eqnarray}
Finally, by cancelling the like terms on the left and right of this equation, dividing the single term that remains on each side by $(Z-s_3)$, and rearranging the terms in $\Delta s_4$ and $\Delta \sigma_4$ that are left, the first-order screening relation of Eq.~(\ref{eq:4eScreeningRelation}) results. 

Below it is shown that the just-proven first order 4-electron screening relation enables derivation of the ionization potential formula for the 4-electron system in Eq.~(\ref{eq:4eExactIPExpansion_Two}).  Before undertaking that derivation, though, we show that another use for the yet-to-be-proven Eq.~(\ref{eq:4eExactIPExpansion_Two}) is to produce a quite accurate second-order formula for the screening parameter $\sigma_4$ in terms of $s_4$---i.e., a \textit{second-order} screening relation for the 4-electron system.  This is most easily shown by using Eq.~(\ref{eq:4eExactIPExpansion_Two}) with Eq.~(\ref{eq:IP_formula}) to write:
\begin{eqnarray}
\lefteqn{\frac{1}{8}(Z - \sigma_4)(Z - s_4)} \\
   & = & \frac{1}{4}(Z - s_4)^2 - \frac{1}{8}(Z - s_3)^2 - \frac{1}{2}s_2^2 + \frac{1}{8}(\Delta s_4)^2	\nonumber
\end{eqnarray}
Into the right side of this equation we substitute:
\begin{eqnarray}
\label{eq:ZMinus_s3_Identity}
(Z - s_3)^2 & = & [(Z - s_4) + \Delta s_4]^2 \\
            & = & (Z - s_4)^2 + 2(Z - s_4)\Delta s_4 + (\Delta s_4)^2 \, . \nonumber
\end{eqnarray}
Then, after expanding the right side using the above and cancelling terms in $(\Delta s_4)^2$ with opposite signs, we multiply both sides by $8/(Z - s_4)$, subtract the term in $Z$ from both sides, and change signs throughout, to obtain:
\begin{equation}
\label{eq:2ndOrder4eScreeningRelation}
\sigma_4 = s_4 + 2 \Delta s_4 + \frac{4 s_2^2}{(Z - s_4)}\,.	
\end{equation}
Substituting values from Table~\ref{tab:ParameterTable} on the right side of this second-order screening relation yields a value $\sigma_4=2.6800$, which differs from the Hartree-Fock-derived value in the table by only $-0.0087$ or only $-0.3\%$.  Because of its accuracy, this relation will  prove useful in the consideration of the 5-electron system, below in Section~\ref{sec:N=5}.

In considering all the steps above that led to the derivation of Eq.~(\ref{eq:2ndOrder4eScreeningRelation}), it can be seen that the first-order screening relation of Eq.~(\ref{eq:4eScreeningRelation}) led to the accurate energy formula Eq.~(\ref{eq:4eExactIPExpansion_Two}) and then that equation was used to derive the second-order screening relation.  Essentially, the first-order screening relation was used in a bootstrap procedure to arrive at the second-order screening relation.

\subsubsection{\label{sec:4eIPformula} Derivation and evaluation of an accurate formula for the first ionization potential of a 4-electron atom}

The first order 4-electron screening relation of Eq.~(\ref{eq:4eScreeningRelation}) may be applied, as asserted above, to derive the simple and accurate formula Eq.~(\ref{eq:4eExactIPExpansion_Two}) for the first ionization potential $I_4$ of the Be atom.  Starting with Eq.~(\ref{eq:4eExactIP}), then applying to the first factor on the right the screening parameter expansions in Eqs.~(\ref{eq:ScreeningParameters_4Expansion}), along with the 3-electron and 4-electron screening relations in Eqs.~(\ref{eq:3eScreeningRelation}) and (\ref{eq:4eScreeningRelation}), one obtains:
\begin{subequations}
\label{eq:Alt4eExactIPExpansion}	
\begin{eqnarray}
I_4 & \approx & \frac{1}{8}\Big[Z - \Big(s_3 + \frac{4s_2^2}{Z - s_3} + 3 \Delta s_4\Big)\Big](Z - s_4)
                                                  \qquad \label{eq:Alt4eExactIPExpansion_a}\\
    & \approx & \frac{1}{8}\Big[(Z - s_4) - \Big(\frac{4s_2^2}{Z - s_3} + 2 \Delta s_4\Big)\Big](Z - s_4)                                  \qquad \quad \label{eq:Alt4eExactIPExpansion_b} \\
    & \approx & \frac{1}{8}(Z - s_4)^2 - \frac{1}{4}(Z - s_4)\Delta s_4 - \frac{1}{2}s_2^2.
                                                  \qquad \label{eq:Alt4eExactIPExpansion_c}
\end{eqnarray}
\end{subequations}

The second line in the set of equations above arises from the first via the application of Eq.~(\ref{eq:s_4Expansion}).  Then, the third line follows from the neglect of all terms higher than second order in the screening parameters or their differences.  This includes the recognition that, through second-order in small quantities, one may write
\begin{equation}
\label{eq:NeglectOfHighOrder4eTerms}
\begin{aligned}
\Big(\frac{4s_2^2}{Z - s_3}\Big)(Z - s_4) & = \Big(\frac{4s_2^2}{Z - s_3}\Big)\big[(Z - s_3) - \Delta s_4 \big] \\
                               & \approx 4s_2^2.	
\end{aligned} 
\end{equation}

To continue the derivation of Eq.~(\ref{eq:4eExactIPExpansion_Two}), we further develop Eq.~(\ref{eq:Alt4eExactIPExpansion_c}) by adding and subtracting $(1/8)(Z-s_3)^2$ on the right side, while expanding the added term using the identity in Eq.~(\ref{eq:ZMinus_s3_Identity}).  After cancellation of two terms in the result that both have the form $(1/8)(Z-s_4)\Delta s_4$, but opposite signs, these steps yield and prove the formula for $I_4$ given in Eq.~(\ref{eq:4eExactIPExpansion_Two}). 

In considering still further this ionization energy formula, we see, once again, a correlation term of the form $(1/2)s_2^2$ correcting the energy, just as such a term did in the $N\!\!=\!\!2$ and $N\!\!=\!\!3$ results.  This correlation correction term is carried forward and penetrates into the $2p$ shell in the $N\!\!=\!\!5$ case, too, as is shown below in Section~\ref{sec:N=5}.

Finally, having proved it, we use the formula for $I_4$ in Eq.~(\ref{eq:4eExactIPExpansion_Two}) above and the formula for $E_3$ in Eq.~(\ref{eq:3eExactEnergy}), derived in Section~\ref{sec:3-electron-energies}, to accurately express the total energy of a 4-electron atom as:
\begin{eqnarray}
\label{eq:4eExactEnergy}
E_4 & = & E_3 -I_4                              \nonumber \\
    & = & -(Z - s_2)^2 - \frac{1}{4}(Z - s_4)^2  \\
    &   & \quad   + \frac{1}{2}s_2^2 - \frac{1}{8}(\Delta s_4)^2 \, .  \nonumber
\end{eqnarray}

\subsubsection{\label{sec:s4-radius-expansion} Radius expansion method \\ for calculating the value of $s_4$}

Using the 4-electron screening relation of Eq.~(\ref{eq:4eScreeningRelation}), it is very easy to employ the radius expansion technique, demonstrated for the $N\!\!=\!\!3$ case in Section~\ref{sec:s3-radius-expansion}, to derive a simple formula that yields a very accurate value for the screening parameter $s_4$, without having to calculate any quantum-mechanical integrals.  In analogy to the 3-electron Eq.~(\ref{eq:3-electron radius-expansion_1}), once again using Eq.~(\ref{eq:s_4Expansion}) along with a first-order binomial expansion, we write for the 4-electron case: 
\begin{equation}
\label{eq:4-electron-radius-expansion_1}	
\begin{aligned}
\frac{1}{Z - s_4} & \approx \frac{1}{(Z - s_3) - \Delta s_4} \\
                  & \approx \frac{1}{(Z - s_3)} \bigg[1 + \frac{\Delta s_4}{Z - s_3} \bigg]
\end{aligned}
\end{equation}

The value of the parameter $s_3$ is assumed to be known, of course, from the application of the radius expansion method in the 3-electron case.  Then, the unknown quantity $\Delta s_4$ can be evaluated approximately using the 4-electron screening relation and Eq.~(\ref{eq:DeltaSigmaApproxn}):
\begin{equation}
\label{eq:Deltas4Approx}
\Delta s_4 \approx \frac{1}{3} \Delta \sigma_4 \approx \frac{1}{3}
\end{equation}
Use of this approximation in the result of Eq.~(\ref{eq:4-electron-radius-expansion_1}) then produces:
\begin{equation}
\label{eq:4-electron radius-expansion_2}
\frac{1}{Z - s_4} \approx \frac{(Z - s_3) + 1/3}{(Z - s_3)^2} \, .\\
\end{equation}

Finally, taking the reciprocal of both sides of the foregoing equation and subtracting both sides from $Z$ produces the radius expansion formula for $s_4$:
\begin{equation}
\label{eq:s4ExpansionFormula}
s_4 = Z - \frac{(Z - s_3)^2}{(Z - s_3) + 1/3} \, . \\
\end{equation}
As suggested above, this formula is amazingly accurate---especially given its simplicity.  For example, using in it $Z\!\!=\!\!4$ and the Hartree-Fock \textit{ab initio} value $s_3\!=\!1.6184$ produces a value $s_4\!=\!1.9108$.  This differs by only $0.05\%$ from the \textit{ab initio} value $s_4\!=\!1.9099$, also shown in Table~\ref{tab:ParameterTable}, which resulted from a complex wavefunction calculation involving many quantum mechanical integrals.

Of course, one does not require the Hartree-Fock value for $s_3$ to use the 4-electron expansion formula in Eq.~(\ref{eq:s4ExpansionFormula}).  One can just use the value $s_3\!=\!1.6110$ derived from the 3-electron expansion formula, Eq.~(\ref{eq:s3ExpansionFormula}), and still get a value $s_4\!=\!1.9035$ that differs by only $-0.33\%$ from the \textit{ab initio} value.  Using these values of $s_3$ and $s_4$ calculated via the radius expansion formulas, along with Eq.~(\ref{eq:2ndOrder4eScreeningRelation}), one also obtains $\Delta s_4\!=\!0.2989$ and $\sigma_4\!=\!2.6751$, very close to the Hartree-Fock-based value $\sigma_4\!=\!2.6887$.

Further, from these QDM values for the screening parameters, it is possible to use Eq.~(\ref{eq:4eExactIPExpansion_Two}) to calculate $I_4\!=\!0.3449$ Hartree.  This value differs from the experimental ionization potential $I_4^{exp'tal}\!=\!0.3426$ Hartree by just $0.7\%$.

\subsection{\label{sec:N=5} Boron ($N\!=\!5$)}

Just as the closed-shell Be atom with two paired electrons was seen in Section~\ref{sec:N=4} to be described with density-mechanical equations analogous to those for the closed shell He atom, the description of the boron atom with one unpaired valence electron is seen to be somewhat analogous to that in Section~\ref{sec:N=3} for the Li atom with a single unpaired electron.  The nonzero angular momentum of the unpaired valence electron is a major difference that must be accounted for in the case of the B atom, however, as is shown in this section.

The primary results, though, still remain simple and reminiscent of those for Li.  The 5-electron screening relation is
\begin{equation}
\label{eq:5eScreeningRelation}
(Z - \sigma_5) \approx \frac{1}{3}(Z - s_5) + \frac{4s_2^2}{(Z - s_5)} \, , 
\end{equation}
where the analogy to the 3-electron relation in Eq.~(\ref{eq:3eScreeningRelation}) is clear if that one from Section~\ref{sec:N=3} is restated as
\begin{equation}
\label{eq:3eScreeningRelationRestated}
(Z - \sigma_3) \approx (Z - s_3) - \frac{4s_2^2}{(Z - s_5)} \, .
\end{equation}

Alternatively, the 5-electron screening relation of Eq.~(\ref{eq:5eScreeningRelation}) can be re-stated to be more in the form of the 3-electron relation of Eq.~(\ref{eq:3eScreeningRelation})
\begin{equation}
\label{eq:Alt5eScreeningRelation}
\sigma_5 = \frac{2}{3} Z + \frac{1}{3} s_5 - \frac{4s_2^2}{Z - s_5}	\, .
\end{equation}
However, the strong, linear $Z$-dependence and the factor of 1/3 modulating the term in $s_5$ make the 5-electron relation different.  Also, within the second form of the screening relation, above in Eq.~(\ref{eq:Alt5eScreeningRelation}), the sign of the term in $s_2^2$ relative to the sign of $s_N$ is different in this $N\!=\!5$ case from that in the $N\!=\!3$ case of Section~\ref{sec:N=3}.

Despite these differences, when the 5-electron screening relation of Eq.~(\ref{eq:5eScreeningRelation}) is substituted in the ionization potential formula of Eq.~(\ref{eq:IP_formula}), with $N\!\!=\!\!5$, principal quantum number $n_5\!\!=\!\!2$, and angular momentum quantum number $\ell_5\!=\!1$, one obtains an accurate second-order formula for the first ionization potential of the boron atom
\begin{subequations}
\label{eq:5eExactIPApproxn}
\begin{eqnarray}
I_5  & = & \frac{1}{8}(Z - \sigma_5)(Z - s_5)  \label{eq:5eExactIPApproxn_a} \\
     & \approx & \frac{1}{24}(Z - s_5)^2 + \frac{1}{2}s_2^2 \,. \label{eq:5eExactIPApproxn_b}
\end{eqnarray}
\end{subequations}
This ionization potential formula is plainly similar in form to the one in Eq.~(\ref{eq:3eExactIPApproxn}) for the first ionization potential of Li.  Once again, the only differences are the coefficient on the first term and the sign on the second.

Note again that the small coefficient (1/24) arises from the nonzero angular momentum of the valence $2p$ electron, as outlined briefly in Section~\ref{sec:Foundations}.  See Appendix~\ref{sec:AppxB} for a more detailed explanation.

As an initial test of the accuracy of the 5-electron ionization potential formula in Eq.~(\ref{eq:5eExactIPApproxn_b}), one can substitute in it the Hartree-Fock values for $s_5$ and $s_2$, from Table~\ref{tab:ParameterTable}.  The result is $I_5\!=\!0.2929$ Hartree, which differs by just $-3.9\%$ from the experimental value $I_5^{exp'tal}\!=\!0.3049$ Hartree.

\subsubsection{\label{sec:5eScreeningRelation} Derivation of the screening relation for a 5-electron atom and a formula for its first ionization potential}

The derivation of the 5-electron screening relation in Eq.~(\ref{eq:5eScreeningRelation}) proceeds in accordance with the approach outlined in Section~\ref{sec:LessonsFm2e}.  Per Eqs.~(\ref{eq:ApproxTotalEnergy}) and (\ref{eq:2pKE}), as well as Appendix~\ref{sec:AppxB}, one writes the approximate total energy of the 5-electron atom
\begin{equation}
\label{eq:5eTotalEnergy}
\widetilde{E}_5 = -(Z - s_2)^2 - \frac{1}{4}(Z - s_4)^2 - \frac{1}{24}(Z - s_5)^2
\end{equation}
and the near-exact total energy of the 4-electron system
\begin{equation}
\label{eq:4eTotalEnergy}
E_4 = -(Z - s_2)^2 - \frac{1}{4}(Z - s_4)^2 - \frac{1}{2}s_2^2 \, ,
\end{equation}
where a small term in $(\Delta s_4)^2$ has been neglected.  Subtracting the former from the latter equation above yields the formula for an approximate ionization potential
\begin{equation}
\widetilde{I}_5 = E_4 - \widetilde{E}_5 = \frac{1}{24}(Z - s_5)^2 + \frac{1}{2}s_2^2	\, .
\end{equation}
This formula is identical to the exact formula for the ionization potential, given in Eq.~(\ref{eq:5eExactIPApproxn_b}), much as the exact and approximate ionization potential expansions are identical in the 3-electron case.  Similarly, the formula for the exact 5-electron total energy $E_5$ is the same as that for the approximate 5-electron total energy $\widetilde{E}_5$ in Eq.~(\ref{eq:5eTotalEnergy}).  That is, $E_5\!=\!\widetilde{E}_5$.

To complete the derivation of the screening relation, we take $I_5\!\approx\!\widetilde{I}_5$, by setting right-hand side of this last expression approximately equal to right side of the expression in Eq.~(\ref{eq:5eExactIPApproxn_a}), to yield the 5-electron governing relation:
\begin{equation}
\label{eq:5electronGovEqn}
\frac{1}{8}(Z - \sigma_5)(Z - s_5) \approx \frac{1}{24}(Z - s_5)^2 + \frac{1}{2}s_2^2
\end{equation}
Finally, multiplication by $8/(Z\!-\!s_5)$ on the left and right side of this governing relation yields the  5-electron screening relation of Eq.~(\ref{eq:5eScreeningRelation}).


\begin{table*}[th]  

\caption{\label{tab:AtomResultsSummaryTable} Table summarizing key QDM results from Section~\ref{sec:Derivations} for the atoms He, Li, Be, and B, which are compared with the corresponding results from the atomic Hartree Fock calculations of Bunge \textit{et al.}~\cite{bunge1993}.  The quantity $\langle 1/r \rangle_{avg}$, the reciprocal radius of the density $\rho(r,Z)$, is calculated from the values of other parameters reported in the table, using Eq.~(\ref{eq:AvgReciprocalRadius1}).  Unlike Table~\ref{tab:IP_Results}, which reports the Hartree-Fock $\Delta$SCF value for the valence ionization potential $I_N$, here the value reported for $I_N$ is the negative of the Hartree-Fock valence orbital energy, because that reflects the long range behavior of the Hartree-Fock density, even though it is usually a less accurate estimator of $I_N$. Arrows indicate that values of $\langle 1/r \rangle_{avg}$ are not for a single shell, but for the density averaged over all shells of the atom.  Energies are given in Hartrees and reciprocal radii in Bohr$^{-1}$.}
\begin{ruledtabular}
\begin{tabular}{c|c|ccc|ccc}
 &   &  \multicolumn{3}{c|}{From QDM (This Work)\, \, \, \,} &  \multicolumn{3}{c}{From Hartree Fock~\cite{bunge1993}\, \, \, \,}    \\
\hline
\noalign{\vskip 0.1cm} 
 &  $i$ \, \, \, &  1s &  2s &  2p \,\, &  1s &  2s &  2p\\
 &  $n_i$ \, \, \, &  1 &  2 &  2 \, \, &  1 &  2 &  2\\
\hline
\noalign{\vskip 0.1cm} 
 &  $N_i$ \, \, \, &  2 &   &   &  2 &    &  \\[0.1cm]
 &  $s_i$ \, \, \, &  0.3125 &   &   &  0.3127 &   &  \\
He &  $\langle 1/r \rangle_i$ \, \, \, &  1.6875 &   &   &  1.6873 &   &  \\
($N\!=\!Z\!=\!2$)\, &  $\langle 1/r \rangle_{avg}$ \, \, \, &   1.6875 &   &   &  1.6873 &   &  \\[0.1cm] 
 &  $I_N$ \, \, \, &   0.9034 &   &   &  0.9180 &   &  \\[0.1cm]
\hline
\noalign{\vskip 0.1cm} 
 &  $N_i$ \, \, \, &   2 &  1 &   &  2 &  1 &  \\[0.1cm]
 &  $s_i$ \, \, \, &   0.3125 &  1.6110 &   &  0.3127 &  1.6184 &  \\
Li &  $\langle 1/r \rangle_i$ \, \, \, &   2.6875 &  0.3473 &   &  2.6873 &  0.3454 &  \\
($N\!=\!Z\!=\!3$)\, &  $\langle 1/r \rangle_{avg}$ \, \, \, &   \multicolumn{2}{c}{\hspace{-0.75cm}$\longleftarrow $ 1.9074 $ \longrightarrow$} &   & \multicolumn{2}{c}{\hspace{-0.75cm}$\longleftarrow $ 1.9067 $\longrightarrow$} &    \\[0.1cm]
 &  $I_N$ \, \, \, &   &  0.1981 &   &   &  0.1963 &  \\[0.1cm]
 \hline
\noalign{\vskip 0.1cm} 
&  $N_i$ \, \, \, & 2 &  2 &   &  2 &  2 &  \\[0.1cm] 
 &  $s_i$ \, \, \, &  0.3125 &  1.9035 &   &  0.3127 &  1.9099 &  \\
Be &  $\langle 1/r \rangle_i$ \, \, \, &  3.6875 &  0.5241 &   &  3.6873 &  0.5225 &  \\
($N\!=\!Z\!=\!4$)\, &  $\langle 1/r \rangle_{avg}$ \, \, \, &  \multicolumn{2}{c}{\hspace{-0.75cm}$\longleftarrow $ 2.1058 $\longrightarrow $} &  &  \multicolumn{2}{c}{\hspace{-0.75cm}$\longleftarrow $ 2.1049 $ \longrightarrow$} &     \\[0.1cm]
 &  $I_N$ \, \, \, &   &  0.3426 &   &   &  0.3093 &  \\[0.1cm]
\hline
\noalign{\vskip 0.1cm} 
 &  $N_i$ \, \, \, &  2 &  2 &  1 \, \, &  2 &  2 &  1\\[0.1cm]
 &  $s_i$ \, \, \, &  0.3125 &  1.9035 &  2.5612 \, \,&  0.3127 &  1.9099 &  2.5800\\
B &  $\langle 1/r \rangle_i$ \, \, \, &  4.6875 &  0.7741 &  0.6097 \, \, &  4.6873 &  0.7725 &  0.6050\\
($N\!=\!Z\!=\!5$)\, &  $\langle 1/r \rangle_{avg}$ \, \, \, &  & $\longleftarrow $ 2.3066  $\longrightarrow$ &  & &  $\longleftarrow $ 2.3049 $ \longrightarrow$  \\[0.1cm]
 &  $I_N$ \, \, \, &   &   &  0.2967 \, \,&   &   &  0.3099\\
\end{tabular}

\end{ruledtabular}
\end{table*}


\subsubsection{\label{sec:s5-radius-expansion} Radius expansion method \\ for calculating the value of $s_5$}

The radius expansion formula for calculating the value of the 5-electron screening parameter $s_5$ from a 4-electron screening parameter is:
\begin{equation}
\label{eq:s5ExpansionFormula}
s_5 \approx	Z - \frac{3}{2}\bigg[\frac{(Z - \sigma_4)^2}{(Z - \sigma_4) + 1}\bigg] \, . \\
\end{equation}
Before discussing its derivation, we first observe that the formula above is amazingly accurate.  For example, using the value $\sigma_4=2.6887$ from Table~\ref{tab:ParameterTable}, this formula produces the result $s_5=2.5801$, which differs by only $1\!\times\!10^{-4}$ from the value $2.5800$ determined from a full wavefunction-based Hartree-Fock calculation.

Further, if one uses the value $\sigma_4\!=\!2.6751$ calculated above from  Eqs.~(\ref{eq:s4ExpansionFormula}) and (\ref{eq:2ndOrder4eScreeningRelation}), one obtains the value $s_5\!=\!2.5615$. This differs by only $-0.7\%$ from the \textit{ab initio} value. 

The derivation of Eq.~(\ref{eq:s5ExpansionFormula}) is given in Appendix~\ref{sec:AppxC}.  It does proceed according to the general approach shown above in the derivations of the 3-electron and 4-electron radius expansion formulas, Eqs.(\ref{eq:s3ExpansionFormula}) and (\ref{eq:s4ExpansionFormula}), within Sections~\ref{sec:s3-radius-expansion} and \ref{sec:s4-radius-expansion}, respectively.  However, the derivation of the 5-electron expansion formula requires some additional explanation reserved for the appendix because it differs in key details from those prior ones.  
That is because of the transition in the 5-electron atom from a valence electron with zero angular momentum, which operates in the 4-electron atom under the influence of a purely coulombic effective potential, to a valence electron with nonzero angular momentum.  The latter operates in the 5-electron system under the influence of a $non$coulombic effective external potential, because of the introduction of a centrifugal potential term in $r^{-2}$.  As shown above in Section~\ref{sec:One-electronEnergies} and explained in detail within Appendix~\ref{sec:AppxB}, that leads to a change in the form of the equation for $\varepsilon_i$.  This is due to a change in the virial relation upon which it depends, and upon which, therefore, the radius expansion formula depends, too.  Again, see Appendix~\ref{sec:AppxC} for details.

\newpage
\subsection{\label{sec:AtomResultsSummary} Summary of Key Results \\
for the Atoms}

In summarizing the key results derived for the He, Li, Be, and B atoms in Section~\ref{sec:Derivations}, we once again observe that the derivations of these results all follow the basic method developed in the treatment of the He atoms, which are summarized in Section~\ref{sec:LessonsFm2e}.  However, for the three larger atoms, the method is enhanced by the application of the radius expansion method that is introduced within Section~\ref{sec:s3-radius-expansion}, in the treatment of Li, and provides a way to calculate the two-electron interaction energies for all the atoms with $N\!>\!2$, without having to calculate any two-electron integrals.  Screening relations and energy formulas, as well as values of the screening parameters $s_i$, the valence ionization potentials $I_N$, and mean reciprocal radii $\langle 1/r \rangle_i$ are thus calculated for all four atoms.

For ease of future reference, in Table~\ref{tab:AtomResultsSummaryTable} we assemble all the key numerical results for each atom treated in the prior four subsections, \ref{sec:N=2} through \ref{sec:N=5}.  In addition, we use these results, which largely were derived for just the valence shells of each atom, to calculate and tabulate an average reciprocal radius for the entire density of each atom:  
\begin{equation}
\label{eq:AvgReciprocalRadius1}
 \langle 1/r \rangle_{avg} 
             = N_{1s} \langle 1/r \rangle_{1s} +  N_{2s} \langle 1/r \rangle_{2s}
               + N_{2p} \langle 1/r \rangle_{2p}. \,
\end{equation}
As seen in the defining equation above, $\langle 1/r \rangle_{avg}$ is given as the weighted sum of the mean reciprocal radii of all the atomic electron shells, designated by the subscripts $1s$, $2s$, and $2p$.  The weights $N_i$ are the numbers of electrons occupying each of those shells.

The values of $\langle 1/r \rangle_{i}$ for each shell, displayed in Table~\ref{tab:AtomicDensityParameters}, have been determined via Eq.~(\ref{eq:MeanRecipRadius}) from the values of the screening parameters $s_i$ for each of the atomic shells.  These are drawn from Table~\ref{tab:ScreeningParameter_Results}, where they are collected from the treatments of the valence shells of each of the four atoms, He, Li, Be, and B.

\section{\label{sec:Densities}Construction \\ of Atomic Electron Densities}

Based upon the foregoing results, a density function $\rho(r,Z)$ now may be readily determined for each of the atoms He, Li, Be, and B treated in the prior section.  Essentially, we construct the density for an atom by taking advantage of the one-to-one correspondence between each density and a one-particle external potential that is defined by our prior determination of its behavior at $r\!=\!(\langle 1/r \rangle_{avg})^{-1}$ and as $r\!\rightarrow\!\infty$.  As a result, we are able to determine atomic densities here $without$ employing any wavefunctions or the variational procedures that are intrinsic to standard methods~\cite{Pauling_and_Wilson1935,roothaan1951,Karplus_and_Porter1970,Parr_and_Yang1989} employed for electron density determination in atomic and molecular quantum mechanics.

Rather, we determine the atomic densities here by first structuring a density $ansatz$ to ensure it meets the well known requirement that its behavior be monotonically decreasing with increasing $r$, but also incorporates terms that account for the behavior of the density in each electron shell.  Then, we insist that a trial density derived from the $ansatz$ will have the correct behavior at its extremes, the cusp and tail, as described in Section~\ref{sec:Constraints}.  Plus, we insist that the trial density for an atom will have the correct behavior in the middle---i.e., it will have the correct value for its average reciprocal radius, as calculated in Section~\ref{sec:AtomResultsSummary}, from results throughout Section~\ref{sec:Derivations} and presented in Table~\ref{tab:AtomResultsSummaryTable}.  We further demonstrate below that this approach accurately determines densities for the He, Li, Be, and B atoms.

To implement the approach outlined above, we employ a simplified version of the density $ansatz$ stated in Eqs.~(\ref{eq:DensityAnsatz}) and (\ref{eq:EBC_sum}):
\begin{eqnarray}
\label{eq:SimplifiedDensityAnsatz} 	
\lefteqn{\rho(r,Z)} \\
& = &\frac{\kappa}{2} \Bigg[ \bigg( \frac{\xi_c^3}{\pi}\bigg) \exp(-2 \xi_c r)
     + \bigg( \frac{\xi_m^3}{\pi}\bigg) \exp(-2 \xi_m r) \nonumber\Bigg] \\
&   &   + \, (1 - \kappa) \bigg( \frac{2\xi_t^4}{3\pi}\bigg) r \exp(-2 \xi_t r)\, . \nonumber 
\end{eqnarray}
The few terms in this simplified $ansatz$ have been chosen to match the elements of the well-known, basic behavior and structure of the atomic density for atoms with $N\!\!=\!\!2$ to 5 electrons, as outlined in Section~\ref{sec:Foundations} above.  The trial function contains two purely exponential $1s$-like components that contribute to the cusp of the density and one $2s$-like component that consists of the product of $r$ with an exponential.  That lattermost term in $r \exp(-2 \xi_t r)$ does not contribute to the cusp at $r\!=\!0$, but ensures that the tail will have the required functional form~\cite{Morrell_etal1975,Katriel_and_Davidson1980} stated above in Eq.~(\ref{eq:LongRangeFunctionOfI}) as it approaches $r\!\rightarrow\!\infty$.  The first two purely exponential terms also help ensure the density's required monotonically decreasing behavior in $r$, while accounting for the key role of the atom's core in building up the density.

The selection of the density basis above on physical grounds is analogous to the orbital basis set selection that occurs as a preliminary step in conventional wavefunction-based \textit{ab initio} calculations~\cite{roothaan1951,Clementi_and_Roetti1974,bunge1993}.  However, one familiar with the usual methods applied for accurate quantum calculations upon atoms might find it disconcerting that there are so few terms in the expansion of $\rho(r,Z)$ within Eq.~(\ref{eq:SimplifiedDensityAnsatz})---i.e., so few functions in the density basis relative to the much larger number of basis functions needed to expand the orbitals in conventional wavefunction or DFT calculations.  There are two primary reasons that we need so few basis functions here.  First, we do not need the flexibility to perform a variational search for a solution.  We already know most of the solution because of the results from Section~\ref{sec:Derivations} that establish the parameter values for the potential and density, and from the constraints upon the form of the density discussed in the preceding paragraphs, as well as in Section~\ref{sec:Constraints}.  Second, fewer basis functions are needed because the density is monotonically decreasing in $r$ and one need not account for the oscillatory behavior of orbitals. 

\begin{figure*}[t]   
\includegraphics[height=6.0in]{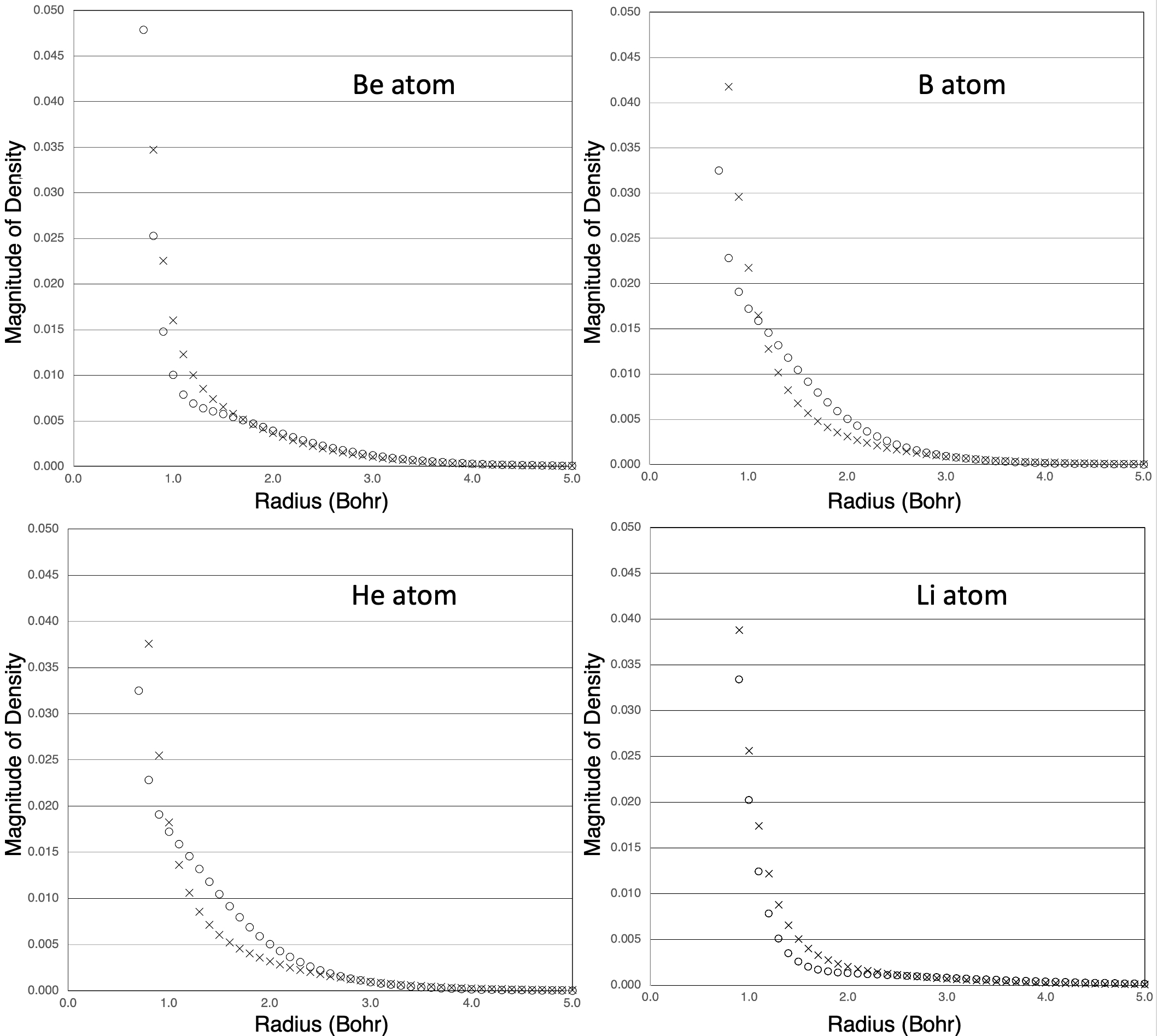}
\caption{\small Atomic electron densities for four atoms calculated via the quantum-density-mechanical (QDM) methods described in this work are plotted (as $\times$'s) versus the atomic radius, using the $ansatz$ of Eq.~(\ref{eq:SimplifiedDensityAnsatz}) and the parameters for it given in Table~\ref{tab:AtomicDensityParameters}.  These are compared in the graphs with the Hartree-Fock (H-F) densities plotted (as $\circ$'s) for the same atoms.  The H-F densities were determined by the author from the H-F calculations of Bunge~\textit{et al.}~\cite{bunge1993}. As seen in the figure and described in the text, all the independently determined QDM densities and H-F densities are in strong agreement, having coefficients of determination (i.e., $R^2$ values) between them in excess of 0.99 in each case, as is also reported in Table~\ref{tab:AtomicDensityParameters}. This is an indicator of the accuracy of the QDM methods for the determination of atomic densities.} 
\label{fig:densities}
\end{figure*}

Using Eq.~(\ref{eq:SimplifiedDensityAnsatz}), QDM atomic densities are calculated for the He, Li, Be, and B atoms, then plotted (as $\times$'s) in Fig.~\ref{fig:densities}.  The plots are made using the parameter values given in Table~\ref{tab:AtomicDensityParameters}.  For comparison, also plotted in the figure (as $\circ$'s) are densities derived from the atomic Hartree-Fock calculations of Bunge \textit{et al.}~\cite{bunge1993}.

As seen in Eq.~(\ref{eq:SimplifiedDensityAnsatz}) and reflected in Table~\ref{tab:AtomicDensityParameters}, for each atom four parameters must have their values determined to specify the behavior of the density using that $ansatz$.  The linear parameter $\kappa$, with its complement, ($1\!-\!\kappa$), maintains the unit-normalization of the density and governs the proportion of $1s$-like versus $2s$-like behavior.  Below, $\kappa$ also is employed to ensure that each trial density developed from a parameterization of the $ansatz$ in Eq.~(\ref{eq:SimplifiedDensityAnsatz}) will have the correct, already determined value for its average reciprocal radius $\langle 1/r \rangle_{avg}$, as given in Tables~\ref{tab:AtomResultsSummaryTable} and \ref{tab:AtomicDensityParameters}.  The set of three nonlinear parameters, $\{\xi_c, \xi_m, \xi_t\}$, specifies the exponential behavior at the \underline{c}usp, \underline{m}iddle, and \underline{t}ail of the density, respectively.  Four conditions are required to specify the values for these four parameters.

%
\begin{table}[t]
\caption{\label{tab:AtomicDensityParameters} Parameter values for calculating QDM atomic electron density expansions via the $ansatz$ of Eq.~(\ref{eq:SimplifiedDensityAnsatz}). For each atom, the values of the nonlinear parameters $\xi_c$, $\xi_m$, $\xi_t$, and that of linear parameter $\kappa$ are determined, as described in the text, from the nuclear charge $Z$, the valence ionization potential $I_N$, and the average reciprocal radius $\langle 1/r \rangle_{avg}$. The latter two of these were evaluated in Section~\ref{sec:Derivations} and summarized for all the atoms in Table~\ref{tab:AtomResultsSummaryTable}, as well as below.  QDM density expansions calculated from the parameter values given in this table are plotted in the graphs shown in Fig.~\ref{fig:densities}.} 
\begin{ruledtabular}
\begin{tabular}{c|c|cccc}
   &     &\multicolumn{4}{c}{\underline{Atom}} \\
Quantity &  Units \, &  He &  Li &  Be &  B\\
\hline 
\noalign{\vskip 0.1cm}
$Z$ &   &  2 &  3 &  4 &  5\\
$I_{N}$ &  Hartree \, &  0.9034 &  0.1981 &  0.3426 &  0.2967\\[0.15cm]
$\langle 1/r \rangle_{avg}$ &  Bohr$^{-1}$ \, &  1.6896 &  1.9074 &  2.1058 &  2.3066\\
 &   &   &   &   &  \\
$\xi_c$ &   &  2.1506 &  3.2259 &  4.2332 &  5.2310 \\
$\xi_m$ &   &  1.6896 &  1.9074 &  2.1058 &  2.3066 \\
$\xi_t$ &   &  1.3439 &  0.6294 &  0.8278 &  0.7703 \\
 &   &   &   &   &   \\
$\kappa$ &   &  0.7749 &  0.6930 &  0.5936 &  0.5508 \\

\end{tabular}
\end{ruledtabular}
\end{table}
%


\begin{table*}[th]  

\caption{\label{tab:DensityComparisonTable} Table comparing results from QDM and Hartree-Fock atomic electron densities for the atoms He, Li, Be, and B.  Values of $R^2$ indicate the degree of agreement between the graphs of the QDM densities shown in Fig.~\ref{fig:densities} with the corresponding graphs of the Hartree-Fock densities shown in that figure.  The fact that the $R^2$ values are so close to 1 (i.e., greater than 0.99 in every case) indicates that the behavior of all the independently calculated QDM densities closely corresponds to that for the  Hartree-Fock densities and therefore is very physically reasonable.  In addition, the fact that the one-electron expectation values shown in the table for the QDM and Hartree-Fock densities have similar values and trends for each atom further validates the simple QDM approach for atomic density determination. QDM expectation values were calculated using parameter values from Table~\ref{tab:AtomicDensityParameters} in the formulas for the expectation values given in Appendix~\ref{sec:AppxD} and in Eq.~(\ref{eq:AvgReciprocalRadius2}).  Hartree-Fock expectation values were determined by the author from occupation-weighted averages of the expectation values for the atomic shells given by Bunge \textit{et al.}~\cite{bunge1993}. All expectation values are given in atomic units.}
\begin{ruledtabular}
\begin{tabular}{c|ccccccccc}
\noalign{\vskip 0.1cm} 
 & \multicolumn{7}{c}{\underline{Comparison of graphs in Fig.~\ref{fig:densities}}}\\ [0.15cm]
 &  He &  Li &  Be &  B \\
$R^2$    &  0.9990 &  0.9990 &  0.9916 &  0.9971 \\
\hline
\noalign{\vskip 0.1cm} 
 & \multicolumn{7}{c}{\, \underline{Comparison of expectation values}}\\[0.15cm] 
 & \multicolumn{4}{c|}{\underline{From QDM Theory}  \, \,}  &\multicolumn{4}{c}{\underline{From Hartree-Fock Theory}\, \, \,} \\[0.1cm]
   &  He &  Li &  Be &  \multicolumn{1}{c|}{B} &  He &  Li &  Be &  B\\
$\langle r^2 \rangle_{avg}$ &  1.2817 &  4.2605 &  3.2157 &  \multicolumn{1}{c|}{3.9703 \,} &  1.1848 &  6.2107 &  4.3296 &  3.1702\\
$\langle r \rangle_{avg}$ &  0.9492 &  1.4092 &  1.2984 &   \multicolumn{1}{c|}{1.4243 \,} &  0.9273 &  1.6733 &  1.5322 &  1.3621\\
$\langle 1/r \rangle_{avg}$ &  1.6896 &  1.9074 &  2.1058 &   \multicolumn{1}{c|}{2.3066 \,} &  1.6873 &  1.9052 &  2.1022 &  2.2759\\
$\langle 1/r^2 \rangle_{avg}$ &  6.0671 &  9.8134 &  13.4563 &   \multicolumn{1}{c|}{18.1804 \, \,}&  5.9955 &  10.0707 &  14.4045 &  18.7310\\
\end{tabular}
\end{ruledtabular}
\end{table*}


The first of these conditions is simply the tail constraint~\cite{Morrell_etal1975,Katriel_and_Davidson1980} of Eq.~(\ref{eq:LongRangeFunctionOfI}), from which the tail exponent $\xi_t$ is given in terms of the valence ionization potential as
\begin{equation}
\label{eq:xitEqn}
	\xi_t = (2I_N)^{1/2}.
\end{equation}
Table~\ref{tab:AtomicDensityParameters} lists the values of $I_N$ determined in Section~\ref{sec:Derivations} for each of the four atoms of interest, which yield via the preceding equation the values of $\xi_t$ that also are shown in the table.

For the second condition, we set exponent $\xi_m$ equal to the average reciprocal radius for the density:
\begin{equation}
\label{eq:ximEqn}
\xi_m = \langle 1/r \rangle_{avg} \, .	
\end{equation}
This condition is the analog of the relationship that applies for the density of a single shell of an atom, such as the $1s$ shell of He, when that density is represented by a single exponential function, much as is discussed in Sections~\ref{sec:Ansatz} and \ref{sec:Screening} above.

The third condition upon the nonlinear parameters follows from the general statement of the Kato cusp constraint~\cite{kato1957,march1986} in Eq.~(\ref{eq:CuspCond}).  When applied to Eq.~(\ref{eq:SimplifiedDensityAnsatz}), it yields
\begin{equation}
	\label{eq:AnsatzCuspCond}
	Z = \frac{\xi_c^4 + \xi_m^4}{\xi_c^3 + \xi_m^3}.
\end{equation}
This condition ties the value of $\xi_c$ to that of $\xi_m$.

Assuming that all three of the nonlinear parameters have been determined from the three preceding conditions, the linear parameter $\kappa$ is determined by the fourth and last condition:

\begin{equation}
\label{eq:kappaEqn}
\begin{aligned}
\kappa & = \frac{\langle 1/r \rangle_{avg} - (2/3)\xi_t}{(1/2)\xi_c + (1/2)\xi_m - (2/3)\xi_t}	\\
       & = \frac{\langle 1/r \rangle_{avg} - (2/3)\xi_t}{(1/2)\xi_c + (1/2)\langle 1/r \rangle_{avg} - (2/3)\xi_t} \, ,
\end{aligned}
\end{equation}
where the second line results from the use of Eq.~(\ref{eq:ximEqn}) within the first.
This condition on $\kappa$ follows from the term-by-term integration of the density $ansatz$ of Eq.~(\ref{eq:SimplifiedDensityAnsatz}) to develop another expression (in addition to that stated above in Eq.~(\ref{eq:AvgReciprocalRadius1})) for the average reciprocal radius of the atomic density:
\begin{equation}
\label{eq:AvgReciprocalRadius2}
\begin{aligned}
\langle 1/r \rangle_{avg} 
            & = 4 \pi \int_0^{\infty} \frac{1}{r} \rho(r,Z) r^2 dr           \\
            & = \frac{\kappa}{2}(\xi_c + \xi_m) + \frac{2}{3}(1 - \kappa)\xi_t \,.
\end{aligned}
\end{equation}
For the integration, one employs the formula~\cite{Johnson_and_Pedersen1974}
\begin{equation}
\label{eq:IntegrationFormula}
\int_0^{\infty}r^k[r^n \exp(-ar)]4\pi r^2dr = 4\pi \frac{(n+k+2)!}{a^{n+k+3}},	
\end{equation}
with $k=-1$, while also taking account of the normalization factors, of the form shown in Eq.~(\ref{eq:NormConst}), that are incorporated in each of the three terms on the right side of Eq.~(\ref{eq:SimplifiedDensityAnsatz}).  Solving Eq.~(\ref{eq:AvgReciprocalRadius2}) for $\kappa$ yields the condition upon that parameter given in the first line of Eq.~(\ref{eq:kappaEqn}).

With these four conditions established in Eqs.~(\ref{eq:xitEqn}) through (\ref{eq:kappaEqn}), in principle it is possible to define $\xi_t$ and $\xi_m$ with Eqs.~(\ref{eq:xitEqn}) and (\ref{eq:ximEqn}), then simultaneously solve Eqs.~(\ref{eq:AnsatzCuspCond}) and (\ref{eq:kappaEqn}) to determine analytically the values of all four parameters that specify the electron density of an atom via Eq.~(\ref{eq:SimplifiedDensityAnsatz}).  In actuality, though, the nonlinearities in Eq.~(\ref{eq:AnsatzCuspCond}) make this one-step analytic solution challenging.

In practice, therefore, it is much easier to employ a simple and rapidly convergent iterative approach.  For each atom we define $\xi_t$ and $\xi_m$ in the manner stated in the preceding paragraph, using their previously determined values, which are given in Table~\ref{tab:AtomicDensityParameters}.  Then, we choose a trial value for the cusp exponent $\xi_c$ that is close to the nuclear charge $Z$ of the atom, but perhaps a bit greater.  This immediately permits calculation of a trial value for $\kappa$ using Eq.~(\ref{eq:kappaEqn}).

Also, the trial value of $\xi_c$, along with the definite value of $\xi_m$ from Eq.~(\ref{eq:ximEqn}) and Table~\ref{tab:AtomicDensityParameters}, permit the use of the cusp condition, Eq.~(\ref{eq:AnsatzCuspCond}), to calculate a trial value of $Z$.  This trial value is compared to the actual value of $Z$ for the atom.  If the two do not quite match, we adjust the chosen value of $\xi_c$ and repeat the procedure until Eq.~(\ref{eq:AnsatzCuspCond}) yields a value that is the same as $Z$ to the desired level of accuracy.

By means of this procedure, we have determined the parameters given in Table~\ref{tab:AtomicDensityParameters} that specify, via Eq.~(\ref{eq:SimplifiedDensityAnsatz}), density functions of all four atoms of interest.  Also, as mentioned above, using those parameter values, we have plotted the QDM density functions.  Those plots appear in Fig.~\ref{fig:densities} in the same graphs with plots of the Hartree-Fock (H-F) densities for the same atoms.

The plots are made for each atom by separately determining discrete values $\rho^{(QDM)}_k$ for the QDM density and values $\rho^{(H-F)}_k$ for the H-F density at each of a set of $k=1,2,\ldots,P$ discrete values $r_k$ of the radius.  These $r_k$ values are established and the points plotted at intervals of 0.1 Bohr from $r_k=0$ to $r_k=5.0$ Bohr, so that $P=51$ points for both graphs in each plot.   

The comparison between the QDM and H-F densities in Fig.~\ref{fig:densities} is significant.  One would want and expect the QDM densities to be better---i.e., more accurate---than those from Hartree-Fock, since the quantities from Section~\ref{sec:Derivations} used in the density determination here were shown in that prior section to be more accurate than those from H-F theory.  However, H-F densities are generally regarded as being fairly accurate representations of the behavior of electrons in atoms.  Thus, if the QDM densities were distributed in space in a manner that differed greatly from the H-F densities, or if the expectation values determined from the two sets of densities had greatly different values and trends, the QDM densities could not be correct.

Simply from a visual inspection of the plots in Fig.~\ref{fig:densities}, however, one can see that the QDM densities closely correspond in their behavior to that of the H-F densities.  Further, to assess quantitatively the accuracy of the QDM densities in terms of the degree of their agreement with the H-F densities, we have calculated the coefficients of determination---i.e., the $R^2$ values---between the points in the two different plots of the density for each atom~\cite{SchaumStatistics}:
\begin{equation}
R^2 = 1 - \frac{\Sigma_k^P \big[\rho^{(QDM)}_k - \rho^{(H-F)}_k\big]^2}
                      {\Sigma_k^P \big[\rho^{(QDM)}_k - \rho^{(H-F)}_{avg}\big]^2} \, ,
\end{equation}
where
\begin{equation}
\rho^{(H-F)}_{avg} = \frac{\Sigma_k^P \rho^{(H-F)}_k}{P} \, .
\end{equation}

As listed in Table~\ref{tab:DensityComparisonTable}, the $R^2$ value for each of the four pairs of plots shown in Fig.~\ref{fig:densities} is in excess of 0.99, indicating a very high degree of agreement between each pair of independently determined densities.  This is an indicator of the validity of the QDM methods for the determination of atomic densities.  (From the use here of the $R^2$ measure, which also is commonly used in regression procedures, the reader should not be confused, though, and led to think that the QDM densities were somehow fit to the H-F densities. They were $not$.  The QDM and H-F densities were determined completely separately and independently from each other, in the manner described above.) 

As still another means for assessing the QDM method for density determination, from both the QDM and H-F atomic electron densities we have calculated numerical values for four atomic expectation values, moments of the density distributions.  This was done using the formula in Eq.~(\ref{eq:AvgReciprocalRadius2}), those in Appendix~\ref{sec:AppxD}, and the parameter values given in Table~\ref{tab:AtomicDensityParameters}.  The results of these calculations are presented in Table~\ref{tab:DensityComparisonTable}.  The similar values and trends in the QDM and H-F results for each atom seen in the table offer further validation for the simple QDM approach for atomic density determination.   

\section{\label{sec:Summary}Summary and Conclusions}

In the preceding sections, for atoms containing two to five electrons, this paper has developed and demonstrated a direct, \textit{ab initio} method for calculating the one-electron density and energy of a many-electron system, without any reference to one-electron or many-electron wavefunctions.  To do so, the work described here has combined in a unique manner diverse ideas~\cite{debroglie1923,thomas1927,fermi1928,dirac1930,Wilson1962}, tools~\cite{pauling1927,slater1930,eckart1930,zener1930,Clementi_and_Raimondi1963,bessis1981,Pilar1990}, and theorems~\cite{kato1957,march1986,Hohenberg_and_Kohn1964,gilbert1975,Morrell_etal1975,Katriel_and_Davidson1980} of quantum mechanics that were developed by prior investigators over nearly a century.  However, in order to accomplish the derivation of this new, purely density-based computational implementation of quantum mechanics it also has proven necessary and advantageous to abandon almost entirely a number of the basic constructs, methods, and artifacts present in the wave-mechanical implementation, as well as in prior density-based approaches, such as DFT.  In addition to wavefunctions, these include the use of the many-electron and one-electron Schr{\"o}dinger equation, variational search for solutions, variational functionals, use of the wavefunctions' second derivatives to calculate kinetic energies, exchange interactions, and almost all other two-electron integrals.

In their place, we have derived and introduced a set of governing equations, one for each number $N$ of electrons, that may be solved for screening relations that connect the behavior of the density far from the nucleus to its behavior at the most probable radius.  (See, for example, Section~\ref{2e-Proof}.) Such relationships were not previously appreciated.  The behavior at those radii and points between is described in terms of the same screening parameters that describe the electrostatic potential experienced by the electrons in those different regions of space.  
Essentially, solving a density-mechanical governing equation analytically identifies in a single step a one-to-one mapping between a screened external potential and the corresponding density (i.e., it constructs a mapping of the type that was proven to exist by Hohenberg and Kohn~\cite{Hohenberg_and_Kohn1964} as the basis for their theorem stating that total energy is a functional of the density).  
This specific realization of the one-to-one match between the parameters governing both the potential and the density enables direct, accurate determination here of both the energies and the density for an atom, without variational search.

The method for formulating, solving, and applying the governing relations to calculate screening relations and energies is developed in the treatment of the He atom, within Section~\ref{sec:N=2} and the basic steps of this method are summarized in Section~\ref{sec:LessonsFm2e}.  The procedure for calculating the densities is given in Section~\ref{sec:Densities}.

Within a governing equation, consistent with the ideas of de Broglie~\cite{debroglie1923}, an electron's momentum and half its square, the kinetic energy, is approximated as a functional of the density via a measure of the degree to which the potential constrains the range of motion of the electron at the most probable radius, or Bohr radius $r=(\langle 1/r \rangle)^{-1}$.  Then, the screening relation derived from solution of that governing equation produces a correlation energy correction which yields an accurate kinetic energy for a valence electron.  Via the virial theorem, that yields an accurate one-electron total energy or valence ionization potential that also includes electron correlation.  See, for example, Section~\ref{sec:BetterE2}.  Further, iterative solution of a governing equation can refine it and thereby improve the energy it delivers, so that the energy approaches the exact, experimental ionization potential, as is demonstrated in Section~\ref{sec:EvenBetterE2}.

In that section, a particularly important achievement is the derivation of a simple, exact or very near-exact analytical formula, seen in Eqs.~(\ref{eq:BetterE2Approxn}) and (\ref{eq:EvenBetter_s2_a}), for the total energy of a two-electron atom.  Restated in one location, that simple analytic formula is:
\begin{subequations}
\label{eq:Exact2eEnergy}
\begin{equation}
E_2 = -(Z - s_2)^2 - \frac{1}{2}s_2^2 \, , \quad \quad \\	
\end{equation}
where
\begin{equation}
s_2  = \frac{1}{3}\sigma_2 - \frac{1}{216}\bigg(\frac{\sigma_2^2}{Z}\bigg) 
                 - \frac{1}{972}\bigg(\frac{\sigma_2^3}{Z^2}\bigg)	
\end{equation}
and
\begin{equation}
\sigma_2 =  15/16 \,. \qquad \qquad \qquad \qquad\ \qquad
\end{equation}
\end{subequations}
While a simple, exact or near-exact formula, much like Eq.~(\ref{eq:E_1}), was derived for the energy of a one-electron atom a century ago by Bohr~\cite{Bohr1913a} and by Schr{\"o}dinger~\cite{Schrodinger1926a}, it had previously been thought and taught that a simple and accurate analytical formula for the energy of any many-electron atom was not possible.

Also very important, this work has introduced a ``radius expansion method" that eliminates the need to calculate any two-electron integrals, beyond a single coulomb integral required for a first approximation to the energy of a two-electron atom.  The radius expansion method, which follows from the screening relations, permits the calculation of the screening parameters and two-electron interaction energy for an $N$-electron atom from those for an $(N\!-\!1)$-electron atom. (See Sections~\ref{sec:s3-radius-expansion}, \ref{sec:s4-radius-expansion}, and \ref{sec:s5-radius-expansion}.) This technique also simply and clearly reveals the manner in which successive electron shells of an atom are coupled together in their behavior and energetics.  Further, the radius expansion technique and the governing equations ensure that the density-mechanical method derived here will scale in complexity only as $N$, not a power of $N$, as do most prior many-electron quantum mechanical methods~\cite{Kohn_etal1996,Strout_and_Scuseria1995}.

Taken together, the simplifications summarized above suggest the possibility that many-electron calculations that now require high performance computers and much processing time, using wavefunction-based or DFT methods, might be reduced to calculations that can be performed using just simple spreadsheets or even using a hand calculator.  That already has been accomplished via the density mechanical treatment described in this work for neutral atoms with $N\!=2$ to 5.

Despite these significant initial achievements here in the development of the \textit{ab initio} density-mechanical approach, it remains to demonstrate its application to atoms for $N\!>\!5$ and also to molecules.  Addressing these problems shall be the subject of further research and publications by the author. 

\begin{appendix}
\section{\label{sec:AppxA} Density-based variational evaluation \\of the screening parameter $s_2$} 

The screening parameter $s_2$ and its value figure prominently in the development within the body of this paper.  Table~\ref{tab:ParameterTable} presents a value $s_2\!=\!0.3127$ that is taken from an accurate, but complex Hartree-Fock \textit{ab initio} wavefunction-based calculation~\cite{Clementi_and_Roetti1974,bunge1993}.  In this appendix, though, we show that a value that is very nearly the same, $s_2\!=\!0.3125$, can be calculated using only quantities derivable from an electron density.

Similar evaluations have been performed by others, previously.  (See, for example, Pilar~\cite{Pilar1990}, pp.\! 160-163.) However, that has been done in the context of a wave-mechanical development, without recognizing or calling attention to the fully density-based nature of the calculation.  Thus, for completeness, here we emphasize that an accurate first approximation to $s_2$ and its value may be determined strictly as a functional of the one-electron density for a 2-electron atom.

For each of the two identical electrons in the atom, taking quantum number $n_2=1$, we employ for $\rho_2(r,Z)$ a single-term exponential one-electron density $ansatz$, as given by Eqs.~(\ref{eq:EBC_sum}) and (\ref{eq:OneTermEBC}).  Then, the coulomb-type two-electron integral between the electrons, having only density functions in its integrand (esp., no one-electron wavefunctions), is evaluated~\cite{Pilar1990,Johnson_and_Pedersen1974}:
\begin{equation}
\label{eq:CoulombIntegral}	
\begin{aligned}
	J_2 & = (4\pi)^2 \int \int \frac{\rho_2(r_1,Z)\rho_2(r_2,Z)}{r_{12}} r_1^2 dr_1 r_2^2 dr_2 \\
	    & = \frac{5}{8}\xi_2.                       
\end{aligned}
\end{equation}
The result above, Eq.~(\ref{eq:ScreenedPotential}), and Eq.~(\ref{eq:BasicKEeqn}) may be used to set up an approximate, density-based variational energy function for the total energy of the two electron atom:
\begin{eqnarray}
\mathcal{E}_2 & = & \langle T \rangle_2 +  \langle V \rangle_2 + J_2    \nonumber  \\
			  & = & (\langle 1/r \rangle_2)^2 -2Z \langle 1/r \rangle_2 + \frac{5}{8}\xi_2 \\
			  & = & (\xi_2)^2 - 2Z \xi_2 + \frac{5}{8}\xi_2,     \nonumber
\end{eqnarray}

Minimizing this function, by setting $(d\mathcal{E}_2/d\xi_2)\!=\!0$, yields an equation
\begin{equation}
	0 = 2\xi_2 - 2Z + 5/8.   \label{eq:xi2eqn}
\end{equation}
Taking, $Z=2$, this has the solution $\xi_2=27/16$, or from Eq.~(\ref{eq:ExponentFormula}):
\begin{equation}
\label{eq:s2valueAppx}
	s_2 = 5/16 = 0.3125.  
\end{equation}

The use of the variational method in this appendix to help calculate the initial value of $s_2$ is the only use of it required throughout the density-mechanical formalism.

\section{\label{sec:AppxB} Derivation of energy formulas for an electron in a shell with nonzero angular momentum quantum number $\ell_i$} 

In this appendix, we derive the formula
\begin{equation}
\label{eq:2pKE_Appx}
	\varepsilon_i = -\frac{1}{24}(Z - s_i)^2 \, ,	                                 
\end{equation}
for the one-electron energy of a $2p$ electron, which was asserted in Eq.~(\ref{eq:2pKE}), within Section~\ref{sec:One-electronEnergies} of the main text, and employed throughout Section~\ref{sec:N=5}.  At first, this formula would seem to be at odds with other development in Section~\ref{sec:One-electronEnergies}, where it was shown that the one-electron energy can be expressed via the formula
\begin{equation}
\label{eq:CoulombOneElectronEnergy}
	\varepsilon_i = -\langle t \rangle_i = -\frac{1}{2n_i^2} (Z - s_i)
\end{equation}
for an electron with principal quantum number $n_i$ operating in a screened Coulomb potential within an atom. 

However, a $2p$ electron, which has nonzero angular momentum and quantum numbers $n_i\!=\!2$ and $\ell_i\!=\!1$, does not operate in a purely Coulomb-type screened potential like that in Eq.~(\ref{eq:ScreenedPotential}).  In general, for any electron with nonzero angular momentum quantum number $\ell_i$, in addition to a Coulombic term that goes as $r^{-1}$, the effective one-particle potential includes centrifugal potential term that goes as $r^{-2}$:
\begin{equation}
\label{eq:EffectivePotential}
	V_i^{eff}(r) = -\frac{Z - s_i}{r} + \frac{\ell_i(\ell_i + 1)}{2r^2}.
\end{equation} 
That is, the effective potential includes a term for which $m\!=\!2$ in Eqs.~(\ref{eq:VirialEq}) and (\ref{eq:GeneralPotl}).

This means we must express the total 1-electron energy as a sum of expectation values like that in Eq.~(\ref{eq:BasicKEeqn}), arising from the $r^{-1}$ Coulomb term in Eq.~(\ref{eq:EffectivePotential}), plus an expectation value arising from the $r^{-2}$ term in the effective potential, as follows:
\begin{equation}
\label{eq:AppxBEqn1}
\begin{aligned}
\varepsilon_i & = \langle t \rangle_i + \langle V_i^{eff} \rangle_i  \\
              & = (1/2)n_i^2  (\langle 1/r \rangle_i)^2              \\ 
              & \quad -(Z - s_i)\langle 1/r \rangle_i                  
			        + (1/2)\ell_i(\ell_i + 1)\langle 1/r^2 \rangle_i               
 \end{aligned}
 \end{equation}
Above, the kinetic energy $\langle t \rangle_i$ has been given in terms of the square of the reciprocal radius by virtue of Eq.~(\ref{eq:BasicKEeqn}).  Also, we may use Eq.~(\ref{eq:MeanRecipRadius}) to recognize that the second term on the right, the screened potential term, may be reexpressed as the square of a reciprocal radius:
\begin{equation}
\label{eq:AppxBEqn2}
-(Z - s_i)\langle 1/r \rangle_i = -n_i^2 (\langle 1/r \rangle_i)^2
\end{equation}
In addition, from the Schr\"odinger solutions for a 1-electron atom with effective nuclear charge $(Z\!-\!s_i)$, we may expand the expectation value for the reciprocal of the $squared$ radius via the formula~\cite{Pauling_and_Wilson1935}:
\begin{equation}	
\label{eq:RecipRadiusSquared}
\begin{aligned}
	\langle 1/r^2 \rangle_i & = \frac{(Z - s_i)^2}{n_i^3 (\ell_i + 1/2)}  \\
	                        & = \frac{n_i}{\ell + 1/2} (\langle 1/r \rangle_i)^2 \, ,
\end{aligned}
\end{equation}
where the second line of the equation follows from the formula for $\langle 1/r \rangle_i$ in Eq.~(\ref{eq:MeanRecipRadius}).

Now, Eqs.~(\ref{eq:AppxBEqn2}), (\ref{eq:RecipRadiusSquared}) permit us to rewrite Eq.~(\ref{eq:AppxBEqn1}) solely in terms of $(\langle 1/r \rangle_i)^2$:
\begin{equation}
\label{eq:AppxBEqn4}
\begin{aligned}
\varepsilon_i & =  \frac{1}{2}n_i^2(\langle 1/r \rangle_i)^2 - n_i^2(\langle 1/r \rangle_i)^2 \\
              &   \quad + \frac{1}{2} n_i \frac{\ell_i(\ell_i + 1)}{\ell + 1/2}(\langle 1/r \rangle_i)^2 \\
              & = -\frac{1}{2}\Big[n_i^2 - n_i \frac{\ell_i(\ell_i + 1)}{\ell + 1/2} \Big](\langle 1/r \rangle_i)^2                           
\end{aligned}
\end{equation}
Then, applying Eq.~(\ref{eq:MeanRecipRadius}) once again, we have the general result:
\begin{equation}
\label{eq:AppxBEqn5}
\varepsilon_i = -\frac{1}{2} \Big[ n_i^2 - n_i \frac{\ell_i(\ell_i + 1)}{\ell + 1/2} \Big] \Big[\frac{(Z-s_i)}{n_i^2} \Big]^2
\end{equation}
For first-row atoms with $2p$ valence shells, having numbers of electrons $N\!\!=\!\!5$ through 10, we substitute in Eq.~(\ref{eq:AppxBEqn5}) the values $n_i\!=\!2$ and $\ell_i\!=\!1$ to obtain the result stated above in Eq.~(\ref{eq:2pKE_Appx}), as well as within Eq.~(\ref{eq:2pKE}) of the main text.

\section{\label{sec:AppxC} Derivation of the 5-electron radius expansion formula, Eq.~(\ref{eq:s5ExpansionFormula})} 

In this appendix the derivation is explained for the 5-electron radius expansion formula, Eq.~(\ref{eq:s5ExpansionFormula}), first stated in Section~\ref{sec:s5-radius-expansion}.
The derivation starts by writing:
\begin{equation}
\label{eq:AppxC-Eqn1}	
I_5 = \frac{1}{8}(Z - \sigma_5)(Z - s_5) \approx \frac{1}{12}(Z - s_5)^2
\end{equation}
with the coefficient (1/12) of the expression on the far right, rather than (1/24), as would be expected from Eqs.~(\ref{eq:5eExactIPApproxn}) in the text.

The reason for this change is that a radius expansion formula is dependent upon the virial theorem in equating the exact expression for $I_N$, involving a factor $(Z - \sigma_N)$ to an approximate one in the square of $(Z - s_N)$.  Thus, in projecting a value for a valence parameter $s_N$ for the $N$-electron system from the values of valence parameters for the $(N\!-\!1)$-electron system, it would be best if the virially-based relations were similar in form for the valence shells of both systems.  That is not the case, though, for the boron atom with $N\!=\!5$.  This is due to the fact that 4-electron atom's valence electrons operate in a purely coulombic $r^{-1}$ potential, but the valence electron in the 5-electron atom does not.  This is prevented by the introduction of the non-coulombic $r^{-2}$ term in the 5-electron valence effective potential to represent the effect of angular momentum, as detailed in Appendix~\ref{sec:AppxB}. 

In the case of the 2, 3, and 4-electron atoms, the virial theorem corresponding to the purely coulombic potential enables the calculation of the total one-electron energy without explicit consideration of the coulombic potential term:
\begin{equation}
\label{eq:AppxC-Eqn2}	
I_N \approx \varepsilon_i = -\langle \hat{t} \rangle_i
\end{equation}
If we similarly delete and remove from consideration the second, noncoulombic potential term on the right in Eq.~(\ref{eq:EffectivePotential}) of Appendix~\ref{sec:AppxB}, we obtain the energetically \textit{incorrect} expression of Eq.~(\ref{eq:AppxC-Eqn1}), which is nonetheless more ``virially" compatible with that for the 4-electron system.

Then, starting with Eq.~(\ref{eq:AppxC-Eqn1}), one can readily state:
\begin{equation}
\label{eq:AppxC-Eqn3}
\begin{aligned}
\frac{1}{(Z - s_5)} & \approx \frac{2}{3}\frac{1}{(Z - \sigma_5)} \\
                    & \approx \frac{2}{3}\frac{1}{(Z - \sigma_4) - \Delta \sigma_5} \, .
\end{aligned}
\end{equation}
From that point, the derivation proceeds much in the way it does for the 3 and 4-electron cases.  One performs a binomial expansion of the expression on the right above and assumes $\Delta \sigma_5\!\approx\!1$.  Finally, after taking the reciprocal of the result on both sides of the equation and subtracting both sides from $Z$, one obtains the highly accurate 5-electron radius expansion formula stated in Eq.~(\ref{eq:s5ExpansionFormula}).

\vspace{0.5cm}
\section{\label{sec:AppxD} Formulas for evaluation of expectation values of the density $ansatz$ in  Eq.~(\ref{eq:SimplifiedDensityAnsatz})} 

Formulas for evaluation of the expectation values shown in Table~\ref{tab:DensityComparisonTable} for the atomic electron density $ansatz$ of Eq.~(\ref{eq:SimplifiedDensityAnsatz}) are:
\begin{subequations}
\label{eq:ExpectationValueFormulas}
\begin{eqnarray}
\langle r^2 \rangle & = & \frac{\kappa}{2}\bigg(\frac{3}{\xi_c^2} + \frac{3}{\xi_m^2}\bigg) + (1 - \kappa)\bigg(\frac{5}{\xi_t^2}\bigg) \label{eq:ExpectationValueFormulas_a} \\
\langle r \rangle & = & \frac{\kappa}{2}\bigg(\frac{3}{\xi_c} + \frac{3}{\xi_m}\bigg) + (1 - \kappa)\bigg(\frac{2}{\xi_t}\bigg) \label{eq:ExpectationValueFormulas_b} \\
\langle 1/r^2 \rangle & = & \kappa\big(\xi_c^2 + \xi_m^2\big) + \frac{2}{3}(1 - \kappa)\ \xi_t^2
         \label{eq:ExpectationValueFormulas_c}
\end{eqnarray}
\end{subequations}
Substitution of the values for the parameters $\xi_c$, $\xi_m$, $\xi_t$, and $\kappa$ given in Table~\ref{tab:AtomicDensityParameters} into these formulas and that of Eq.~(\ref{eq:AvgReciprocalRadius2}) yield the QDM results presented in the last four rows of Table~\ref{tab:DensityComparisonTable}.
Equations~(\ref{eq:ExpectationValueFormulas}) are derived by applying the integral formula in Eq.~(\ref{eq:IntegrationFormula}) for the term-by-term integration of
\begin{equation}
\langle r^k\rangle = 4 \pi \int_0^{\infty}r^{k}\rho(r,Z)	r^2dr \, ,
\end{equation}
having used Eq.~(\ref{eq:SimplifiedDensityAnsatz}) to expand $\rho(r,Z)$, and taking $k=2,1,-2$, respectively.

\end{appendix}

\begin{acknowledgments}
\vspace{-3mm}
The author gratefully acknowledges valuable comments on the manuscript by Drs. Brett Dunlap of the U.S. Naval Research Laboratory, Alex Atanasov of Harvard University, and Carl Picconatto of the MITRE Corporation.
\end{acknowledgments}


\bibliography{QDMbib07-24v08}

\begin{thebibliography}{50}%
\makeatletter
\providecommand \@ifxundefined [1]{%
 \@ifx{#1\undefined}
}%
\providecommand \@ifnum [1]{%
 \ifnum #1\expandafter \@firstoftwo
 \else \expandafter \@secondoftwo
 \fi
}%
\providecommand \@ifx [1]{%
 \ifx #1\expandafter \@firstoftwo
 \else \expandafter \@secondoftwo
 \fi
}%
\providecommand \natexlab [1]{#1}%
\providecommand \enquote  [1]{``#1''}%
\providecommand \bibnamefont  [1]{#1}%
\providecommand \bibfnamefont [1]{#1}%
\providecommand \citenamefont [1]{#1}%
\providecommand \href@noop [0]{\@secondoftwo}%
\providecommand \href [0]{\begingroup \@sanitize@url \@href}%
\providecommand \@href[1]{\@@startlink{#1}\@@href}%
\providecommand \@@href[1]{\endgroup#1\@@endlink}%
\providecommand \@sanitize@url [0]{\catcode `\\12\catcode `\$12\catcode
  `\&12\catcode `\#12\catcode `\^12\catcode `\_12\catcode `\%12\relax}%
\providecommand \@@startlink[1]{}%
\providecommand \@@endlink[0]{}%
\providecommand \url  [0]{\begingroup\@sanitize@url \@url }%
\providecommand \@url [1]{\endgroup\@href {#1}{\urlprefix }}%
\providecommand \urlprefix  [0]{URL }%
\providecommand \Eprint [0]{\href }%
\providecommand \doibase [0]{http://dx.doi.org/}%
\providecommand \selectlanguage [0]{\@gobble}%
\providecommand \bibinfo  [0]{\@secondoftwo}%
\providecommand \bibfield  [0]{\@secondoftwo}%
\providecommand \translation [1]{[#1]}%
\providecommand \BibitemOpen [0]{}%
\providecommand \bibitemStop [0]{}%
\providecommand \bibitemNoStop [0]{.\EOS\space}%
\providecommand \EOS [0]{\spacefactor3000\relax}%
\providecommand \BibitemShut  [1]{\csname bibitem#1\endcsname}%
\let\auto@bib@innerbib\@empty
\bibitem [{\citenamefont {de~Broglie}(1923)}]{debroglie1923}%
  \BibitemOpen
  \bibfield  {author} {\bibinfo {author} {\bibfnamefont {L.}~\bibnamefont
  {de~Broglie}},\ }\bibfield  {title} {\enquote {\bibinfo {title} {Onde et
  quanta},}\ }\href@noop {} {\bibfield  {journal} {\bibinfo  {journal} {Comptes
  rendus}\ }\textbf {\bibinfo {volume} {177}},\ \bibinfo {pages} {507}
  (\bibinfo {year} {1923})},\ \bibinfo {note} {\protect{An English translation
  of this paper can be found on the internet at the URL
  http://www.davis-inc.com/physics/rendus-e.pdf}}\BibitemShut {NoStop}%
\bibitem [{\citenamefont {\protect{E.
  Schr\"{o}dinger}}(1926)}]{Schrodinger1926a}%
  \BibitemOpen
  \bibfield  {author} {\bibinfo {author} {\bibnamefont {\protect{E.
  Schr\"{o}dinger}}},\ }\bibfield  {title} {\enquote {\bibinfo {title}
  {Quantizierung als eigenwertproblem (\protect{Erste Mitteilung})},}\
  }\href@noop {} {\bibfield  {journal} {\bibinfo  {journal} {Ann. Phys}\
  }\textbf {\bibinfo {volume} {79}},\ \bibinfo {pages} {361--376} (\bibinfo
  {year} {1926})}\BibitemShut {NoStop}%
\bibitem [{\citenamefont {Thomas}(1927)}]{thomas1927}%
  \BibitemOpen
  \bibfield  {author} {\bibinfo {author} {\bibfnamefont {Llewellyn~H}\
  \bibnamefont {Thomas}},\ }\bibfield  {title} {\enquote {\bibinfo {title} {The
  calculation of atomic fields},}\ }\href@noop {} {\bibfield  {journal}
  {\bibinfo  {journal} {Proc. Camb. Phil. Soc}\ }\textbf {\bibinfo {volume}
  {23}},\ \bibinfo {pages} {542--548} (\bibinfo {year} {1927})}\BibitemShut
  {NoStop}%
\bibitem [{\citenamefont {Fermi}(1928)}]{fermi1928}%
  \BibitemOpen
  \bibfield  {author} {\bibinfo {author} {\bibfnamefont {E}~\bibnamefont
  {Fermi}},\ }\bibfield  {title} {\enquote {\bibinfo {title} {A statistical
  method for determining some properties of the atoms and its application to
  the theory of the periodic table of elements},}\ }\href@noop {} {\bibfield
  {journal} {\bibinfo  {journal} {Z. Phys}\ }\textbf {\bibinfo {volume} {48}},\
  \bibinfo {pages} {73--79} (\bibinfo {year} {1928})}\BibitemShut {NoStop}%
\bibitem [{\citenamefont {Dirac}(1930)}]{dirac1930}%
  \BibitemOpen
  \bibfield  {author} {\bibinfo {author} {\bibfnamefont {P.~A.~M.}\
  \bibnamefont {Dirac}},\ }\bibfield  {title} {\enquote {\bibinfo {title} {Note
  on exchange phenomena in the \protect{Thomas} atom},}\ }\href@noop {}
  {\bibfield  {journal} {\bibinfo  {journal} {Proc. Camb. Phil. Soc}\ }\textbf
  {\bibinfo {volume} {26}},\ \bibinfo {pages} {376--385} (\bibinfo {year}
  {1930})}\BibitemShut {NoStop}%
\bibitem [{\citenamefont {Wilson}(1962)}]{Wilson1962}%
  \BibitemOpen
  \bibfield  {author} {\bibinfo {author} {\bibfnamefont {E.~B.}\ \bibnamefont
  {Wilson}},\ }\bibfield  {title} {\enquote {\bibinfo {title} {Four-dimensional
  electron density function},}\ }\href@noop {} {\bibfield  {journal} {\bibinfo
  {journal} {J. Chem. Phys.}\ }\textbf {\bibinfo {volume} {36}},\ \bibinfo
  {pages} {2232--2233} (\bibinfo {year} {1962})}\BibitemShut {NoStop}%
\bibitem [{\citenamefont {Hohenberg}\ and\ \citenamefont
  {Kohn}(1964)}]{Hohenberg_and_Kohn1964}%
  \BibitemOpen
  \bibfield  {author} {\bibinfo {author} {\bibfnamefont {P.}~\bibnamefont
  {Hohenberg}}\ and\ \bibinfo {author} {\bibfnamefont {W.}~\bibnamefont
  {Kohn}},\ }\bibfield  {title} {\enquote {\bibinfo {title} {Inhomogeneous
  electron gas},}\ }\href@noop {} {\bibfield  {journal} {\bibinfo  {journal}
  {Phys. Rev.}\ }\textbf {\bibinfo {volume} {136}},\ \bibinfo {pages} {B864}
  (\bibinfo {year} {1964})}\BibitemShut {NoStop}%
\bibitem [{\citenamefont {Epstein}\ and\ \citenamefont
  {Rosenthal}(1976)}]{Epstein_and_Rosenthal1976}%
  \BibitemOpen
  \bibfield  {author} {\bibinfo {author} {\bibfnamefont {S.~T.}\ \bibnamefont
  {Epstein}}\ and\ \bibinfo {author} {\bibfnamefont {C.~M.}\ \bibnamefont
  {Rosenthal}},\ }\bibfield  {title} {\enquote {\bibinfo {title} {The
  \protect{Hohenberg-Kohn} theorem},}\ }\href@noop {} {\bibfield  {journal}
  {\bibinfo  {journal} {J. Chem. Phys.}\ ,\ \bibinfo {pages} {247--249}}
  (\bibinfo {year} {1976})}\BibitemShut {NoStop}%
\bibitem [{\citenamefont {Kohn}\ and\ \citenamefont
  {Sham}(1965)}]{Kohn_and_Sham1965}%
  \BibitemOpen
  \bibfield  {author} {\bibinfo {author} {\bibfnamefont {W.}~\bibnamefont
  {Kohn}}\ and\ \bibinfo {author} {\bibfnamefont {L.~J.}\ \bibnamefont
  {Sham}},\ }\bibfield  {title} {\enquote {\bibinfo {title} {Self-consistent
  equations including exchange and correlation effects},}\ }\href@noop {}
  {\bibfield  {journal} {\bibinfo  {journal} {Phys. Rev.}\ ,\ \bibinfo {pages}
  {A1133}} (\bibinfo {year} {1965})}\BibitemShut {NoStop}%
\bibitem [{\citenamefont {Parr}\ and\ \citenamefont
  {Yang}(1989)}]{Parr_and_Yang1989}%
  \BibitemOpen
  \bibfield  {author} {\bibinfo {author} {\bibfnamefont {R.~G.}\ \bibnamefont
  {Parr}}\ and\ \bibinfo {author} {\bibfnamefont {W.}~\bibnamefont {Yang}},\
  }\href@noop {} {\emph {\bibinfo {title} {Density-Functional Theory of Atoms
  and Molecules}}}\ (\bibinfo  {publisher} {Oxford U. Press},\ \bibinfo
  {address} {New York, NY},\ \bibinfo {year} {1989})\BibitemShut {NoStop}%
\bibitem [{\citenamefont {Kohn}\ \emph {et~al.}(1996)\citenamefont {Kohn},
  \citenamefont {Becke},\ and\ \citenamefont {Parr}}]{Kohn_etal1996}%
  \BibitemOpen
  \bibfield  {author} {\bibinfo {author} {\bibfnamefont {W.}~\bibnamefont
  {Kohn}}, \bibinfo {author} {\bibfnamefont {A.~D.}\ \bibnamefont {Becke}}, \
  and\ \bibinfo {author} {\bibfnamefont {R.~G.}\ \bibnamefont {Parr}},\
  }\bibfield  {title} {\enquote {\bibinfo {title} {Density functional theory of
  electronic structure},}\ }\href@noop {} {\bibfield  {journal} {\bibinfo
  {journal} {J. Phys. Chem.}\ }\textbf {\bibinfo {volume} {100}},\ \bibinfo
  {pages} {12974--12980} (\bibinfo {year} {1996})},\ \bibinfo {note} {and
  references cited therein.}\BibitemShut {Stop}%
\bibitem [{\citenamefont {Becke}(2014)}]{becke2014}%
  \BibitemOpen
  \bibfield  {author} {\bibinfo {author} {\bibfnamefont {Axel~D}\ \bibnamefont
  {Becke}},\ }\bibfield  {title} {\enquote {\bibinfo {title} {Perspective:
  Fifty years of density-functional theory in chemical physics},}\ }\href@noop
  {} {\bibfield  {journal} {\bibinfo  {journal} {J. Chem. Phys.}\ }\textbf
  {\bibinfo {volume} {140}},\ \bibinfo {pages} {18A301} (\bibinfo {year}
  {2014})}\BibitemShut {NoStop}%
\bibitem [{\citenamefont {Perdew}\ \emph {et~al.}(2009)\citenamefont {Perdew},
  \citenamefont {Ruzsinszky}, \citenamefont {Constantin}, \citenamefont {Sun},\
  and\ \citenamefont {Csonka}}]{perdew2009}%
  \BibitemOpen
  \bibfield  {author} {\bibinfo {author} {\bibfnamefont {John~P}\ \bibnamefont
  {Perdew}}, \bibinfo {author} {\bibfnamefont {Adrienn}\ \bibnamefont
  {Ruzsinszky}}, \bibinfo {author} {\bibfnamefont {Lucian~A}\ \bibnamefont
  {Constantin}}, \bibinfo {author} {\bibfnamefont {Jianwei}\ \bibnamefont
  {Sun}}, \ and\ \bibinfo {author} {\bibfnamefont {G{\'a}bor~I}\ \bibnamefont
  {Csonka}},\ }\bibfield  {title} {\enquote {\bibinfo {title} {Some fundamental
  issues in ground-state density functional theory: A guide for the
  perplexed},}\ }\href@noop {} {\bibfield  {journal} {\bibinfo  {journal} {J.
  Chem. Theory and Computation}\ }\textbf {\bibinfo {volume} {5}},\ \bibinfo
  {pages} {902--908} (\bibinfo {year} {2009})}\BibitemShut {NoStop}%
\bibitem [{\citenamefont {Roothaan}(1951)}]{roothaan1951}%
  \BibitemOpen
  \bibfield  {author} {\bibinfo {author} {\bibfnamefont {C.~C.~J.}\
  \bibnamefont {Roothaan}},\ }\bibfield  {title} {\enquote {\bibinfo {title}
  {New developments in molecular orbital theory},}\ }\href@noop {} {\bibfield
  {journal} {\bibinfo  {journal} {Rev. Mod. Phys.}\ }\textbf {\bibinfo {volume}
  {23}},\ \bibinfo {pages} {69} (\bibinfo {year} {1951})}\BibitemShut {NoStop}%
\bibitem [{\citenamefont {Froese-Fischer}(1977)}]{froese-fischer1977}%
  \BibitemOpen
  \bibfield  {author} {\bibinfo {author} {\bibfnamefont {Charlotte}\
  \bibnamefont {Froese-Fischer}},\ }\href@noop {} {\emph {\bibinfo {title} {The
  Hartree-Fock method for atoms: a numerical approach}}}\ (\bibinfo
  {publisher} {Wiley},\ \bibinfo {year} {1977})\BibitemShut {NoStop}%
\bibitem [{\citenamefont {Lign{\`e}res}\ and\ \citenamefont
  {Carter}(2005)}]{ligneres_and_carter2005}%
  \BibitemOpen
  \bibfield  {author} {\bibinfo {author} {\bibfnamefont {Vincent~L}\
  \bibnamefont {Lign{\`e}res}}\ and\ \bibinfo {author} {\bibfnamefont
  {Emily~A}\ \bibnamefont {Carter}},\ }\bibfield  {title} {\enquote {\bibinfo
  {title} {An introduction to orbital-free density functional theory},}\ }in\
  \href@noop {} {\emph {\bibinfo {booktitle} {Handbook of Materials
  Modeling}}}\ (\bibinfo  {publisher} {Springer},\ \bibinfo {year} {2005})\
  pp.\ \bibinfo {pages} {137--148}\BibitemShut {NoStop}%
\bibitem [{\citenamefont {Slater}(1929)}]{slater1929}%
  \BibitemOpen
  \bibfield  {author} {\bibinfo {author} {\bibfnamefont {John~C}\ \bibnamefont
  {Slater}},\ }\bibfield  {title} {\enquote {\bibinfo {title} {The theory of
  complex spectra},}\ }\href@noop {} {\bibfield  {journal} {\bibinfo  {journal}
  {Phys. Rev.}\ }\textbf {\bibinfo {volume} {34}},\ \bibinfo {pages} {1293}
  (\bibinfo {year} {1929})}\BibitemShut {NoStop}%
\bibitem [{\citenamefont {Becke}(1988)}]{becke1988density}%
  \BibitemOpen
  \bibfield  {author} {\bibinfo {author} {\bibfnamefont {Axel~D}\ \bibnamefont
  {Becke}},\ }\bibfield  {title} {\enquote {\bibinfo {title}
  {Density-functional exchange-energy approximation with correct asymptotic
  behavior},}\ }\href@noop {} {\bibfield  {journal} {\bibinfo  {journal} {Phys.
  Rev. A}\ }\textbf {\bibinfo {volume} {38}},\ \bibinfo {pages} {3098}
  (\bibinfo {year} {1988})}\BibitemShut {NoStop}%
\bibitem [{\citenamefont {Perdew}\ \emph {et~al.}(1992)\citenamefont {Perdew},
  \citenamefont {Chevary}, \citenamefont {Vosko}, \citenamefont {Jackson},
  \citenamefont {Pederson}, \citenamefont {Singh},\ and\ \citenamefont
  {Fiolhais}}]{Perdew_et_al1992}%
  \BibitemOpen
  \bibfield  {author} {\bibinfo {author} {\bibfnamefont {John~P}\ \bibnamefont
  {Perdew}}, \bibinfo {author} {\bibfnamefont {John~A}\ \bibnamefont
  {Chevary}}, \bibinfo {author} {\bibfnamefont {Sy~H}\ \bibnamefont {Vosko}},
  \bibinfo {author} {\bibfnamefont {Koblar~A}\ \bibnamefont {Jackson}},
  \bibinfo {author} {\bibfnamefont {Mark~R}\ \bibnamefont {Pederson}}, \bibinfo
  {author} {\bibfnamefont {Dig~J}\ \bibnamefont {Singh}}, \ and\ \bibinfo
  {author} {\bibfnamefont {Carlos}\ \bibnamefont {Fiolhais}},\ }\bibfield
  {title} {\enquote {\bibinfo {title} {Atoms, molecules, solids, and surfaces:
  Applications of the generalized gradient approximation for exchange and
  correlation},}\ }\href@noop {} {\bibfield  {journal} {\bibinfo  {journal}
  {Phys. Rev. B}\ }\textbf {\bibinfo {volume} {46}},\ \bibinfo {pages} {6671}
  (\bibinfo {year} {1992})}\BibitemShut {NoStop}%
\bibitem [{\citenamefont {Perdew}\ \emph {et~al.}(1996)\citenamefont {Perdew},
  \citenamefont {Burke},\ and\ \citenamefont {Ernzerhof}}]{Perdew_et_al1996}%
  \BibitemOpen
  \bibfield  {author} {\bibinfo {author} {\bibfnamefont {John~P}\ \bibnamefont
  {Perdew}}, \bibinfo {author} {\bibfnamefont {Kieron}\ \bibnamefont {Burke}},
  \ and\ \bibinfo {author} {\bibfnamefont {Matthias}\ \bibnamefont
  {Ernzerhof}},\ }\bibfield  {title} {\enquote {\bibinfo {title} {Generalized
  gradient approximation made simple},}\ }\href@noop {} {\bibfield  {journal}
  {\bibinfo  {journal} {Phys. Rev. Lett.}\ }\textbf {\bibinfo {volume} {77}},\
  \bibinfo {pages} {3865} (\bibinfo {year} {1996})}\BibitemShut {NoStop}%
\bibitem [{\citenamefont {Pauling}\ and\ \citenamefont
  {Wilson}(1935)}]{Pauling_and_Wilson1935}%
  \BibitemOpen
  \bibfield  {author} {\bibinfo {author} {\bibfnamefont {L.}~\bibnamefont
  {Pauling}}\ and\ \bibinfo {author} {\bibfnamefont {E.~B.}\ \bibnamefont
  {Wilson}},\ }\href@noop {} {\emph {\bibinfo {title} {Introduction to Quantum
  Mechanics with Applications to Chemistry}}}\ (\bibinfo  {publisher}
  {McGraw-Hill},\ \bibinfo {address} {New York},\ \bibinfo {year} {1935})\
  \bibinfo {note} {\protect{republished by Dover, New York, 1985.}}\BibitemShut
  {Stop}%
\bibitem [{\citenamefont {Karplus}\ and\ \citenamefont
  {Porter}(1970)}]{Karplus_and_Porter1970}%
  \BibitemOpen
  \bibfield  {author} {\bibinfo {author} {\bibfnamefont {M.}~\bibnamefont
  {Karplus}}\ and\ \bibinfo {author} {\bibfnamefont {R.~N.}\ \bibnamefont
  {Porter}},\ }\href@noop {} {\emph {\bibinfo {title} {Atoms and Molecules}}}\
  (\bibinfo  {publisher} {Addison-Wesley},\ \bibinfo {address} {New York, NY},\
  \bibinfo {year} {1970})\BibitemShut {NoStop}%
\bibitem [{\citenamefont {Pilar}(1990)}]{Pilar1990}%
  \BibitemOpen
  \bibfield  {author} {\bibinfo {author} {\bibfnamefont {F.~L.}\ \bibnamefont
  {Pilar}},\ }\href@noop {} {\emph {\bibinfo {title} {Elementary Quantum
  Chemistry}}}\ (\bibinfo  {publisher} {Dover},\ \bibinfo {address} {New York,
  NY},\ \bibinfo {year} {1990})\ \bibinfo {note} {\protect{S}ee esp. pp.
  161-163.}\BibitemShut {Stop}%
\bibitem [{\citenamefont {Kramida}\ \emph {et~al.}(2023)\citenamefont
  {Kramida}, \citenamefont {Ralchenko},\ and\ \citenamefont
  {Reader}}]{NIST_AtomicIPs}%
  \BibitemOpen
  \bibinfo {editor} {\bibfnamefont {A.}~\bibnamefont {Kramida}}, \bibinfo
  {editor} {\bibfnamefont {Y.}~\bibnamefont {Ralchenko}}, \ and\ \bibinfo
  {editor} {\bibfnamefont {J.}~\bibnamefont {Reader}},\ eds.,\ \href@noop {}
  {\emph {\bibinfo {title} {\uppercase{NIST} \uppercase{A}tomic
  \uppercase{S}pectra \uppercase{D}atabase}}},\ NIST Standard Reference
  Database No. 78\ (\bibinfo  {publisher} {\protect{Natl. Inst. of Standards
  and Technology}},\ \bibinfo {year} {2023})\ \bibinfo {note}
  {\url{https://physics.nist.gov/asd}}\BibitemShut {NoStop}%
\bibitem [{\citenamefont {Clementi}\ and\ \citenamefont
  {Roetti}(1974)}]{Clementi_and_Roetti1974}%
  \BibitemOpen
  \bibfield  {author} {\bibinfo {author} {\bibfnamefont {Enrico}\ \bibnamefont
  {Clementi}}\ and\ \bibinfo {author} {\bibfnamefont {Carla}\ \bibnamefont
  {Roetti}},\ }\bibfield  {title} {\enquote {\bibinfo {title}
  {\protect{Roothaan-Hartree-Fock} atomic wavefunctions: Basis functions and
  their coefficients for ground and certain excited states of neutral and
  ionized atoms, \protect{Z}$\ensuremath{\leq}$54},}\ }\href@noop {} {\bibfield
   {journal} {\bibinfo  {journal} {At. Data Nucl. Data Tables}\ }\textbf
  {\bibinfo {volume} {14}},\ \bibinfo {pages} {177--478} (\bibinfo {year}
  {1974})}\BibitemShut {NoStop}%
\bibitem [{\citenamefont {Bunge}\ \emph {et~al.}(1993)\citenamefont {Bunge},
  \citenamefont {Barrientos},\ and\ \citenamefont {Bunge}}]{bunge1993}%
  \BibitemOpen
  \bibfield  {author} {\bibinfo {author} {\bibfnamefont {Carlos~F}\
  \bibnamefont {Bunge}}, \bibinfo {author} {\bibfnamefont {Jose~A}\
  \bibnamefont {Barrientos}}, \ and\ \bibinfo {author} {\bibfnamefont
  {A~Vivier}\ \bibnamefont {Bunge}},\ }\bibfield  {title} {\enquote {\bibinfo
  {title} {\protect{Roothaan-Hartree-Fock} ground-state atomic wave functions:
  Slater-type orbital expansions and expectation values for
  \protect{Z=2-54}},}\ }\href@noop {} {\bibfield  {journal} {\bibinfo
  {journal} {At. Data Nucl. Data Tables}\ }\textbf {\bibinfo {volume} {53}},\
  \bibinfo {pages} {113--162} (\bibinfo {year} {1993})},\ \bibinfo {note}
  {\protect{Data from this paper is available online at:
  \url{http://server.ccl.net/cca/data/atomic-RHF-wavefunctions/tables}}}\BibitemShut
  {NoStop}%
\bibitem [{\citenamefont {Pauling}(1927)}]{pauling1927}%
  \BibitemOpen
  \bibfield  {author} {\bibinfo {author} {\bibfnamefont {Linus}\ \bibnamefont
  {Pauling}},\ }\bibfield  {title} {\enquote {\bibinfo {title} {Die
  abschirmungskonstanten der relativistischen oder magnetischen
  r{\"o}ntgenstrahlendubletts},}\ }\href@noop {} {\bibfield  {journal}
  {\bibinfo  {journal} {Z. Phys.}\ }\textbf {\bibinfo {volume} {40}},\ \bibinfo
  {pages} {344--350} (\bibinfo {year} {1927})}\BibitemShut {NoStop}%
\bibitem [{\citenamefont {Slater}(1930)}]{slater1930}%
  \BibitemOpen
  \bibfield  {author} {\bibinfo {author} {\bibfnamefont {John~C}\ \bibnamefont
  {Slater}},\ }\bibfield  {title} {\enquote {\bibinfo {title} {Atomic shielding
  constants},}\ }\href@noop {} {\bibfield  {journal} {\bibinfo  {journal}
  {Phys. Rev.}\ }\textbf {\bibinfo {volume} {36}},\ \bibinfo {pages} {57}
  (\bibinfo {year} {1930})}\BibitemShut {NoStop}%
\bibitem [{\citenamefont {Eckart}(1930)}]{eckart1930}%
  \BibitemOpen
  \bibfield  {author} {\bibinfo {author} {\bibfnamefont {Carl}\ \bibnamefont
  {Eckart}},\ }\bibfield  {title} {\enquote {\bibinfo {title} {The theory and
  calculation of screening constants},}\ }\href@noop {} {\bibfield  {journal}
  {\bibinfo  {journal} {Phys. Rev.}\ }\textbf {\bibinfo {volume} {36}},\
  \bibinfo {pages} {878} (\bibinfo {year} {1930})}\BibitemShut {NoStop}%
\bibitem [{\citenamefont {Zener}(1930)}]{zener1930}%
  \BibitemOpen
  \bibfield  {author} {\bibinfo {author} {\bibfnamefont {Clarence}\
  \bibnamefont {Zener}},\ }\bibfield  {title} {\enquote {\bibinfo {title}
  {Analytic atomic wave functions},}\ }\href@noop {} {\bibfield  {journal}
  {\bibinfo  {journal} {Phys. Rev.}\ }\textbf {\bibinfo {volume} {36}},\
  \bibinfo {pages} {51} (\bibinfo {year} {1930})}\BibitemShut {NoStop}%
\bibitem [{\citenamefont {Clementi}\ and\ \citenamefont
  {Raimondi}(1963)}]{Clementi_and_Raimondi1963}%
  \BibitemOpen
  \bibfield  {author} {\bibinfo {author} {\bibfnamefont {E.}~\bibnamefont
  {Clementi}}\ and\ \bibinfo {author} {\bibfnamefont {D.~L.}\ \bibnamefont
  {Raimondi}},\ }\bibfield  {title} {\enquote {\bibinfo {title} {Atomic
  screening constants from \protect{SCF} functions},}\ }\href@noop {}
  {\bibfield  {journal} {\bibinfo  {journal} {J. Chem. Phys.}\ }\textbf
  {\bibinfo {volume} {38}},\ \bibinfo {pages} {2686--2689} (\bibinfo {year}
  {1963})}\BibitemShut {NoStop}%
\bibitem [{\citenamefont {Bessis}\ and\ \citenamefont
  {Bessis}(1981)}]{bessis1981}%
  \BibitemOpen
  \bibfield  {author} {\bibinfo {author} {\bibfnamefont {N}~\bibnamefont
  {Bessis}}\ and\ \bibinfo {author} {\bibfnamefont {G}~\bibnamefont {Bessis}},\
  }\bibfield  {title} {\enquote {\bibinfo {title} {Analytic atomic shielding
  parameters},}\ }\href@noop {} {\bibfield  {journal} {\bibinfo  {journal} {J.
  Chem. Phys.}\ }\textbf {\bibinfo {volume} {74}},\ \bibinfo {pages}
  {3628--3630} (\bibinfo {year} {1981})}\BibitemShut {NoStop}%
\bibitem [{\citenamefont {Halliday}\ \emph {et~al.}(2001)\citenamefont
  {Halliday}, \citenamefont {Resnick},\ and\ \citenamefont
  {Walker}}]{Halliday_and_Resnick2001}%
  \BibitemOpen
  \bibfield  {author} {\bibinfo {author} {\bibfnamefont {D.}~\bibnamefont
  {Halliday}}, \bibinfo {author} {\bibfnamefont {R.}~\bibnamefont {Resnick}}, \
  and\ \bibinfo {author} {\bibfnamefont {J.}~\bibnamefont {Walker}},\
  }\href@noop {} {\emph {\bibinfo {title} {Fundamentals of Physics}}},\
  \bibinfo {edition} {sixth}\ ed.\ (\bibinfo  {publisher} {Wiley},\ \bibinfo
  {address} {New York, NY},\ \bibinfo {year} {2001})\BibitemShut {NoStop}%
\bibitem [{\citenamefont {Morrell}\ \emph {et~al.}(1975)\citenamefont
  {Morrell}, \citenamefont {Parr},\ and\ \citenamefont
  {Levy}}]{Morrell_etal1975}%
  \BibitemOpen
  \bibfield  {author} {\bibinfo {author} {\bibfnamefont {M.~M.}\ \bibnamefont
  {Morrell}}, \bibinfo {author} {\bibfnamefont {R.~G.}\ \bibnamefont {Parr}}, \
  and\ \bibinfo {author} {\bibfnamefont {M.}~\bibnamefont {Levy}},\ }\bibfield
  {title} {\enquote {\bibinfo {title} {Calculation of ionization potentials
  from density matrices and natural functions, and the long-range behavior of
  natural orbitals and electron density},}\ }\href@noop {} {\bibfield
  {journal} {\bibinfo  {journal} {J. Chem. Phys}\ }\textbf {\bibinfo {volume}
  {62}},\ \bibinfo {pages} {549--554} (\bibinfo {year} {1975})}\BibitemShut
  {NoStop}%
\bibitem [{\citenamefont {Katriel}\ and\ \citenamefont
  {Davidson}(1980)}]{Katriel_and_Davidson1980}%
  \BibitemOpen
  \bibfield  {author} {\bibinfo {author} {\bibfnamefont {J.}~\bibnamefont
  {Katriel}}\ and\ \bibinfo {author} {\bibfnamefont {E.~R.}\ \bibnamefont
  {Davidson}},\ }\bibfield  {title} {\enquote {\bibinfo {title} {Asymptotic
  behavior of atomic and molecular wave functions},}\ }\href@noop {} {\bibfield
   {journal} {\bibinfo  {journal} {Proc. Natl. Acad. Sci.}\ }\textbf {\bibinfo
  {volume} {77}},\ \bibinfo {pages} {4403--4406} (\bibinfo {year}
  {1980})}\BibitemShut {NoStop}%
\bibitem [{\citenamefont {Kato}(1957)}]{kato1957}%
  \BibitemOpen
  \bibfield  {author} {\bibinfo {author} {\bibfnamefont {Tosio}\ \bibnamefont
  {Kato}},\ }\bibfield  {title} {\enquote {\bibinfo {title} {On the
  eigenfunctions of many-particle systems in quantum mechanics},}\ }\href@noop
  {} {\bibfield  {journal} {\bibinfo  {journal} {Commun. Pure and Appl. Math.}\
  }\textbf {\bibinfo {volume} {10}},\ \bibinfo {pages} {151--177} (\bibinfo
  {year} {1957})}\BibitemShut {NoStop}%
\bibitem [{\citenamefont {March}(1986)}]{march1986}%
  \BibitemOpen
  \bibfield  {author} {\bibinfo {author} {\bibfnamefont {NH}~\bibnamefont
  {March}},\ }\bibfield  {title} {\enquote {\bibinfo {title} {Spatially
  dependent generalization of \protect{Kato's} theorem for atomic closed shells
  in a bare coulomb field},}\ }\href@noop {} {\bibfield  {journal} {\bibinfo
  {journal} {Phys. Rev. A}\ }\textbf {\bibinfo {volume} {33}},\ \bibinfo
  {pages} {88} (\bibinfo {year} {1986})}\BibitemShut {NoStop}%
\bibitem [{\citenamefont {Coleman}(1963)}]{Coleman1963}%
  \BibitemOpen
  \bibfield  {author} {\bibinfo {author} {\bibfnamefont {A.~J.}\ \bibnamefont
  {Coleman}},\ }\bibfield  {title} {\enquote {\bibinfo {title} {Structure of
  fermion density matrices},}\ }\href@noop {} {\bibfield  {journal} {\bibinfo
  {journal} {Rev. Mod. Phys.}\ }\textbf {\bibinfo {volume} {35}},\ \bibinfo
  {pages} {668} (\bibinfo {year} {1963})}\BibitemShut {NoStop}%
\bibitem [{\citenamefont {Davidson}(1976)}]{Davidson1976}%
  \BibitemOpen
  \bibfield  {author} {\bibinfo {author} {\bibfnamefont {E.~R.}\ \bibnamefont
  {Davidson}},\ }\href@noop {} {\emph {\bibinfo {title} {Reduced Density
  Matrices in Quantum Chemistry}}}\ (\bibinfo  {publisher} {Academic Press},\
  \bibinfo {address} {New York, NY},\ \bibinfo {year} {1976})\BibitemShut
  {NoStop}%
\bibitem [{\citenamefont {Gilbert}(1975)}]{gilbert1975}%
  \BibitemOpen
  \bibfield  {author} {\bibinfo {author} {\bibfnamefont {Thomas~L}\
  \bibnamefont {Gilbert}},\ }\bibfield  {title} {\enquote {\bibinfo {title}
  {\protect{Hohenberg-Kohn} theorem for nonlocal external potentials},}\
  }\href@noop {} {\bibfield  {journal} {\bibinfo  {journal} {Phys. Rev. B}\
  }\textbf {\bibinfo {volume} {12}},\ \bibinfo {pages} {2111} (\bibinfo {year}
  {1975})}\BibitemShut {NoStop}%
\bibitem [{\citenamefont {\protect{C. S. Johnson, Jr.}}\ and\ \citenamefont
  {Pedersen}(1974)}]{Johnson_and_Pedersen1974}%
  \BibitemOpen
  \bibfield  {author} {\bibinfo {author} {\bibnamefont {\protect{C. S. Johnson,
  Jr.}}}\ and\ \bibinfo {author} {\bibfnamefont {L.~G.}\ \bibnamefont
  {Pedersen}},\ }\href@noop {} {\emph {\bibinfo {title} {Problems and Solutions
  in Quantum Chemistry and Physics}}}\ (\bibinfo  {publisher}
  {Addison-Wesley},\ \bibinfo {address} {New York, NY},\ \bibinfo {year}
  {1974})\ \bibinfo {note} {\protect{republished by Dover, New York,
  1986.}}\BibitemShut {Stop}%
\bibitem [{\citenamefont {Lide}(2004)}]{CRC_Handbook2004}%
  \BibitemOpen
  \bibinfo {editor} {\bibfnamefont {D.~R.}\ \bibnamefont {Lide}},\ ed.,\
  \href@noop {} {\emph {\bibinfo {title} {Handbook of Chemistry and
  Physics}}},\ \bibinfo {edition} {85th}\ ed.\ (\bibinfo  {publisher} {CRC
  Press},\ \bibinfo {address} {Boca Raton, FL},\ \bibinfo {year} {2004})\
  \bibinfo {note} {\protect{See esp. tables of ionization potentials in
  \uppercase{S}ection 10.}}\BibitemShut {Stop}%
\bibitem [{\citenamefont {Ellenbogen}(2006)}]{Ellenbogen_PRA2006}%
  \BibitemOpen
  \bibfield  {author} {\bibinfo {author} {\bibfnamefont {J.~C.}\ \bibnamefont
  {Ellenbogen}},\ }\bibfield  {title} {\enquote {\bibinfo {title} {Neutral
  atoms behave much like classical spherical capacitors},}\ }\href@noop {}
  {\bibfield  {journal} {\bibinfo  {journal} {Phys. Rev. A}\ }\textbf {\bibinfo
  {volume} {74}},\ \bibinfo {pages} {034501} (\bibinfo {year}
  {2006})}\BibitemShut {NoStop}%
\bibitem [{\citenamefont {Bohr}(1913)}]{Bohr1913a}%
  \BibitemOpen
  \bibfield  {author} {\bibinfo {author} {\bibfnamefont {N.}~\bibnamefont
  {Bohr}},\ }\bibfield  {title} {\enquote {\bibinfo {title} {I. \protect{On}
  the constitution of atoms and molecules},}\ }\href@noop {} {\bibfield
  {journal} {\bibinfo  {journal} {The London, Edinburgh, and Dublin
  Philosophical Magazine and Journal of Science}\ }\textbf {\bibinfo {volume}
  {26}},\ \bibinfo {pages} {1--25} (\bibinfo {year} {1913})}\BibitemShut
  {NoStop}%
\bibitem [{\citenamefont {Taylor}\ and\ \citenamefont
  {Parr}(1952)}]{Taylor_and_Parr1952}%
  \BibitemOpen
  \bibfield  {author} {\bibinfo {author} {\bibfnamefont {G.R.}\ \bibnamefont
  {Taylor}}\ and\ \bibinfo {author} {\bibfnamefont {R.G.}\ \bibnamefont
  {Parr}},\ }\bibfield  {title} {\enquote {\bibinfo {title} {Superposition of
  configurations: the helium atom},}\ }\href@noop {} {\bibfield  {journal}
  {\bibinfo  {journal} {Proc. Natl. Acad. Sci.}\ }\textbf {\bibinfo {volume}
  {38}},\ \bibinfo {pages} {154--160} (\bibinfo {year} {1952})}\BibitemShut
  {NoStop}%
\bibitem [{\citenamefont {Weiss}(1961)}]{Weiss1961}%
  \BibitemOpen
  \bibfield  {author} {\bibinfo {author} {\bibfnamefont {A.~W.}\ \bibnamefont
  {Weiss}},\ }\bibfield  {title} {\enquote {\bibinfo {title} {Configuration
  interaction in simple atomic systems},}\ }\href@noop {} {\bibfield  {journal}
  {\bibinfo  {journal} {Phys. Rev.}\ }\textbf {\bibinfo {volume} {122}},\
  \bibinfo {pages} {1826} (\bibinfo {year} {1961})}\BibitemShut {NoStop}%
\bibitem [{\citenamefont {Pekeris}(1959)}]{pekeris1959}%
  \BibitemOpen
  \bibfield  {author} {\bibinfo {author} {\bibfnamefont {C.~L.}\ \bibnamefont
  {Pekeris}},\ }\bibfield  {title} {\enquote {\bibinfo {title} {$\ensuremath{1
  ^1S}$ and $\ensuremath{2 ^3S$ States of Helium}},}\ }\href@noop {} {\bibfield
   {journal} {\bibinfo  {journal} {Physical Review}\ }\textbf {\bibinfo
  {volume} {115}},\ \bibinfo {pages} {1216} (\bibinfo {year}
  {1959})}\BibitemShut {NoStop}%
\bibitem [{\citenamefont {Aznabaev}\ \emph {et~al.}(2018)\citenamefont
  {Aznabaev}, \citenamefont {Bekbaev},\ and\ \citenamefont
  {Korobov}}]{aznabaev_etal2018}%
  \BibitemOpen
  \bibfield  {author} {\bibinfo {author} {\bibfnamefont {DT}~\bibnamefont
  {Aznabaev}}, \bibinfo {author} {\bibfnamefont {AK}~\bibnamefont {Bekbaev}}, \
  and\ \bibinfo {author} {\bibfnamefont {Vladimir~I}\ \bibnamefont {Korobov}},\
  }\bibfield  {title} {\enquote {\bibinfo {title} {Nonrelativistic energy
  levels of helium atoms},}\ }\href@noop {} {\bibfield  {journal} {\bibinfo
  {journal} {Physical Review A}\ }\textbf {\bibinfo {volume} {98}},\ \bibinfo
  {pages} {012510} (\bibinfo {year} {2018})}\BibitemShut {NoStop}%
\bibitem [{\citenamefont {Strout}\ and\ \citenamefont
  {Scuseria}(1995)}]{Strout_and_Scuseria1995}%
  \BibitemOpen
  \bibfield  {author} {\bibinfo {author} {\bibfnamefont {D.~L.}\ \bibnamefont
  {Strout}}\ and\ \bibinfo {author} {\bibfnamefont {G.~E.}\ \bibnamefont
  {Scuseria}},\ }\bibfield  {title} {\enquote {\bibinfo {title} {A quantitative
  study of the scaling properties of the \protect{Hartree-Fock} method},}\
  }\href@noop {} {\bibfield  {journal} {\bibinfo  {journal} {J. Chem. Phys.}\
  }\textbf {\bibinfo {volume} {102}},\ \bibinfo {pages} {8448--8452} (\bibinfo
  {year} {1995})}\BibitemShut {NoStop}%
\bibitem [{\citenamefont {Spiegel}\ and\ \citenamefont
  {Stephens}(2008)}]{SchaumStatistics}%
  \BibitemOpen
  \bibfield  {author} {\bibinfo {author} {\bibfnamefont {Murray~R}\
  \bibnamefont {Spiegel}}\ and\ \bibinfo {author} {\bibfnamefont {Larry~J}\
  \bibnamefont {Stephens}},\ }\href@noop {} {\emph {\bibinfo {title}
  {Schaum’s Outline of Theory and Problems of Statistics}}}\ (\bibinfo
  {publisher} {McGraw-Hill: New York},\ \bibinfo {year} {2008})\ \bibinfo
  {note} {\protect{\!\!, see esp. Chapt. 14, p. 348-349.}}\BibitemShut {Stop}%
\end{thebibliography}%


\end{document}